\documentclass[hyper,11pt,paper]{article}
\usepackage{jheppub}
\usepackage[usenames,dvipsnames,table]{xcolor}
\usepackage{graphicx,amsmath,amssymb,multirow,array,bm,mathrsfs,bbm}
\usepackage{epsf,amsfonts,slashed,soul}
\usepackage[numbers,sort&compress]{natbib}
\usepackage{comment}
\usepackage[bbgreekl]{mathbbol}
\usepackage{xfrac,faktor}
\newcommand{\beq}{\begin{equation}}
\newcommand{\eeq}{\end{equation}}

\title{
A theory of reparameterizations for AdS$_3$ gravity
}

\author[a]{Jordan Cotler}
\author[b]{and Kristan Jensen}
\affiliation[a]{Stanford Institute for Theoretical Physics, Stanford University, Stanford, CA 94305, USA}
\affiliation[b]{Department of Physics and Astronomy, San Francisco State University, San Francisco, CA 94132, USA}

\emailAdd{jcotler@stanford.edu}
\emailAdd{kristanj@sfsu.edu}

\abstract{
We rewrite the Chern-Simons description of pure gravity on global AdS$_3$ and on Euclidean BTZ black holes as a quantum field theory on the AdS boundary. The resulting theory is (two copies of) the path integral quantization of a certain coadjoint orbit of the Virasoro group, and it should be regarded as the quantum field theory of the boundary gravitons. This theory respects all of the conformal field theory axioms except one: it is not modular invariant. The coupling constant is $1/c$ with $c$ the central charge, and perturbation theory in $1/c$ encodes loop contributions in the gravity dual. The QFT is a theory of reparametrizations analogous to the Schwarzian description of nearly AdS$_2$ gravity, and has several features including: (i) it is ultraviolet-complete; (ii) the torus partition function is the vacuum Virasoro character, which is one-loop exact by a localization argument; (iii) it reduces to the Schwarzian theory upon compactification; (iv) it provides a powerful new tool for computing Virasoro blocks at large $c$ via a diagrammatic expansion. We use the theory to compute several observables to one-loop order in the bulk, including the ``heavy-light'' limit of the identity block. We also work out some generalizations of this theory, including the boundary theory which describes fluctuations around two-sided eternal black holes.
}

\begin{document}

\maketitle

\section{Introduction}

There has been significant effort devoted to answer the question of whether pure three-dimensional gravity with a negative cosmological constant (``pure AdS$_3$ gravity'') is dual to a unitary two-dimensional conformal field theory (CFT). Relatedly, is pure AdS$_3$ gravity a consistent quantum theory? Strong evidence suggests that the answer to both questions is no. For example, consider the Euclidean partition function $\mathcal{Z}$ of pure gravity with torus boundary. If pure AdS$_3$ gravity has a CFT dual, then $\mathcal{Z}$ is the torus partition function of the CFT, and it must satisfy all of the axioms of CFT. In particular, $\mathcal{Z}$ should admit a power series expansion at small $q$ with positive integer coefficients which count the number of states with a given energy and momentum.

To compute this partition function one must make an ansatz for the integration contour of the path integral of three-dimensional quantum gravity. For a torus boundary of complex structure $\tau$, Maloney and Witten made a reasonable proposal~\cite{Maloney:2007ud}, which amounts to summing over fluctuations around real, non-singular, Euclidean saddles with boundary complex structure $\tau$. This prescription includes a sum over ``topologies'': in order to specify a Euclidean AdS$_3$ geometry with torus boundary one must choose a cycle of the boundary torus which is contractible in the bulk, and one sums over this choice. By a suitable modular transformation $\tau \to \tau'$ one may always take this contractible circle to be the spatial one. Within that convention the sum over ``topologies'' in the Maloney-Witten prescription becomes a modular sum.

There is a Euclidean BTZ geometry for any such $\tau'$. Its contribution to the bulk partition function (including the saddle and fluctuations around it) is the Virasoro character of the vacuum module at complex structure $\tau'$~\cite{Maloney:2007ud,Giombi:2008vd},
\beq
\label{E:ZMW1}
	\mathcal{Z}_{\rm vac}(\tau',\bar{\tau}') =\left| q'^{-k} \prod_{n=2}^{\infty} \frac{1}{1-q'^n}\right|^2\,, \qquad q' = e^{2\pi i \tau'}\,, \qquad k =\frac{c}{24}\,.
\eeq 
The states counted are the boundary gravitons and multiparticle states thereof. Here $c$ is the Brown-Henneaux central charge~\cite{Brown:1986nw}. So the ``Maloney-Witten'' partition function roughly takes the form
\beq
\label{E:ZMW2}
	\mathcal{Z}_{\rm MW}(\tau,\bar{\tau}) = \sum_{g\in PSL(2;\mathbb{Z})} \mathcal{Z}_{\rm vac}(g\tau,g\bar{\tau})\,,
\eeq
i.e., it is a sum of modular images of the vacuum character. This sum is divergent and one must regulate it. After employing a modular invariant regularization the resulting partition function no longer has a well-behaved small $q$ expansion: there are logarithmic terms in the expansion, in addition to coefficients which are neither positive nor integer. This cannot be the partition function of a unitary 2d CFT.

There are some loopholes to this argument.  For instance, the modular sum is sensible for extremely small values of the central charge~\cite{Castro:2011zq}. If one considers quantized $k=1,2,...$ and includes complex saddles, then the regularized sum is sensible as a CFT partition function and in fact is equal~\cite{Maloney:2009ck} to the ``extremal CFT'' partition function conjectured by Witten~\cite{Witten:2007kt}. For $k=1$ there is a known CFT with this partition function, the ``Monster CFT'' of Frenkel, Lepowsky, and Meurman~\cite{Frenkel:1988xz}, but there are no known examples of extremal CFT with $k\geq 2$ (and in fact there are good reasons to believe that they do not exist~\cite{Gaberdiel:2007ve,Gaiotto:2008jt}). However, these complex BTZ solutions do not have a good physical interpretation and it seems incorrect to include them.

Sometimes there is a rather different answer given to the question of whether AdS$_3$ gravity has a CFT dual. Namely, there is a dual, and that dual is Liouville theory. See~\cite{Coussaert:1995zp}, where a boundary Liouville theory was directly obtained from pure AdS$_3$ gravity on a Lorentzian cylinder. Of course this answer cannot be correct as stated -- Liouville theory has a continuous spectrum without a normalizable vacuum, while AdS$_3$ gravity has a normalizable vacuum and a one-loop computation~\cite{Giombi:2008vd} reveals a discrete spectrum of excitations around global AdS$_3$ given by the boundary gravitons and multiparticle states thereof. Moreover, there does not seem to be a clear relation between Liouville theory and the Euclidean results outlined above.

In this manuscript we show that, perhaps surprisingly, these two approaches to AdS$_3$ gravity are intimately related, and along the way clarify a number of extant puzzles.

Let us first recall the results of~\cite{Coussaert:1995zp}. Those authors begin with the Chern-Simons formulation of pure AdS$_3$ gravity on a cylinder, i.e.\!\! on global AdS$_3$. Crucially, the bulk Chern-Simons gauge fields are constrained by the boundary condition that the bulk spacetime is asymptotically AdS$_3$. Thus, rather than obtaining an ordinary chiral Wess-Zumino-Witten (WZW) model, one finds that the Chern-Simons theory is equivalent to a constrained $SO(2,1)\times SO(2,1)$ chiral WZW model on the boundary. The first factor is left-moving and the second right-moving. After a suitable field redefinition this theory can be written as an ordinary, albeit constrained $SO(2,1)$ WZW model, and this in turn can be rewritten as a Liouville theory.

However, in this derivation there is an important step which, while familiar from the canonical quantization of Chern-Simons theory~\cite{Elitzur:1989nr}, seems to have gone unnoticed in the AdS$_3$ literature. In the quantization of Chern-Simons theory with gauge group $G$ on a cylinder, one often separates the time direction from the spatial directions, $A = A_0 dt + \tilde{A}_i dx^i$, gauge-fixes $A_0 = 0$, and then the Chern-Simons path integral reduces to an integral over flat spatial connections $\tilde{A}(\vec{x},t)$. These may be parameterized in terms of a gauge transformation parameter $g(\vec{x},t)$ as
\beq
\label{E:decomposition}
 	\tilde{A}_i = g^{-1}\partial_i g\,,
\eeq
and the Chern-Simons action becomes a boundary chiral WZW action where the boundary value of $g$ is the WZW field. However, this is not the end of the story, as the decomposition~\eqref{E:decomposition} is redundant: both $g(\vec{x},t)$ and $h(t)g(\vec{x},t)$ give the same bulk field for any $h\in G$, and so we must identify these two field configurations in the path integral. In this way the boundary field $g$ is subject to a quasi-local quotient. This quotient is in some ways like a gauge symmetry, although without an accompanying gauge field.

We have performed the corresponding analysis for AdS$_3$ gravity. This analysis is slightly trickier on account of the asymptotically AdS$_3$ boundary conditions, which among other things forbid gauge-fixing $A_0$ to vanish. Nevertheless, we find that the boundary model is subject to a quasi-local $SO(2,1)\times SO(2,1)$ quotient.

After solving the constraints, the boundary model breaks apart into chiral halves, and we can discuss each separately. The left-moving part is an unconventional theory of a single scalar field $\phi$ with an action
\beq
\label{E:AStheory}
	S = -\frac{C}{24\pi}\int d^2x \left( \frac{(\partial_+ \phi') \phi''}{\phi'^2}-(\partial_+\phi)\phi'\right)\,, \qquad \partial_+= \frac{1}{2}(\partial_{\theta} + \partial_t)\,,
\eeq
where $\theta$ is an angular coordinate, the prime $'$ denotes an angular derivative, and $C=24k$ is the bare central charge. The field $\phi$ obeys an unconventional boundary condition $\phi(\theta+2\pi,t) = \phi(\theta,t)+2\pi$, and is subject to the quasi-local $SO(2,1) = PSL(2;\mathbb{R})$ redundancy
\beq
\label{E:ExcQuotient}
	\tan \left(\frac{\phi(\theta,t)}{2}\right)\sim \frac{a(t) \tan\left( \frac{\phi(\theta,t)}{2}\right)+b(t)}{c(t)\tan\left(\frac{\phi(\theta,t)}{2}\right) + d(t)}\,, \qquad \begin{pmatrix} a(t) & b(t) \\ c(t) & d(t) \end{pmatrix} \in PSL(2;\mathbb{R})\,.
\eeq
The boundary condition is a consequence of the spatial circle being contractible in the bulk. Taken together, the boundary condition and quotient on $\phi$ imply that at any fixed time, $\phi$ is an element of the quotient space $\text{Diff}(\mathbb{S}^1)/PSL(2;\mathbb{R})$. The right-moving half has a second scalar field $\tilde{\phi}$ whose action takes the same form with $\partial_+ \to \partial_-$. The left- and right-moving halves are tied together only by virtue of sharing the same boundary conditions -- namely, they both wind once around the spatial circle.

What is this boundary model? Eq.~\eqref{E:AStheory} is the quantum field theory of the (left-moving) boundary gravitons of pure AdS$_3$ gravity. Because the fundamental field $\phi$ is an element of Diff$(\mathbb{S}^1)$ (modulo the quotient) it may be understood as a theory of reparameterizations, thereby justifying the title of this manuscript.

This theory has appeared in the literature before. It is the Alekseev-Shatashvili path integral quantization of a certain coadjoint orbit, Diff$(\mathbb{S}^1)/PSL(2;\mathbb{R})$, of the Virasoro group~\cite{Alekseev:1988ce} (with a particular choice of Hamiltonian, as we explain later). In fact, this boundary quantum field theory is commensurate with the work of Maloney and Witten. In the course of arguing for~\eqref{E:ZMW1}, they showed that the classical phase space of smooth AdS$_3$ metrics continuously connected to global AdS$_3$ is precisely (two copies of) Diff$(\mathbb{S}^1)/PSL(2;\mathbb{R})$, which admits a geometric quantization described in~\cite{Witten:1987ty}. Their work establishes this geometric quantization as a boundary description of quantum gravity on global AdS$_3$. Our derivation of~\eqref{E:AStheory} is complementary. Whereas Maloney and Witten relied on Hamiltonian methods and geometric quantization, we use path integral methods throughout.

We also note that other authors have recently stated in words that a more careful version of the analysis of~\cite{Coussaert:1995zp} leads to the boundary model~\eqref{E:AStheory} -- see especially~\cite{Barnich:2017jgw,Mertens:2018fds}. However to our knowledge we are the first to give a path integral derivation of the boundary theory~\eqref{E:AStheory} \emph{in toto} from AdS$_3$ gravity, including the quotient and boundary conditions. We stress that these ingredients are crucial. As we discuss below, they render the boundary theory well-defined as a quantum mechanical model, with a normalizable vacuum and discrete spectrum.

For those readers unfamiliar with the quantization of coadjoint orbits, we refer them to Section~\ref{S:review} for a short review. The punchline is that coadjoint orbits of a Lie group lead to elegant, geometric Hamiltonian systems. The orbits are symplectic spaces~\cite{kirillov} and one may associate a Hamiltonian to any element of the algebra. Under favorable circumstances these systems may be quantized, either via geometric quantization as in e.g.~\cite{Witten:1987ty} or via a phase space path integral~\cite{Alekseev:1988vx}. The Chern-Simons formulation of AdS$_3$ gravity gives us the latter.

While the model~\eqref{E:AStheory} has been in the literature for nearly thirty years, it seems that relatively little has been computed with it using standard path integral techniques. Besides carefully obtaining the model from AdS$_3$ gravity, a central focus  of this manuscript is to establish various computational results for this theory. Some of our findings are:
\begin{enumerate}
	\item Eq.~\eqref{E:AStheory} is an ultraviolet-complete two-dimensional theory in its own right. 	It is a perfectly sensible CFT in all respects except one: it is not modular invariant. In fact, the Hilbert space of the model is just the vacuum module, which can be established by computing the torus partition function. The latter is one-loop exact on account of the localization formula, and is simply the vacuum character. (Maloney and Witten obtained the same Hilbert space from their geometric quantization.)
	\item The action~\eqref{E:AStheory} may be understood as a Wess-Zumino term for the Weyl and gravitational anomalies of a chiral theory with chiral central charge $C$. The field $\phi$ also contributes $13$ to the anomaly (rather than $1$, due to the $PSL(2;\mathbb{R})$ quotient, which effectively removes the first oscillator), and we find a total chiral central charge $c = C +13$ as we explain in Section~\ref{S:torus}. Moreover, $1/C$ plays the role of the coupling constant of the model, so that it is weakly coupled at large central charge. Excitations of the $\phi$ field correspond to boundary gravitons, and so the model computes loop-level Witten diagrams with exchanges of the gravitational field. The shift $c = C+13$ may then be understood as a one-loop renormalization of the central charge, which in fact is already visible in the direct gravitational computation of~\cite{Giombi:2008vd}. (There is no contradiction with the statement that the cosmological constant of AdS$_3$ gravity is unnrenormalized in the Chern-Simons formulation~\cite{Witten:2007kt}. The shift comes from the boundary excitations rather than the renormalization of a bulk divergence.)
	\item The only $PSL(2;\mathbb{R})$-invariant local operators of the model are built from products and derivatives of the stress tensor. This is commensurate with the fact that the Hilbert space only contains the vacuum and its Virasoro descendants.
	\item There is also a path integral quantization of other coadjoint orbits of the Virasoro group, which corresponds to the theory of boundary gravitons in the presence of a massive particle. In these models the quasi-local $PSL(2;\mathbb{R})$ redundancy is reduced to a quasi-local $U(1)$ redundancy $\phi(\theta,t) \sim \phi(\theta,t) + a(t)$. Many computational features carry over. The Hilbert space remains a single Verma module, the torus partition function may be computed exactly, and so on.
	\item In addition to the stress tensor, the model contains bilocal operators invariant under the $PSL(2;\mathbb{R})$ redundancy, which may be understood as reparameterized two-point functions of primary operators. The correlation functions of these bilocal operators encode Virasoro blocks, and the diagrammatic expansion in $1/C$ computes the $1/c$ expansion for the blocks at large central charge. For example, we compute the so-called ``heavy-light'' limit of the identity block to $O(1/c)$ via a one-loop computation, reproducing the result of ~\cite{Fitzpatrick:2015dlt} (see also~\cite{Beccaria:2015shq}) which was computed by algebraic methods.  Our computation is much simpler.
	\item The action~\eqref{E:AStheory} is Lorentz-invariant, although not manifestly so. This should not be too alarming since the theory is chiral. However, the lack of manifest Lorentz-invariance makes it tricky to couple the theory to a background gravitational field, as one would need to do in order to study higher-genus observables. Using the bulk Chern-Simons description we solve this problem and couple the theory~\eqref{E:AStheory} to a background geometry. We then compute the sphere partition function of the model, which gives us a consistency check: the sphere partition function encodes the central charge $c$ which we know by other methods.
	 	\item Bulk spacetimes of different topology lead to different boundary graviton actions. We deduce the boundary action for fluctuations around a two-sided eternal BTZ black hole, which roughly speaking is two copies of the model~\eqref{E:AStheory} subject to a single (rather than a doubled) $PSL(2;\mathbb{R})$ quotient. This result paves the way for future discussions~\cite{ourWormholes} of traversable wormholes~\cite{Gao:2016bin,Maldacena:2017axo} and perhaps even a semiclassical ``ramp''~\cite{Cotler:2016fpe,Saad:2018bqo} in AdS$_3$ gravity using this boundary field theory.
\end{enumerate}

Further comments in order. The Schwarzian model that arises in the Sachdev-Ye-Kitaev model~\cite{kitaev,Polchinski:2016xgd,Maldacena:2016hyu} and nearly AdS$_2$ gravity~\cite{Almheiri:2014cka,Jensen:2016pah,Maldacena:2016upp,Engelsoy:2016xyb} is also embedded within the theory~\eqref{E:AStheory}. The Euclidean version of the Schwarzian model is the theory of a single real scalar $\phi \in \text{Diff}(\mathbb{S}^1)/PSL(2;\mathbb{R})$ and its Lagrangian is the Schwarzian derivative of $\tan\left(\frac{\phi}{2}\right)$. It describes the boundary degree of freedom of Jackiw-Teitelboim gravity on Euclidean AdS$_2$ with a linear dilaton. The relation between~\eqref{E:AStheory} and the Schwarzian model is the following. (Other work relating Liouville theory to the Schwarzian model can be found in~\cite{Cvetic:2016eiv,Mertens:2017mtv,Mertens:2018fds}.) Projecting onto the left-moving sector~\eqref{E:AStheory}\footnote{Observe that, without the projection, a simple dimensional reduction of the Euclideanized theory would result in two copies of the Schwarzian theory.}, Wick-rotating to Euclidean time $t = - i y$ and taking the time circle to be very small, namely $\Delta y\to 0$ with $C' = C\Delta y$ fixed, the boundary graviton action reduces to that of the Schwarzian model
	\beq
		S \to \frac{C'}{48\pi} \int_0^{2\pi} d\theta \, \left( \frac{\phi''(\theta)^2}{\phi'(\theta)^2} - \phi'(\theta)^2\right) = - \frac{C'}{24\pi}\int_0^{2\pi} d\theta \left\{ \tan\left(\frac{\phi(\theta)}{2}\right),\theta\right\}\,,
	\eeq
	where $\{f(x),x\}$ is the Schwarzian derivative. To complete the identification, we note that the quasi-local quotient and boundary condition on $\phi(\theta,y)$ imply that the field $\phi(\theta)$ is now an element of Diff$(\mathbb{S}^1)/PSL(2;\mathbb{R})$. This embedding addresses various speculations that AdS$_2$ gravity is the chiral half of AdS$_3$ gravity, e.g.~\cite{Strominger:1998yg}. 

In fact, Jackiw-Teitelboim gravity may be recast as a $PSL(2;\mathbb{R})$ BF theory, and by using nearly identical methods as our AdS$_3$ analysis one can derive the boundary Schwarzian model. 
The Lorentzian case is rather interesting. The bulk geometry in that case is a non-traversable wormhole with two asymptotic boundaries. Using the usual rules of the AdS/CFT correspondence one would expect gravity on that spacetime to be dual to two copies of a single boundary theory in an entangled thermofield double state~\cite{Maldacena:2001kr}. However, this is not the case~\cite{Harlow:2018tqv}. The boundary model has non-local constraints tying the two boundaries together. This is the ``non-factorization'' property of Jackiw-Teitelboim gravity. 

In Section~\ref{S:BTZ} we obtain the boundary description of the two-sided BTZ black hole. This boundary description is not a doubling of the theory on the boundary of global AdS$_3$. It is instead, roughly speaking, two copies of the theory on the boundary of global AdS$_3$, subject to non-local constraints between the degrees of freedom on the two asymptotic boundaries.  Therefore pure AdS$_3$ gravity exhibits a similar non-factorization property, as anticipated by~\cite{Harlow:2018tqv}.

Finally, we are able to make a precise connection with the findings of Maloney and Witten. For Euclidean AdS$_3$ geometries with torus boundary, the same logic that leads to~\eqref{E:AStheory} leads to a boundary action where the field $\phi$ is periodic around the cycle which is non-contractible in the bulk, but winds once around the other cycle. In this way one obtains a different theory of boundary gravitons describing fluctuations around each bulk ``topology.'' For instance, the theory appropriate for thermal global AdS$_3$ has a partition function given by the vacuum character at complex structure $\tau$, while the theory appropriate for the Euclidean BTZ black hole has a partition function given by its $S$-transformation. The modular sum amounts to summing up the partition functions for these different CFTs. From this point of view, the Maloney-Witten prescription is sensible as a path integral instruction but has no Hamiltonian interpretation within the theory~\eqref{E:AStheory}. (One might think that summing over these different geometric models amounts to a sum over boundary conditions for the Diff$(\mathbb{S}^1)$ field $\phi$, analogous to a sum over winding states or over spin structures. However this is not the case.) 

The remainder of this manuscript is organized as follows. In Section~\ref{S:review} we review basic properties of the coadjoint orbits of the Virasoro group along with their path integral quantization. Our gravitational analysis is contained in Sections~\ref{S:AdS} and~\ref{S:BTZ}. We compute the torus partition function of the model~\eqref{E:AStheory} and show that it is one-loop exact in Section~\ref{S:torus}, and the identity Virasoro block in the so-called ``light-light'' and ``heavy-light'' limits in Section~\ref{S:blocks}. We discuss the minimally supersymmetric generalization of the boundary graviton action in Section~\ref{S:generalizations}. We wrap up with a discussion and some future directions in Section~\ref{S:discussion}. 

\emph{Note:} As this work was nearing completion we were made aware of~\cite{Haehl:2018izb} which has some overlap with our manuscript. The action obtained by those authors is (two copies of) the quadratic approximation to the quantization of Diff$(\mathbb{S}^1)/PSL(2;\mathbb{R})$ discussed at length below.

\section{Review}
\label{S:review}

Consider a Lie group $G$ with algebra $\mathfrak{g}$, and let $b$ be a coadjoint vector, $b\in \mathfrak{g}^*$. Coadjoint vectors are linear maps $b:\mathfrak{g}\to \mathbb{R}$ which take adjoint vectors $v\in \mathfrak{g}$ to numbers. We denote $b$ evaluated on $v$ by $b(v)$. There is a natural $G$-action on both adjoint and coadjoint vectors which leaves the pairing $b(v)$ invariant.  Specifically, we consider the adjoint action given by $g \cdot v = g v g^{-1}$, and the coadjoint action given by $g \cdot b(\,\cdot\,) = b(g^{-1} \, \cdot \, g)$. Under this $G$-action, $b$ sweeps out a coadjoint orbit, namely $\mathcal{M}_b = \{ g\cdot b: g\in G\}$. The coadjoint orbit of $b$ is a symplectic manifold, equipped with the Kirillov-Kostant symplectic form $\omega$ which may be simply written as follows. At the point $\tilde{b}\in \mathcal{M}_b$, any tangent vector $X$ to the coadjoint orbit can be associated with a vector $v\in \mathfrak{g}$ which generates translation in the $X$-direction. Then the Kirillov-Kostant form~\cite{kirillov} reads
\beq
\label{E:KKform}
	\omega(X_1,X_2) = -\tilde{b}([v_1,v_2])\,,
\eeq
where $[\cdot,\cdot]$ is the commutator of adjoint vectors. It is straightforward to show that this two-form is non-degenerate and closed, and so a symplectic form. Furthermore, $\omega$ is $G$-invariant since $\tilde{b}([v_1,v_2])$ is invariant under the group action.

Recalling that the coadjoint orbit $\mathcal{M}_b$ is swept out under the action of the group $G$, consider any adjoint vector $v$. The vector $v$ generates translation along the orbit via $\exp(t v)$, which we take to be in the $X$-direction. This translation is enacted by a Hamiltonian function on the orbit, $H_X:\mathcal{M}_b \to \mathbb{R}$, defined via $H_X(\tilde{b}) = -i\tilde{b}(v)$.  In particular, the differential of $H_X$ produces the constant Hamiltonian vector field $v$, with flows corresponding to $\exp(t v)$.

With a symplectic space and Hamiltonian in hand, we have the necessary ingredients for a Hamiltonian system whose phase space is given by the coadjoint orbit $\mathcal{M}_b$\,. Quantizing these orbits leads to elegant quantum mechanical models with a Hilbert space that is typically a single irreducible representation of $G$. Correlation functions and partition functions are simply group-theoretic functions (like characters) of this representation.

Historically, the quantization of coadjoint orbits has been performed using two different methods which are not obviously equivalent: geometric quantization, and the phase space path integral quantization first noted by Alekseev, Fadeev, and Shatashvili~\cite{Alekseev:1988vx}.  In the following we will exclusively use path integrals. The starting point for path integral quantization is to find a presymplectic potential $\alpha$ satisfying
\beq
\label{E:omega}
	\omega = d\alpha\,.
\eeq
Denoting coordinates on the coadjoint orbit $\mathcal{M}_b$ as $x^i$, one then promotes the coordinates to functions of time and defines an action functional
\beq
\label{E:action}
	S_X = -\int dt \left( \dot{x}^i \alpha_i + H_X\right)\,.
\eeq
Because this action is first-order in time derivatives, upon Legendre transformation we recover the Hamiltonian system we started with, with phase space $\mathcal{M}_b$, symplectic form $\omega$, and Hamiltonian $H_X$. To see this, note that the canonical momenta $\pi_i = -\alpha_i(x^j)$ depend on the $x^j$s and so are not independent coordinates, and thus the phase space is still $\mathcal{M}_b$\,. The Hamiltonian $\pi_i \dot{x}^i - L$ is just $H_X$. And finally, the canonical symplectic form is
\beq
	\omega_0= dx^i \wedge d\pi_i = -dx^i \wedge dx^j \partial_j  \alpha_i = \frac{1}{2}(\partial_i \alpha_j - \partial_j \alpha_i)dx^i \wedge dx^j = \omega\,.
\eeq
Quantizing, the relevant path integral is given by
\beq
\label{E:pathIntegral}
	\int [dx^i(t)] \text{Pf}(\omega)e^{i S_X}\,,
\eeq
where Pf$(\omega)$ is the Pfaffian of the symplectic form, which provides a suitable measure on the phase space $\mathcal{M}_b$.

This procedure is sensible only if a cousin of the Dirac quantization condition holds. The presymplectic potential $\alpha$ is defined modulo a ``gauge redundancy,'' $\alpha \to \alpha + d \Lambda$.  (These ``gauge'' transformations are ``symplectomorphisms'' -- coordinate transformations on phase space which leave the symplectic form invariant.) Thus the phase factor $\exp(i S_X)$ is well-defined only if the symplectic form $\omega$ has quantized periods,
\beq
	\oint \omega = 2\pi \mathbb{Z}\,,
\eeq
i.e. if the presymplectic potential $\alpha$ is a connection on a complex line bundle.

In the remainder of this Section we review the classification of coadjoint orbits of the Virasoro group and their path integral quantization. We extensively cull the results of Witten~\cite{Witten:1987ty} and Alekseev and Shatashvili~\cite{Alekseev:1988ce,Alekseev:1990mp}. See also e.g.~\cite{khesin,Verlinde:1989ua,Verlinde:1989hv} which we found useful. Many results concerning the coadjoint orbits of the Virasoro group are also found in~\cite{Kirillov1981,Bowick:1987pw}. Lastly, there are several recent papers related to coadjoint orbits, the Schwarzian quantum mechanics, and nearly AdS$_2$ gravity~\cite{Mandal:2017thl,Stanford:2017thb,Alekseev:2018pbv,Mertens:2018fds}.

\subsection{Coadjoint orbits of the Virasoro group}

$\text{Diff}(\mathbb{S}^1)$ is the diffeomorphism group of the circle. Elements of $\text{Diff}(\mathbb{S}^1)$ are monotone, single-valued functions $\phi : [0,2\pi) \to [0,2\pi)$ satisfying $\phi(\theta+2\pi) = \phi(\theta)+2\pi$. The Virasoro group is the central extension of $\text{Diff}(\mathbb{S}^1)$, which we write as $\widehat{\text{Diff}}(\mathbb{S}^1)$. Elements of $\widehat{\text{Diff}}(\mathbb{S}^1)$ are pairs $(\phi(\theta),a)$ of a diffeomorphism $\phi(\theta) \in \text{Diff}(\mathbb{S}^1)$ and a number $a$ which multiplies the central element $c$. Vectors of the algebra of $\widehat{\text{Diff}}(\mathbb{S}^1)$ are pairs $(f(\theta),a)$ of a vector field $f(\theta) \frac{\partial}{\partial\theta}$ and a number $a$ which multiplies the central element $c$. We also denote $(f(\theta),a)$ by $f(\theta) \frac{\partial}{\partial\theta}-i a c$. The algebra is given by
\beq
\label{E:comm1}
	\left[ f_1(\theta)\frac{\partial}{\partial\theta}-i a_1 c, f_2(\theta) \frac{\partial}{\partial\theta} - i a_2 c\right] = (f_1 f_2' - f_2 f_1')\frac{\partial}{\partial\theta} + \frac{i c}{48\pi}\int_0^{2\pi}d\theta (f_1 f_2''' - f_2 f_1''')\,.
\eeq
The generator $L_n$ corresponds to the vector field $i e^{i n \theta} \frac{\partial}{\partial\theta}$, and so this algebra leads to a slightly unconventional parameterization of the Virasoro algebra,
\beq
\label{E:VirasoroAlgebra}
	[L_n,L_m] = (n-m)L_{n+m}  + \frac{c}{12}n^3 \delta_{n+m}\,,
\eeq
which differs from the usual one by $L_{0,\rm usual}= L_{0, \rm here} +\frac{c}{24}$. The commutator in Eq.~\eqref{E:comm1} may be understood as the action of the vector $(f_1(\theta),a_1)$ on the vector $(f_2(\theta),a_2)$,
\beq
	\delta_{(f_1,a_1)}(f_2,a_2) = \left[ f_1\frac{\partial}{\partial\theta}\,,f_2\frac{\partial}{\partial\theta}\right]\,.
\eeq
We henceforth denote the action $\delta_{(f_1,a_1)}$ simply as $\delta_1$.

A coadjoint vector is a pair $(b(\theta),t)$ of a quadratic differential $b(\theta) d\theta^2$ and a number $t$ which multiplies $\tilde{c}$, the dual of $c$ satisfying $\tilde{c}(c) = 1$. The pairing of $(b,t)$ with $(f,a)$ is
\beq
\label{E:pairing}
	\langle (b,t),(f,a)\rangle = \int_0^{2\pi} d\theta \, b f+ta\,.
\eeq
The pairing is Virasoro-invariant, which fixes the transformation law of $(b,t)$. Using the variation of the vector $(f_2,a_2)$ under the transformation generated by another vector $(f_1,a_1)$, we compute
\begin{align}
	\nonumber
	\delta_1 \langle (b,t),(f_2,a_2)\rangle &= \int_0^{2\pi}d\theta \, \left( f_2 \delta_1 b + (f_1 f_2' - f_2 f_1')b \right) + a \delta_1 t + t \left( - \frac{1}{48\pi}\int_0^{2\pi} d\theta (f_1f_2'''-f_2f_1''')\right)
	\\
	& = a_2 \delta_1 t + \int_0^{2\pi} d\theta \, f_2 \left( \delta_1 b - \left( f_1 b' + 2 b f_1' -\frac{t f_1'''}{24\pi}\right)\right)
	\\
	\nonumber
	& = 0\,,
\end{align}
from which we deduce
\begin{align}
\begin{split}
\label{E:deltaCoadjoint}
	\delta_1 b & = f_1 b' + 2 b f_1' - \frac{t f_1'''}{24\pi}\,,
	\\
	\delta_1 t & = 0\,.
\end{split}
\end{align}

Orbits of the Virasoro group can be divided into two types: those which coincide with the orbit of a constant coadjoint vector $(b_0,C)$ for some $b_0$, and those which do not. The orbits of a constant vector may be labeled by $b_0$ and $C$. The finite form of the transformation~\eqref{E:deltaCoadjoint}  by $(\phi(\theta),a)\in \widehat{\text{Diff}}(\mathbb{S}^1)$ reads
\beq
\label{E:coadjointOrbit}
	(b_0,C) \to (b(\phi),C) = \left( b_0 \phi'^2 - \frac{C \{\phi,\theta\}}{24\pi},C\right)\,,
\eeq
where
\beq
	\{ f,\theta\} = \frac{f'''}{f'}-\frac{3}{2}\left(\frac{f''}{f'}\right)^2\,,
\eeq
is the Schwarzian derivative of $f$ with respect to $\theta$. So points on the coadjoint orbit may be parameterized in terms of $\phi(\theta)$, modulo identifications depending on the stabilizer $S$ of the transformation. The coadjoint orbit is then isomorphic to $\text{Diff}(\mathbb{S}^1)/S$. The stabilizer depends on the value of $b_0$, and there are three distinct cases:
\begin{enumerate}
\item $b_0 \neq - \frac{Cn^2}{48\pi}$. These are the ``ordinary'' coadjoint orbits of the Virasoro group. The transformation $\phi(\theta)$ and $\phi(\theta) + a$ lead to the same coadjoint vector for any $a$. Since $\phi$ is itself an angular variable, so is $a$. Thus the stabilizer is $U(1)$, generated by $L_0$.
\item $b_0 = - \frac{C}{48\pi}$. This is the ``first exceptional orbit.'' For this orbit the transformation~\eqref{E:coadjointOrbit} may be rewritten as
\beq
\label{E:firstExceptional}
	b(\phi) = - \frac{C}{24\pi}\left\{ \tan \left(\frac{\phi}{2}\right),\theta\right\}\,.
\eeq
Using the fact that $\phi$ is a Diff($\mathbb{S}^1)$ field, i.e. $\phi(\theta+2\pi) = \phi(\theta)+2\pi$, we see that the stabilizer is $PSL(2;\mathbb{R})$. The stabilizer acts on $\phi$ as
\beq
	\tan\left(\frac{\phi}{2}\right)\to \frac{a \tan\left(\frac{\phi}{2}\right)+b}{c\tan\left(\frac{\phi}{2}\right)+d}\,, \qquad \begin{pmatrix} a & b \\ c & d \end{pmatrix} \in PSL(2;\mathbb{R})\,,
\eeq
under which the transformation~\eqref{E:firstExceptional} is invariant. These transformations are generated by the vectors $f(\theta) = a_0 + a_1 e^{i \theta} + a_{-1} e^{-i \theta}$, i.e. by $L_0$ and $L_{\pm 1}$. The orbit, Diff$(\mathbb{S}^1)/PSL(2;\mathbb{R})$, is also the Teichm\"uller space of the disk.
\item $b_0 = - \frac{C n^2}{48\pi}, n>1$. These are the ``higher exceptional orbits.'' Their respective stabilizers are sometimes called $PSL^{(n)}(2;\mathbb{R})$.  Here, $PSL^{(n)}(2;\mathbb{R})$ is the group generated by $L_0$ together with $L_{\pm n}$. For the higher exceptional orbits the transformation~\eqref{E:coadjointOrbit} may be written as
\beq
	b(\phi) = - \frac{C}{24\pi}\left\{ \tan\left(\frac{n\phi}{2}\right),\theta\right\}\,,
\eeq
which is invariant under
\beq
	\tan\left( \frac{n\phi}{2}\right) \to \frac{a \tan\left( \frac{n \phi}{2}\right)+b}{c\tan\left(\frac{n\phi}{2}\right)+d}\,.
\eeq
\end{enumerate}

There are also orbits which are not generated by the group action on a constant coadjoint vector. These are, in a sense, perturbations of the exceptional orbits. See Witten~\cite{Witten:1987ty} for further discussion. We do not consider such orbits further.

While all coadjoint orbits are symplectic, the first exceptional orbit and the normal orbits for $b_0 > - \frac{C}{48\pi}$ and $C>0$ are special since they are K\"ahler, and furthermore possess a Virasoro-invariant K\"ahler form which is in fact the Kirillov-Kostant form. Let us review the argument for this structure given by Witten.

The first step in the argument is to determine whether an orbit $G/H$ admits a $G$-invariant complex structure. $G$-invariant complex structures on $G/H$ are in one-to-one correspondence with $H$-invariant complex structures at $[\mathbf{1}]$ with $\mathbf{1}$ the origin of $G$. The idea is that one uses the $G$-action to transport the complex structure at the origin to the rest of the orbit, and $H$-invariance at the origin guarantees that the resulting complex structure is well-defined everywhere on the quotient space $G/H$. At the origin, one may classify such complex structures at the level of Lie algebras rather than Lie groups. Let $\mathfrak{g}$ be the algebra of $G$, $\mathfrak{h}$ the algebra of $H$, and $\mathfrak{q}$ the complement of $\mathfrak{h}$ in $\mathfrak{g}$ satisfying $\mathfrak{g} = \mathfrak{h} \oplus \mathfrak{q}$. Here $\mathfrak{q}$ is isomorphic to the tangent space of $G/H$ at $[\mathbf{1}]$. An almost complex structure at $[\mathbf{1}]$ is a decomposition of $\mathfrak{q}$ into holomorphic and anti-holomorphic directions,
\beq
	\mathfrak{q} = \mathfrak{q}_+ \oplus \mathfrak{q}_-\,.
\eeq
The almost complex structure is $H$-invariant on $G$ if the decomposition at $[\mathbf{1}]$ is $\mathfrak{h}$-closed, meaning that
\beq
	[h,q_+] \in \mathfrak{h}\oplus \mathfrak{q}_+\,, \qquad \forall h\in \mathfrak{h}\,, \, q_+\in \mathfrak{q}_+\,.
\eeq
The almost complex structure is a complex structure if the holomorphic tangent vectors are also closed under commutation,
\beq
	[q_+,q_+'] \in \mathfrak{h} \oplus \mathfrak{q}_+\,, \qquad \forall q_+,q_+'\in \mathfrak{q}_+\,.
\eeq

For the normal orbits of the Virasoro group, we have $G=\widehat{\text{Diff}}(\mathbb{S}^1)$, and $H=U(1)$ is the one-parameter subgroup generated by $L_0$. So $\mathfrak{q}$ is generated by $\{L_{n\neq 0}\}$ and the natural guess for an almost complex structure is to designate the $L_{n>0}$ as holomorphic vectors and the $L_{n<0}$ as anti-holomorphic. This guess works, and leads to a Virasoro-invariant complex structure. The argument is essentially the same for the first exceptional orbit, for which $H = PSL(2;\mathbb{R})$ and then $\mathfrak{q}$ is generated by the $\{L_{n>1}\}$ and $\{L_{n<-1}\}$. Again, this separation leads to an invariant complex structure, on account of the Virasoro algebra~\eqref{E:VirasoroAlgebra}. However for the higher exceptional orbits, where $H=PSL^{(n)}(2;\mathbb{R})$ is generated by $L_0$ and $L_{\pm n}$, a separation into positive and negative $L_m$'s is not $H$-closed, e.g. $[L_{-n},L_1] \propto L_{-n+1}$, and so does not lead to an invariant complex structure.

The second part of the argument is to show that not only are the orbits with $b_0\geq -\frac{C}{48\pi}$ complex manifolds, but further that they possess an invariant K\"ahler form. On a complex manifold a K\"ahler form is a symplectic form of type $(1,1)$ (meaning one leg is in the holomorphic direction and one in the anti-holomorphic direction). As we will see momentarily, the Kirillov-Kostant symplectic form gives exactly such a $(1,1)$ form on these orbits.

\subsection{Phase space path integrals}
\label{S:phasespace}

To construct a path integral quantization of a coadjoint orbit we require two ingredients: (i) a presymplectic potential $\alpha$ corresponding to the Kirillov-Kostant symplectic form, and (ii) a suitable Hamiltonian.

Let us now compute the Kirillov-Kostant symplectic form~\eqref{E:KKform} on the coadjoint orbits. Evaluated on a pair of vectors $F_1=(f_1(\theta),a_1)$ and $F_2=(f_2(\theta),a_2)$, it is
\begin{align}
\begin{split}
	\omega(X_1,X_2) &= -\langle (b,C),[F_1,F_2]\rangle =- \int_0^{2\pi} d\theta\left( b (f_1f_2'-f_2f_1') - \frac{C}{48\pi}(f_1f_2'''-f_2f_1''')\right)
	\\
	& =- \int_0^{2\pi} d\theta \left\{\frac{C}{48\pi}( f_1'f_2''-f_2'f_1'') +\left( b_0\phi'^2 - \frac{C}{24\pi}\{\phi,\theta\}\right)(f_1f_2'-f_2f_1')\right\}\,,
\end{split}
\end{align}
where we have used~\eqref{E:coadjointOrbit} to write $b=b(\phi) = b_0 \phi'^2 - \frac{C}{24\pi}\{\phi,\theta\}$. Under an infinitesimal transformation $F=(f,a)$, the variation of $\phi$ is $\delta_F \phi = f \phi'$.  Then writing the symplectic form in terms of a formal one-form $d\phi$ such that $d\phi(F) = \delta_F \phi' = f\phi'$, the symplectic form becomes
\begin{align}
\begin{split}
\label{E:symplecticForm}
	\omega &=- \int_0^{2\pi} d\theta\left\{ \frac{C}{48\pi} \left( \frac{d\phi}{\phi'}\right)'\wedge \left( \frac{d\phi}{\phi'}\right)'' + \left( b_0 -\frac{C\{\phi,\theta\}}{24\pi\phi'^2}\right)d\phi \wedge d\phi'\right\}
	\\
	&= -\int_0^{2\pi}d\theta \left\{ \frac{C}{48\pi}\frac{d\phi'\wedge d\phi''}{\phi'^2} + b_0 d\phi\wedge d\phi'\right\}\,.
\end{split}
\end{align}
In going from the first line to the second we have integrated by parts and used identities like $(d\phi)' = d\phi'$ and $d\phi' \wedge d\phi' = 0$. The presymplectic potential corresponding to $\omega$ is simply
\beq
	\alpha =  \int_0^{2\pi} d\theta\left( \frac{C}{48\pi} \frac{\phi''d\phi' }{\phi'^2}+ b_0\phi' d\phi\right)\,.
\eeq

Let us briefly return to the question of K\"ahler structures for the normal orbits as well as for the first exceptional orbit. At the origin, where $\phi(\theta)=\theta$, we use the complex structure to separate vectors into holomorphic and anti-holomorphic parts, $f =  \alpha_0 L_0+ \sum_{n>0}^{\infty} (\alpha_n L_n +\bar{\alpha}_n L_{-n})$. The symplectic form at the origin is then
\beq
\label{E:kahler}
	i\omega =  \sum_{n>0} \frac{C}{12}n\left( n^2+\frac{48\pi b_0}{C}\right) d\alpha_n \wedge d\bar{\alpha}_n\,,
\eeq
and the sum is taken over $n>0$ for normal orbits and over $n>1$ for the first exceptional orbit $b_0 = - \frac{C}{48\pi}$. Clearly $i\omega$ is of type $(1,1)$ and moreover is positive-definite provided that $b_0 \geq - \frac{C}{48\pi}$. We then conclude that normal orbits obeying this bound possess a Virasoro-invariant K\"ahler form. This fact will be useful later in Section~\ref{S:torus} when we compute the torus partition function of the quantization of Diff$(\mathbb{S}^1)/U(1)$ and Diff$(\mathbb{S}^1)/PSL(2;\mathbb{R})$. 

There are two natural Hamiltonians to consider. The first is simply $H=0$, which was used extensively by Alekseev and Shatashvili. Promoting $\phi(\theta)$ to a function of time, $\phi(\theta,t)$, the action functional~\eqref{E:action} becomes the Alekseev-Shatashvili action
\beq
\label{E:H=0}
	S_{AS} = -\int d^2x \left( \frac{C}{48\pi} \frac{\dot{\phi}' \phi''}{\phi'^2}+b_0 \dot{\phi} \phi'\right)\,.
\eeq
Here $\phi$ is an element of the loop space $L\left(\text{Diff}(\mathbb{S}^1)/S\right)$, and so $\phi(\theta,t)$ is monotone in $\theta$ at fixed $t$ and respects the boundary condition $\phi(\theta+2\pi,t) = \phi(\theta,t)+2\pi$. The field $\phi(\theta,t)$ is also subject to a redundancy that depends on the orbit, where now the parameters appearing in the redundancy can have any dependence on time. For an ordinary orbit, we identify
\beq
	\phi(\theta,t) \sim \phi(\theta,t) + a(t)\,,
\eeq
and for the first exceptional orbit, we identify
\beq
	\tan\left( \frac{\phi(\theta,t)}{2} \right)\sim \frac{a(t) \tan\left( \frac{\phi(\theta,t)}{2}\right)+b(t)}{c(t)\tan\left( \frac{\phi(\theta,t)}{2}\right)+d(t)}\,, \qquad \begin{pmatrix} a(t)&b(t) \\ c(t) & d(t) \end{pmatrix} \in PSL(2;\mathbb{R})\,.
\eeq

 The other natural choice for the Hamiltonian is for it to be the function associated with $L_0$. This choice is in fact mandated by Lorentz invariance as we presently explain. The action $S_{AS}$ in~\eqref{E:H=0} has a conserved momentum, given by
 \beq
 \label{E:P}
 	P = -\int_0^{2\pi} d\theta\left( \frac{C}{24}\{\phi,\theta\} -b_0\phi'^2\right)\,,
 \eeq
along with zero Hamiltonian. This momentum is in fact bounded below for $b_0\geq -\frac{C}{48\pi}$, as we will see below. To restore a chiral Lorentz invariance we need to deform by the Hamiltonian $H=P$. Using $L_0 = i \frac{\partial}{\partial\theta}$ and the pairing~\eqref{E:pairing}, we construct the corresponding finite transformation and find that the Hamiltonian corresponding to $L_0$ is the desired one,
\beq
\label{E:L0}
	H = -\int_0^{2\pi} d\theta \left( \frac{C}{24\pi}\{\phi,\theta\} - b_0 \phi'^2\right) = P\,.
\eeq
The corresponding action functional will be central in the remainder of this work, namely
\beq
\label{E:ourTheory}
	S_+ =- \frac{C}{24\pi}\int d^2x \left( \frac{(\partial_+\phi')\phi''}{\phi'^2}+B (\partial_+\phi)\phi'\right)\,, \qquad B = \frac{b_0}{48\pi C}\,,
\eeq
with $\partial_+ = \frac{1}{2}(\partial_{\theta} +\partial_t)$. This theory is entirely left-moving, with $H=P$.

The construction above can also be used to obtain a right-moving theory. In~\eqref{E:KKform} we could have chosen the opposite sign for the symplectic form, flipping the sign of the presymplectic potential $\alpha$. This in turn would lead, for $H=0$, to the same action in~\eqref{E:H=0} but with an overall $+$ sign. That theory has a conserved momentum equal to minus the expression in~\eqref{E:P} with $P_{\rm here} = - P_{\rm there}$. To restore Lorentz invariance one again deforms by the Hamiltonian corresponding to $L_0$ so that one has $H=-P_{\rm here}$. The resulting theory is right-moving and its action is given by
\beq
	S_- = -\frac{C}{24\pi}\int d^2x \left( \frac{(\partial_-\phi')\phi''}{\phi'^2}+B (\partial_-\phi)\phi'\right)\,, \qquad \partial_- = \frac{1}{2}(\partial_{\theta}-\partial_t)\,.
\eeq

We henceforth restrict our attention to the left-moving theory in~\eqref{E:ourTheory} with $C>0$ and $B \geq -1$, i.e.\! to $b_0 \geq - \frac{C}{48\pi}$. For these cases the Hamiltonian~\eqref{E:L0} is bounded below, as argued by Witten. A variational argument shows us why the Hamiltonian is bounded below. The variation of $H$ is
\beq
	\frac{\delta H}{\delta \phi(\theta)} = \frac{1}{\phi'}\left( \frac{C}{24\pi}\{\phi,\theta\} - b_0\phi'^2\right)'  \,.
\eeq
For $b_0 > - \frac{C}{48\pi}$, there is a unique critical point of $H$ modulo the $U(1)$ redundancy $\phi \sim \phi+a$ which obeys the boundary conditions $\phi(\theta+2\pi) = \phi(\theta)+2\pi$ and $\phi'(\theta)\geq 0$, namely $\phi = \theta$. For $b_0=-\frac{C}{48\pi}$, this is again the unique critical point modulo the redundancy. Expanding $H$ near the critical point we find
\beq
	H[\phi = \theta + \sum_n \phi_n e^{i n \theta}] = 2\pi b_0 + \frac{C}{12}\sum_n n^2 (n^2+B) |\phi_n|^2 + O(\phi_n^3)\,,
\eeq
where the sum runs over $n\neq 0$ for $B >-1$ and $|n| >1$ for the first exceptional orbit $B=-1$. Thus, in each case, $\phi = \theta$ is a local minimum. That $H$ has a unique critical point with a positive Hessian strongly suggests that $H$ is bounded below. To complete the argument one needs that the orbits with $B\geq -1$ are connected and contractible. It follows that $\phi = \theta$ is a global minimum and $H$ is bounded below as
\beq
H \geq 2\pi b_0\,.
\eeq

Wick-rotating to Euclidean time $t=-i y$, the Euclidean action corresponding to~\eqref{E:ourTheory} is
\beq
\label{E:Seuc}
	S_E = \frac{C}{24\pi}\int d^2x \left( \frac{(\bar{\partial} \phi')\phi''}{\phi'^2}+B (\bar{\partial}\phi)\phi'\right)\,,
\eeq
with $\bar{\partial} = \frac{1}{2}(\partial_{\theta}+i \partial_y)$. Its real part is
\beq
	\text{Re}(S_E) = \int dy \, H(y)\,.
\eeq
So not only is $H$ bounded below, but so is the real part of the Euclidean action. Thus it is at least plausible that the functional integral corresponding to the theory~\eqref{E:ourTheory} is well-defined.

A similar variational argument demonstrates that for the other orbits $B<-1$, $H$ still has a single critical point but its Hessian is not positive definite: the $n=\pm 1$ fluctuations decrease the value of $H$. It is also easy to exhibit kink-like configurations for $\phi(\theta)$ which have arbitrarily large and negative values of $H$. Thus we expect the path integral to not exist for orbits with $B<-1$, i.e. for $b_0 < - \frac{C}{48\pi}$, for which the real part of the Euclidean action is not bounded below.

The action~\eqref{E:ourTheory} describes a two-dimensional chiral quantum field theory. It is nearly a chiral CFT since it satisfies all of the CFT axioms except modular invariance. Indeed, we will see later that the spectrum is a single Verma module.  In order to avoid any ambiguity, we will call the theory a QFT rather than a CFT.

The left-moving stress tensor is just the Hamiltonian density
\beq
\label{E:stress}
	T =2\pi T_{--} = -\frac{C}{12} \{\phi,\theta\} +2\pi b_0 \phi'^2\,,
\eeq
and in fact the field equation for $\phi$ is just the conservation of this stress tensor,
\beq
	\partial_+ T = 0\,.
\eeq
In this sense, the theory~\eqref{E:ourTheory} is just a theory of the stress tensor, and so is an unconventional kind of hydrodynamics.\footnote{A na\"{i}ve computation also gives a left-moving stress tensor~\cite{Alekseev:1988ce}. However, as we will see in Subsection~\ref{S:curved}, this stress tensor is identically zero after the addition of a suitable improvement term.}

The theory~\eqref{E:ourTheory} is not manifestly Lorentz-invariant. However, this should not be surprising: the theory~\eqref{E:ourTheory} is chiral, and there is no manifestly Lorentz-invariant Lagrangian for a single chiral boson. What is this chiral theory? The action~\eqref{E:ourTheory} clearly admits a weak coupling limit $C \gg 1$, and in that limit the semiclassical $L_n$'s
\beq
L_n = -\int_0^{2\pi} d\theta\left( \frac{C}{24\pi}\{\phi,\theta\}-b_0 \phi'^2\right)e^{i n \theta}\,,
\eeq
have Poisson brackets
\beq
\{ L_n,L_m\}_{P.B.} = (n-m)L_{n+m} + \frac{C}{12}n^3 \delta_{n+m}\,,
\eeq
indicating that $C$ plays the role of the chiral central charge at large $C$. At finite $C$ we will see that $C$ has a one-loop exact renormalization. Moreover at the origin of the orbit, $\phi = \theta$, we have $L_0 = 2\pi b_0$. So at large $C$ we expect that the Hilbert space of the quantized theory is a single Verma module with lowest weight state $|h\rangle$ where $h$ has the classical value $h= 2\pi b_0+\frac{C}{24}$, which may receive quantum corrections at finite $C$. For the first exceptional orbit, the classical value of $h$ is $h = 0$, and in this case we expect the Hilbert space to be the vacuum Verma module. Roughly speaking, the Fourier modes of the field $\phi(\theta) = \theta + \delta \phi$ correspond to the Virasoro generators $L_n$, and the Virasoro descendant states are built up by dressing the ground state with excitations of the $\phi$ field. Later in Section~\ref{S:torus} we will see that these expectations are correct.

In the classical theory it is easy to enforce that at any fixed $t$, $\phi(\theta,t)$ is a Diff$(\mathbb{S}^1)/S$ field, i.e. that it is monotone and obeys $\phi(\theta+2\pi,t) = \phi(\theta,t)+2\pi$. What about in the quantum theory? The boundary condition and quotient can be accounted for with standard techniques, but \emph{a priori} one might be concerned about enforcing that $\phi$ is monotone at fixed $t$. It turns out that, at least for the Euclidean theory, we can extend the domain of path integration from monotone $\phi$ to all $\phi$: the real part of the action~\eqref{E:Seuc} diverges for non-monotone $\phi$, so that the region of field space with non-monotone $\phi$ contributes zero to the Euclidean path integral.

In the Introduction we claimed that the theory~\eqref{E:ourTheory} is ultraviolet-complete. We presently justify this claim. In this theory one assigns $\phi$ dimension $-1$, so that $\phi'$ and $\dot{\phi}$ are dimensionless. For the first exceptional orbit, it is a straightforward exercise to demonstrate that there are no candidate relevant local counterterms invariant under the $PSL(2;\mathbb{R})$ symmetry. Further, there is a unique dimension--$2$ operator consistent with the $PSL(2;\mathbb{R})$ and Lorentz symmetries that may be generated, namely the Lagrangian itself. That is, quantum corrections may only renormalize the bare central charge $C$. In Section~\ref{S:torus} we demonstrate that this renormalization is one-loop exact and under it only $C$ is shifted.

The argument for the normal orbits is similar, although the quotient is smaller and so there are more invariant local operators. For example, any polynomial in $\phi'$ is $U(1)$-invariant and has mass dimension 0. However, Lorentz invariance forbids almost all relevant and almost all marginal operators from being generated. Indeed the combination of the Lorentz and $U(1)$ symmetries only allow $C$ and $b_0$ to be renormalized, and as above their renormalization is one-loop exact, shifting $C$.

The fact that the theory in Eq.~\eqref{E:ourTheory} has a single real field with a tunable central charge is reminiscent of Liouville theory. Of course the two are not the same. The theory in Eq.~\eqref{E:ourTheory} is chiral whereas Liouville is not. The chiral theory has a Hilbert space with a single Verma module, while Liouville theory has a continuous spectrum. Nevertheless, there is a precise relation~\cite{Zamolodchikov:2001ah,Mertens:2018fds}, namely that Liouville theory between so-called ZZ-branes has been argued to reduce to the Alekseev-Shatashvili theory considered here.

At the beginning of this Section we mentioned that the path integral quantization of a coadjoint orbit only exists when the symplectic flux is quantized. This issue is slightly more subtle when quantizing the orbits of the infinite-dimensional Virasoro group, as the measure may be anomalous. One requires that $[d\phi] e^{i S}$ is well-defined. However, this does not appear to lead to a quantization condition on $C$.

\section{From AdS$_3$ gravity to coadjoint orbits}
\label{S:AdS}

Pure AdS$_3$ gravity is described by the action
\beq
\label{S:einstein}
	S = \frac{1}{16\pi G}\int d^3x \sqrt{-g} ( R +2)\,,
\eeq
plus suitable boundary terms. (Here and throughout we set the AdS radius to unity.) Famously, it may be rewritten as Chern-Simons theory at the classical level~\cite{Achucarro:1987vz} (see~\cite{Witten:1988hc,Witten:2007kt} for a discussion of quantum aspects thereof). One groups the dreibein $e^A_{M}$ and spin connection $\omega^A{}_{BM}$ into one-forms
\beq
	A^A =  \omega^A+e^A\,, \qquad \bar{A}^A =  \omega^A-e^A\,,
\eeq
where 
\beq
	 \omega^A = \frac{1}{2}\epsilon^{ABC} \omega_{BC}\,,
\eeq
and flat indices $A,B,C$ are raised and lowered with the Minkowski metric $\eta_{AB}$. Taking $J_A$ and $\bar{J}_A$ to be generators in the fundamental representation of the $\mathfrak{so}(2,1)$ algebra,
\beq
	[J_A,J_B]= \epsilon_{ABC} J^C\,, \qquad \text{tr}(J_AJ_B) =2\eta_{AB}\,,
\eeq
we define the algebra-valued one-forms $A = A^AJ_A$ and $\bar{A} = \bar{A}^A \bar{J}_A$. The Einstein-Hilbert action may be rewritten as a difference of Chern-Simons actions for $A$ and $\bar{A}$,
\beq
	S = \frac{1}{64\pi G}\int\left(  I[A]-I[\bar{A}]\right)\,, \qquad I[A] =\text{tr}\left(  A\wedge dA + \frac{2}{3}A\wedge A \wedge A\right)\,,
\eeq
again up to a boundary term. On a classical solution, the one-forms $A$ and $\bar{A}$ transform as independent $SO(2,1)$ connections under the combination of infinitesimal diffeomorphisms and local Lorentz transformations. In this way, classical pure AdS$_3$ gravity can be mapped to Chern-Simons theory. 

As stressed in~\cite{Witten:2007kt}, the Chern-Simons fields are the natural variables in which to quantize three-dimensional gravity in perturbation theory around a classical geometry. Indeed, in the next Subsection we carefully quantize around global AdS$_3$, accounting for the asymptotically AdS$_3$ boundary conditions. However, there are various non-perturbative puzzles as discussed in detail in~\cite{Witten:2007kt}. Most importantly, one expects a theory of quantum gravity to include a sum over topologies, whereas the natural procedure in Chern-Simons theory is to quantize on a constant-time slice of fixed topology. Also, while infinitesimal diffeomorphisms and local Lorentz rotations correspond to infinitesimal $SO(2,1)\times SO(2,1)$ gauge transformations for $A$ and $\bar{A}$, the same is not necessarily true for globally non-trivial transformations. Furthermore, this classical map only establishes the local form of the gauge group of the Chern-Simons theory, but not its global topology. The minimal possibility is $SO(2,1)\times SO(2,1)$, but \emph{a priori} perhaps one should instead consider a cover thereof. (Although as we will discuss shortly, physical considerations fix that the group is $SO(2,1)\times SO(2,1)$~\cite{Castro:2011iw}.)

In the remainder of this Section we work out various aspects of the Chern-Simons formulation. In the next Subsection we carefully derive the boundary graviton action~\eqref{E:AStheory} advertised in the Introduction from the Chern-Simons theory on the cylinder. Related computations give the appropriate theory when the bulk spacetime is a Euclidean BTZ geometry. We also couple the theory~\eqref{E:AStheory} to a background geometry, which allows us to work toward a description of Chern-Simons theory on handlebody Euclidean AdS$_3$ geometries with higher-genus boundary. 

\subsection{Cylinder boundary}

We begin with global AdS$_3$, described by the metric
\beq
	g = -(r^2+1)dt^2+r^2d\theta^2 + \frac{dr^2}{r^2+1}\,,
\eeq
where $\theta $ is an angular variable with periodicity $2\pi$. The conformal boundary is located as $r\to\infty$. Constant-time slices are disks, so that the total space is a solid cylinder. Describing the bulk metric with the dreibein
\beq
	e^0 = \sqrt{r^2+1}\,dt\,, \qquad e^1 = r d\theta\,, \qquad e^r = \frac{dr}{\sqrt{r^2+1}}\,,
\eeq
this geometry corresponds to
\begin{align}
\begin{split}
	A &= \sqrt{r^2+1} \,dx^+ \,J_0 + r dx^+ \,J_1 + \frac{dr }{\sqrt{r^2+1}}J_2\,,
	\\
	\bar{A} &= \sqrt{r^2+1}\, dx^-\,\bar{J}_0 + r dx^- \,\bar{J}_1 - \frac{dr}{\sqrt{r^2+1}}\bar{J}_2\,,
\end{split}
\end{align}
where $x^{\pm} =\pm  t+ \theta$.

In the AdS$_3$ literature it is common to work in a different convention than the one we used in the beginning of this Section. In this more common convention one writes out the Chern-Simons gauge fields as $2\times 2$ matrices, i.e. takes the $J_A$ and $\bar{J}_A$ to be the generators of the fundamental representation of $\mathfrak{sl}(2;\mathbb{R})$. In this choice, we write out $A$ and $\bar{A}$ as 
\beq
\label{E:globalAdS3}
	A = \begin{pmatrix} \frac{dr}{2\sqrt{r^2+1}} & -\frac{(\sqrt{r^2+1}-r)dx^+}{2} \\ \frac{(\sqrt{r^2+1}+r)dx^+}{2} & -\frac{dr}{2\sqrt{r^2+1}}\end{pmatrix}\,,
	\qquad
	\bar{A} = \begin{pmatrix} -\frac{dr}{2\sqrt{r^2+1}} & -\frac{(\sqrt{r^2+1}+r)dx^-}{2} \\ \frac{(\sqrt{r^2+1}-r)dx^-}{2} & \frac{dr}{2\sqrt{r^2+1}}\end{pmatrix}\,.
\eeq
In the Chern-Simons formulation the Einstein's equations and torsion-free constraint correspond to $A$ and $\bar{A}$ being flat. So locally we have
\beq
\label{E:flat}
	A = g^{-1} dg\,, \qquad \bar{A} = \bar{g}^{-1}d\bar{g}^{-1}\,,
\eeq
and one may find representatives for $g$ and $\bar{g}$,
\beq
	g = \begin{pmatrix}\rho\,\cos\left(\frac{x^+}{2}\right) & -\rho^{-1}\sin\left(\frac{x^+}{2}\right) \\ \rho \sin\left(\frac{x^+}{2}\right) & \rho^{-1}\cos\left(\frac{x^+}{2}\right)\end{pmatrix}\,,
	\qquad
	\bar{g} = \begin{pmatrix} \rho^{-1} \cos\left( \frac{x^-}{2}\right) & -\rho\sin\left(\frac{x^-}{2}\right) \\ \rho^{-1} \sin\left(\frac{x^-}{2}\right) & \rho \cos\left(\frac{x^-}{2}\right)\end{pmatrix}\,,
\eeq
where
\beq
	\rho = \sqrt{\sqrt{r^2+1}+r}\,.
\eeq

These representatives illustrate two facts which will be useful. First, these representatives are double-valued: going around the circle $\theta \to \theta+2\pi$ takes us from $g \to -g$ and $\bar{g}\to -\bar{g}$. Relatedly, the $SL(2;\mathbb{R})\times SL(2;\mathbb{R})$ holonomy around the spatial circle is non-trivial:
\beq
	\mathcal{P}e^{\int_0^{2\pi} d\phi\,A_{\phi}} = -I\,, \qquad \mathcal{P}e^{\int_0^{2\pi}d\phi\,\bar{A}_{\phi}} = -I\,.
\eeq
It follows that $A$ and $\bar{A}$ are singular as $SL(2;\mathbb{R})$ connections.

This seems like an undesirable state of affairs. Global AdS$_3$ ought to be a non-singular configuration. Fortunately there is a simple remedy: rather than working with $SL(2;\mathbb{R})$ we quotient it by the $\mathbb{Z}_2$ subgroup $\{I,-I\}$, thereby identifying $g\in SL(2;\mathbb{R})$ with $-g$. The resulting group is $SL(2;\mathbb{R})/\mathbb{Z}_2 = PSL(2;\mathbb{R})=SO(2,1)$. It follows that $A$ and $\bar{A}$ are perfectly non-singular as $SO(2,1)$ connections. A corollary is that $A$ and $\bar{A}$ are singular as connections for any cover of $SO(2,1)\times SO(2,1)$. 

That is, if we demand that global AdS$_3$ is a non-singular configuration, then the gauge group of the Chern-Simons theory is fixed to be $SO(2,1)\times SO(2,1)$. To our knowledge this was first noted in~\cite{Castro:2011iw}.

The second fact is the following. $SO(2,1)$ is contractible to its maximum compact subgroup $SO(2)$. At fixed $(t,r)$, $g$ and $\bar{g}$ are maps from the spatial circle $\theta$ into this subgroup, winding the group circle exactly once. Because any classical configuration is locally AdS$_3$, this is an example of a more general lesson: given any solution to Einstein's equations with a contractible cycle $C$, writing the connections $A$ and $\bar{A}$ as~\eqref{E:flat}, the gauge transformation parameters wind the group circle once around $C$.

We proceed to quantize on this space following~\cite{Coussaert:1995zp,Elitzur:1989nr}. (For work related to~\cite{Coussaert:1995zp} and the WZW description of Chern-Simons  theory on AdS$_3$ see~\cite{Carlip:1996yb,Martinec:1998wm}.) We separate $A$ and $\bar{A}$ into temporal and spatial parts,
\beq
	A = A_0 dt + \tilde{A}_i dx^i\,, \qquad \bar{A} = \bar{A}_0 dt + \tilde{\bar{A}}_i dx^i\,.
\eeq
The total action, including boundary terms is
\beq
\label{E:CSaction}
	S_{\rm grav} =  S[A]-S[\bar{A}] + S_{\rm bdy}\,, 
\eeq
with\footnote{We choose an orientation so that $\epsilon^{t\theta r}=\frac{1}{\sqrt{-g}}$. Also, in~\eqref{E:canonicalCS}, we are using a different convention for $k$ than in the Introduction, with $k_{\rm there} = 4 k_{\rm here}$.}
\begin{align}
\begin{split}
\label{E:canonicalCS}
	 S[A] &= -\frac{k}{2\pi}  \int_{\mathcal{M}} dt \wedge  \text{tr}'\left(- \frac{1}{2}\tilde{A} \wedge \dot{\tilde{A}} + A_0 \tilde{F}\right)\,,
	\\
	S_{\rm bdy} &= - \frac{k}{4\pi}\int_{\partial\mathcal{M}} d^2x \left( \text{tr}'(A_{\theta}^2) + \text{tr}'(\bar{A}_{\theta}^2)\right)\,.
\end{split}
\end{align}
Here $k = \frac{1}{4G}$, $\tilde{F}$ is the spatial field strength $\tilde{F} = \tilde{d}\tilde{A} + \tilde{A}\wedge \tilde{A}$ with $\tilde{d}$ the spatial exterior derivative, and tr$'$ denotes the trace in the fundamental representation of $SL(2;\mathbb{R})$. The unconventional boundary term is required~\cite{Coussaert:1995zp} in order to enforce a variational principle consistent with the AdS$_3$ boundary conditions,\footnote{A similar boundary term appears in the treatment of ordinary Chern-Simons theory on AdS$_3$~\cite{Kraus:2006wn,Jensen:2010em}.} which are that $A$ and $\bar{A}$ asymptote as $r\to\infty$ as~\cite{Brown:1986nw,Coussaert:1995zp}
\beq
\label{E:AdSBC}
	A = \begin{pmatrix} \frac{dr}{2r}  + O(r^{-2}) & O(r^{-1}) \\ r dx^+ +O(r^{-1}) & - \frac{dr}{2r} + O(r^{-2})\end{pmatrix}\,, 
	\quad 
	\bar{A} = \begin{pmatrix} -\frac{dr}{2r}+O(r^{-2}) & -r dx^- + O(r^{-1}) \\ O(r^{-1}) & \frac{dr}{2r}+O(r^{-2})\end{pmatrix}\,,
\eeq
and $A$ and $\bar{A}$ are allowed to fluctuate at the indicated powers in $1/r$. The connections describing global AdS$_3$~\eqref{E:globalAdS3} clearly respect these boundary conditions. Consistent with these boundary conditions, we only allow bulk gauge transformations $\Lambda$ and $\bar{\Lambda}$ which die off near the AdS boundary.

The on-shell variation of the action, including the boundary term, is
\beq
\label{E:deltaS}
	\delta S_{\rm grav} =- \frac{k}{\pi}\int_{\partial\mathcal{M}} d^2x \left\{\text{tr}'\left(A_-\delta A_{\theta} \right)+\text{tr}'\left(\bar{A}_+\delta \bar{A}_{\theta}\right)\right\}\,,
\eeq
which vanishes by the boundary conditions~\eqref{E:AdSBC} as advertised.

The fields $A_0$ and $\bar{A}_0$ appear as Lagrange multiplier fields inside the action~\eqref{E:CSaction}, enforcing the constraint that the spatial field strength vanishes. We observe that although $A_0$ and $\bar{A}_0$ drop out of the action, they can be gauge-fixed in a way that is consistent with the asymptotically AdS$_3$ boundary conditions. Namely, we may fix them to their values for global AdS$_3$ in~\eqref{E:globalAdS3} as 
\beq
	A_0 = \begin{pmatrix} 0 & -\frac{\sqrt{r^2+1}-r}{2} \\ \frac{\sqrt{r^2+1}+r}{2} & 0 \end{pmatrix}\,, \qquad \bar{A}_0 = \begin{pmatrix} 0 & \frac{\sqrt{r^2+1}+r}{2} \\ -\frac{\sqrt{r^2+1}-r}{2} & 0 \end{pmatrix}\,.
\eeq

Before going on, some words are in order regarding gauge-fixing and ghosts. In the usual equivalence between CS theory on a cylinder and a chiral WZW model on its boundary, one allows for bulk gauge transformations that vanish at the boundary of the cylinder. The details of the bulk gauge fixing are then immaterial from the point of view of the WZW model, as any ghosts introduced by the gauge-fixing decouple on the boundary. Indeed, in the natural gauge choice $A_0=0$, the Fadeev-Popov determinant does not introduce a coupling between the ghosts and the physical degrees of freedom. Similar considerations hold for our gauge theory in AdS, as we require that all bulk gauge transformations fall off near the AdS boundary.

The remaining functional integral is taken over the moduli space of flat connections on the disk, which we parameterize as
\beq
	\tilde{A} = g^{-1} \tilde{d} g\,, \qquad \tilde{\bar{A}} = \bar{g}^{-1} \tilde{d} \bar{g}\,.
\eeq
This decomposition is redundant: we get the same spatial connection $A$ for $g(\vec{x},t)$ and $h(t)g(\vec{x},t)$ for any $h(t) \in PSL(2;\mathbb{R})$, and similarly for $\tilde{\bar{A}}$. We identify these configurations in the residual integral over $g$, so that the field $g$ is subject to a quasi-local $SO(2,1)$ quotient. 

In terms of $g$ the original action~\eqref{E:CSaction} becomes a difference of chiral WZW actions,
\beq
	S = S_-[g]+S_+[\bar{g}]\,, 
\eeq
with
\beq
\label{E:chiralWZW}
	 S_{\pm}[g]=\frac{k}{2\pi}\left( \int_{\partial\mathcal{M}} d^2x \,\text{tr}' \big((g^{-1})'\partial_{\pm} g\big)  \mp \frac{1}{6}\int_{\mathcal{M}}\text{tr}'(g^{-1}dg \wedge g^{-1}dg \wedge g^{-1}dg)\right)\,.
\eeq
Here $\partial_{\pm} = \frac{1}{2}( \partial_{\theta}\pm \partial_t)$ and \,$'$\, indicates an angular derivative. Now recall the topological property of the map $g$ we noted above for global AdS$_3$. In the quantum theory we enforce the same property: around the contractible $\theta$-circle, we integrate over $g$ and $\bar{g}$ which wind once around the circle in $SO(2,1)$. It remains to enforce this boundary condition as well as to translate the AdS$_3$ boundary conditions~\eqref{E:AdSBC} into boundary conditions on $g$ and $\bar{g}$.

To proceed we find it helpful to use a Gauss parameterization of $SL(2;\mathbb{R})$ group elements as in~\cite{Alekseev:1988ce},\footnote{The Gauss parameterization, with $F,\Psi\in \mathbb{R}$ and $\lambda>0$, covers $PSL(2;\mathbb{R})$ but not $SL(2;\mathbb{R})$. This is sufficient for our purposes.}
\beq
\label{E:Gauss}
	g = \begin{pmatrix} 1 & 0 \\ F& 1\end{pmatrix} \begin{pmatrix}\lambda & 0 \\ 0 & \lambda^{-1} \end{pmatrix} \begin{pmatrix} 1 & \Psi \\ 0 & 1 \end{pmatrix}\,.
\eeq
The corresponding spatial connection is
\beq
\label{E:gaussA}
	\tilde{A} =\begin{pmatrix} \tilde{d}\ln \lambda -\Psi(\lambda^2\tilde{d}F)\,\,\,\,\,\,\,\,\, & 2\Psi \tilde{d}\ln\lambda +\tilde{d}\Psi- \Psi^2 (\lambda^2\tilde{d}F) \\ \lambda^2 \tilde{d}F \,\,\,\,\,\,\,\,\, &-\tilde{d}\ln\lambda+ \Psi (\lambda^2\tilde{d}F) \end{pmatrix}\,.
\eeq
Comparing with~\eqref{E:AdSBC} we see that the fields $\lambda$ and $\Psi$ are fixed as $r\to\infty$ in terms of $F$ as
\beq
\label{E:constraint}
	\lambda = \sqrt{\frac{r}{F'}} \,, \qquad \Psi = - \frac{F''}{2rF'}\,,  
\eeq
with $F$ finite. 
The constraint on $\lambda$ comes from matching the bottom left components of~\eqref{E:AdSBC} and~\eqref{E:gaussA}, and the constraint on $\Psi$ comes from matching the diagonal components. 
There are similar relations for the barred fields, for which it is convenient to parameterize $\bar{g}$ as
\beq
\label{E:gauss2}
	\bar{g} = \begin{pmatrix} 1 & -\bar{F} \\ 0 & 1 \end{pmatrix} \begin{pmatrix} \bar{\lambda}^{-1} & 0 \\ 0 & \bar{\lambda}\end{pmatrix} \begin{pmatrix} 1 & 0 \\  \bar{\Psi} & 1\end{pmatrix}\,.
\eeq
The asymptotically AdS boundary conditions enforce that as $r\to\infty$
\beq
	\bar{\lambda} = \sqrt{\frac{r}{\bar{F}'}}\,, \qquad \bar{\Psi} = - \frac{\bar{F}''}{2r\bar{F}'}\,.
\eeq
Parameterizing the boundary value of $F$ as $F|_{\partial} = \tan \left(\frac{\phi}{2}\right)$ so that $\phi$ is an angular variable, the single-valuedness of $g$ implies that $\phi'>0$ and the winding property amounts to the boundary condition $\phi(\theta+2\pi,t)=\phi(\theta,t)+2\pi$. That is, at fixed time, $\phi$ is an element of Diff$(\mathbb{S}^1)$.

That is not the end of the story, as we have to account for the quasi-local $SO(2,1)$ quotient. In terms of $(\lambda,\Psi,F)$, the action $g(\vec{x},t) \to h(t) g(\vec{x},t)$ is
\beq
\label{E:Raction}
	\lambda \to (cF+d)\lambda\,, \qquad \Psi \to \Psi + \frac{c \lambda^{-2}}{cF+d}\,, \qquad F \to \frac{a F+b}{cF+d}\,, \qquad h = \begin{pmatrix} d & c \\ b & a \end{pmatrix}\,.
\eeq
This action is consistent with the constraints~\eqref{E:constraint}. It then follows that in the residual path integral over $\phi$ we identify
\beq
\label{E:theQuotient}
	\tan\left( \frac{\phi(\theta,t)}{2}\right)\sim\frac{a(t) \tan\left(\frac{\phi(\theta,t)}{2}\right)+b(t) }{c(t)\tan\left(\frac{\phi(\theta,t)}{2}\right)+d(t)}\,.
\eeq
So the precise statement is that at any fixed time $\phi$ is an element of Diff$(\mathbb{S}^1)/PSL(2;\mathbb{R})$.

In the Gauss parameterization~\eqref{E:Gauss} the kinetic and WZW terms in~\eqref{E:chiralWZW} evaluate to
\begin{align}
\begin{split}
	\int_{\partial\mathcal{M}}d^2x \,\text{tr}'((g^{-1})'\partial_- g) & = -\int_{\partial\mathcal{M}} d^2x \left( \frac{2\lambda' \partial_- \lambda}{\lambda^2}+\lambda^2 F' \Psi'- \frac{\lambda^2}{2}(\dot{\Psi}F'+\Psi'\dot{F})\right) \,,
	\\
	\frac{1}{3}\int_{\mathcal{M}} \text{tr}'\big((g^{-1}dg )^3\big) &=\int_{\mathcal{M}} d\lambda^2 \wedge d\Psi \wedge dF = \int_{\partial\mathcal{M}} \lambda^2 d\Psi \wedge dF\,,
\end{split}
\end{align}
with similar relations for the barred fields. The end result is that
\beq
\label{E:almostThere}
	S_{\pm}[g] = -\frac{k}{\pi}\int_{\partial\mathcal{M}} d^2x \left( \frac{\lambda' (\partial_{\pm}\lambda)}{\lambda^2} +\lambda^2F' (\partial_{\pm}\Psi)\right)\,.
\eeq
Plugging in~\eqref{E:constraint}, which may be viewed as a constraint on the WZW model, and using $F|_{\partial} = \tan\left( \frac{\phi}{2}\right)$, we obtain the boundary action
\beq
\label{E:finalS}
	S_{\pm}[\phi] = -\frac{C}{24\pi}\int d^2x \left( \frac{\phi'' \partial_{\pm} \phi'}{\phi'^2} - \phi' \partial_{\pm} \phi\right)\,, \qquad C = 6k = \frac{3}{2G}\,.
\eeq
Combined with the boundary condition on $\phi$ and the quasi-local quotient, we recognize the left-moving half $S_+$ as exactly the Alekseev-Shatashvili quantization of Diff$(\mathbb{S}^1)/PSL(2;\mathbb{R})$ in~\eqref{E:ourTheory}. The coupling constant $C$ is fixed to the Brown-Henneaux central charge.

What is the path integral measure for $\phi$? The bulk measure for $A$ leads to the usual Haar measure for the boundary degree of freedom $g$, which in terms of the Gauss parameterization~\eqref{E:Gauss} is
\beq
	[dg] = \prod_{\theta,t} d\lambda d\Psi dF \, \lambda\,.
\eeq
It is invariant under both the right-action~\eqref{E:Raction} and a left-action $g\to gh^{-1}$. Now we account for the constraints on $\lambda$ and $\Psi$. Comparing~\eqref{E:AdSBC} with~\eqref{E:gaussA}, we see that the bottom left entry reads $\lambda^2 F' = r$, and the diagonal entries lead to a linear constraint on $\Psi$. The Haar measure after the constraints then reads
\beq
\label{E:measure1}
	\prod_{\theta,t}dF\int d\lambda d\Psi \,\lambda\, \delta(\lambda^2F'-r)\delta\left(\Psi+\frac{F''}{2rF'}\right)=\prod_{\theta,t} \frac{dF}{F'}=\prod_{\theta,t} \frac{d\phi}{\phi'}\,,
\eeq
which is invariant under the quasi-local quotient~\eqref{E:theQuotient}. (We note that Alekseev and Shatashvili obtained the same measure for the quantization of Diff$(\mathbb{S}^1)/PSL(2;\mathbb{R})$~\cite{Alekseev:1988ce}.) In Section~\ref{S:localization} we discuss the measure for the quantization of Diff$(\mathbb{S}^1)/PSL(2;\mathbb{R})$. At the end of that Subsection we note that the natural measure for the quantization, which we notate as $[d\phi] \text{Pf}(\omega)$, is equivalent to~\eqref{E:measure1}.

In summary, we have shown that the Chern-Simons description of gravity on global AdS$_3$ is quantum mechanically equivalent to the path integral quantization of (two copies) of Diff$(\mathbb{S}^1)/PSL(2;\mathbb{R})$.

Recall that in the quantization of coadjoint orbits one may associate a Hamiltonian with any element of the Virasoro algebra. We observe that AdS$_3$ gravity hands us the Lorentz-invariant quantization, with Hamiltonian corresponding to $L_0$. This is a consequence of the boundary term in~\eqref{E:CSaction}. Had we neglected it, we would have instead obtained an action
\beq
	\mp \frac{C}{48\pi}\int d^2x \left( \frac{\dot{\phi}' \phi''}{\phi'^2}-\dot{\phi}\phi'\right)\,,
\eeq
i.e. the appropriate action with $H=0$~\eqref{E:H=0}. 

As we mentioned in the Introduction, our analysis here corrects that of~\cite{Coussaert:1995zp}. The new ingredients are: (i) finding the constraint that fixes the field $\Psi$, and more importantly (ii) accounting for the quasi-local $PSL(2;\mathbb{R})\times PSL(2;\mathbb{R})$ quotient and deducing the winding boundary condition. That is, the new work above demonstrates that the boundary fields $\phi$ and $\bar{\phi}$ are, at fixed time, elements of Diff$(\mathbb{S}^1)/PSL(2;\mathbb{R})$. Following similar arguments as~\cite{Alekseev:1988ce}, we have also shown that the gravitational description leads to the correct measure.

Our analysis immediately extends to Euclidean global AdS$_3$, i.e.\! to the quantization of $SO(2,1) \times SO(2,1)$ Chern-Simons theory around the space 
\beq
\label{E:euclideanGlobalAdS}
	 (r^2+1)dy^2 + r^2 d\theta^2 + \frac{dr^2}{r^2+1}\,, \qquad \theta \sim \theta+2\pi\,.
\eeq 
To get to here from global AdS$_3$ we have simply Wick-rotated $t = - i y$. Repeating our analysis above leads to the Wick-rotation of~\eqref{E:finalS}, i.e. to the action
\beq
\label{E:euclideanS}
	S_+[\phi] = \frac{C}{24\pi}\int d^2x \left( \frac{\phi'' \bar{\partial} \phi'}{\phi'^2}-\phi'\bar{\partial}\phi\right)\,, \qquad \bar{\partial} = \frac{1}{2}(\partial_{\theta}+i\partial_y)\,.
\eeq

We may also obtain the boundary action for fluctuations around Euclidean BTZ black holes. In these geometries the bulk space asymptotes to a boundary torus, and a cycle of the boundary torus is contractible in the bulk. When one speaks of a Euclidean black hole, one refers to a geometry for which the Euclidean time circle is contractible, while ``thermal global AdS$_3$'' refers to another geometry where the spatial circle is contractible.

For fixed boundary complex structure $\tau$ there are infinitely many smooth solutions to Einstein's equations which asymptote to the same boundary torus. These geometries are distinguished by which cycle is contractible in the bulk. There is a unique geometry for which the spatial circle is contractible. The metric on this space is~\eqref{E:euclideanGlobalAdS}, subject to the further identification
\beq
	y \sim y+2\pi \,\text{Im}(\tau)\,, \qquad \theta \sim \theta + 2\pi \,\text{Re}(\tau)\,.
\eeq
Following our algorithm above, the action which describes fluctuations around this space is the same as for Euclidean global AdS$_3$,~\eqref{E:euclideanS}. The only difference is that now the boundary conditions are modified to be
\begin{align}
\begin{split}
\label{E:euclideanBC}
	\phi(\theta+2\pi,y) & = \phi(\theta,y)+2\pi\,,
	\\
	\phi(\theta+2\pi \text{Re}(\tau),y+2\pi \text{Im}(\tau)) & = \phi(\theta,y)\,.
\end{split}
\end{align}
We further investigate the properties of these Euclidean black holes in the next Subsection.

The derivation above for global AdS$_3$ bears a striking similarity to the so-called Drinfeld-Sokolov reduction of a chiral $SL(2;\mathbb{R})$ WZW model to the quantization of Diff$(\mathbb{S}^1)/PSL(2;\mathbb{R})$ described by Alekseev and Shatashvili~\cite{Alekseev:1988ce}. The two seem to be almost, but not quite the same. In the Drinkfeld-Sokolov reduction one starts with a $SL(2;\mathbb{R})$ model and imposes the constraint on the $SL(2;\mathbb{R})$ current 
\beq
	\text{tr} (U\cdot g^{-1} g') = 1 \,, \qquad U = \begin{pmatrix} 0 & 1 \\ 0 & 0 \end{pmatrix}\,.
\eeq
Parameterizing $g$ using Gauss parameters as in~\eqref{E:Gauss}, this constraint fixes $\lambda = \frac{1}{\sqrt{F'}}$. It is essentially the same as the AdS$_3$ boundary condition that fixes $\lambda$ in terms of $F$ in~\eqref{E:constraint}. However $\Psi$ remains unfixed. Fortunately, $\Psi$ only appears in the constrained chiral WZW action~\eqref{E:almostThere} through the last term, which is a total derivative upon plugging in the constrained value of $\lambda$. Ignoring that term, the action of the constrained model is exactly the quantization with $H=0$. 

However, it is not enough to obtain the action of the quantization. The field $\phi$ must also be an element of Diff$(\mathbb{S}^1)/PSL(2;\mathbb{R})$. The quotient is accounted for in the same way as in the derivation from gravity, but we have not seen an argument that the boundary condition $\phi(\theta+2\pi,t) = \phi(\theta,t)+2\pi$ arises in the reduction. 

\subsection{More on Euclidean black holes}

In the last Subsection we derived the boundary model appropriate for global AdS$_3$ as well as for Euclidean black holes. Our discussion of the Euclidean models was rather brief. Here we expound on their physics.

Consider a Euclidean BTZ geometry with torus boundary of complex structure $\tau$, whose spatial circle is contractible in the bulk. For pure imaginary $\tau$, such a geometry is often called ``thermal global AdS,'' since it is global AdS with compactified Euclidean time. The relevant boundary action for the holomorphic half was given in~\eqref{E:euclideanS} with a corresponding action for the anti-holomorphic half, subject to the boundary conditions~\eqref{E:euclideanBC} and a quasi-local quotient. This theory is weakly coupled as $C\gg 1$. Let us consider the holomorphic half. The field equation of the model is
\beq
	\bar{\partial} T = 0\,, \qquad T = -\frac{C}{12}\left\{\tan\left(\frac{\phi}{2}\right),\theta\right\}\,.
\eeq
Up to the time-dependent $PSL(2;\mathbb{R})$ redundancy, there is a unique solution consistent with the boundary conditions, given by
\beq
	\phi_0 = \theta - \frac{\text{Re}(\tau)}{\text{Im}(\tau)}y\,.
\eeq
The saddle point approximation to the stress tensor is then
\beq
	\langle T\rangle  = -\frac{C}{24} + O(C^0)\,,
\eeq
i.e. the Casimir energy on the circle. Now accounting for the anti-holomorphic half, the saddle point approximation to the torus partition function reads
\beq
\label{E:torusSaddleZ}
 -\ln \mathcal{Z} = S_+[\phi_0] +S_-[\phi_0] +O(C^0) = \frac{ C}{24}2\pi i (\tau-\bar{\tau}) + O(C^0)\,.
\eeq

This partition function is obviously not modular invariant. Among other things, it does not exhibit Cardy-like behavior at high temperature, as $\tau\to 0$. As we will uncover in Section~\ref{S:torus}, the torus partition function may be computed exactly via localization. The result is the vacuum character, of which~\eqref{E:torusSaddleZ} is its large $C$ approximation. This is consistent with the fact that the geometric model does not have infinitely many saddles, corresponding to the infinite number of bulk saddles asymptoting to the same boundary torus. Instead, the model accounts for fluctuations around a single saddle, without knowledge of the others. For this reason the geometric model does not exhibit a Hawking-Page transition.

If the spatial circle is the $A$ cycle of the torus, then what one usually means by a Euclidean black hole is a space for which the $B$ cycle is contractible in the bulk. The holomorphic half of the action one obtains from such a background is~\eqref{E:euclideanS}, with angular derivatives replaced by a normalized derivative along the $B$ cycle. The boundary conditions become
\begin{align}
\begin{split}
	\phi(\theta+2\pi,y) & = \phi(\theta,y)\,,
	\\
	\phi(\theta+2\pi \text{Re}(\tau),y+2\pi \text{Im}(\tau)) & = \phi(\theta,y)+2\pi\,,
\end{split}
\end{align}
and the $PSL(2;\mathbb{R})$ quotient identifies
\beq
	\tan\left(\frac{\phi(\theta,y)}{2}\right) \sim \frac{a \tan\left(\frac{\phi(\theta,y)}{2}\right)+b}{c \tan\left(\frac{\phi(\theta,y)}{2}\right)+d}\,,
\eeq
where $a,b,c,d$ are functions of $\theta-\frac{\text{Re}(\tau)}{\text{Im}(\tau)}y$ satisfying $ad-bc=1$. There is again a unique solution to the equation of motion for $\phi$ modulo the quotient,
\beq
	\phi_0 = \frac{y}{\text{Im}(\tau)}\,.
\eeq
The saddle point approximation to the partition function is now given by
\beq
	-\ln \mathcal{Z} = -\frac{C}{24}2\pi i \frac{\tau-\bar{\tau}}{|\tau|^2} + O(C^0)\,.
\eeq
This is the semiclassical approximation to the Euclidean BTZ partition function (see e.g.~\cite{Kraus:2006wn}), which exhibits the requisite Cardy-like growth at high temperature. It is also the image of the partition function~\eqref{E:torusSaddleZ} under the modular $S$-transformation $\tau\to - \frac{1}{\tau}$, which swaps the $A$ and $B$ cycles. 

\subsection{$PSL(2;\mathbb{R})$ currents}
\label{S:noether}

Let us go back to the boundary action describing the boundary gravitons of global AdS$_3$, and in particular its left-moving half. We identify field configurations under the quasi-local $PSL(2;\mathbb{R})$ redundancy in~\eqref{E:theQuotient}. It is implicit in this statement that the action for the quantization, Eq.~\eqref{E:finalS}, is invariant under 
\beq
	\tan\left( \frac{\phi(\theta,t)}{2}\right) \to \frac{a(t) \tan\left( \frac{\phi(\theta,t)}{2}\right)+b(t)}{c(t)\tan\left(\frac{\phi(\theta,t)}{2}\right)+d(t)}\,,
\eeq
although this is not manifest. It is straightforward to obtain the corresponding Noether currents. An infinitesimal $h \in PSL(2;\mathbb{R})$ transformation is parameterized by
\beq
	h(t) = I_2 +\varepsilon  + O(\varepsilon^2)  \,, \qquad \varepsilon = \begin{pmatrix} \frac{\epsilon_0(t)}{2} & \epsilon_+(t) \\ \epsilon_-(t) & - \frac{\epsilon_0(t)}{2}\end{pmatrix}\,,
\eeq
under which $F = \tan\left(\frac{\phi}{2}\right)$ varies as
\beq
	\delta F =   \epsilon_+ +\epsilon_0 F- \epsilon_-\,F^2\,.
\eeq
After the addition of a suitable improvement term, the corresponding conserved current is
\beq
\label{E:noether}
	J^0 =0\,, 
	\quad
	J^{\theta}  = \frac{C}{12\pi}\left\{ 
\frac{\partial_+F}{F'^3}\left( F''\delta F'-F'\delta F''\right)+\frac{\partial_+F'}{F'^3}(F'\delta F)' -\frac{\partial_+F''}{F'^2}\delta F\right\}\,.
\eeq
It follows that, at each time, $J^{\theta}$ is independent of $\theta$
\beq
	\frac{\partial J^{\theta}}{\partial\theta}=0\,.
\eeq
From $J^\theta$ we define three families of ``conserved charges'' by
\beq
	Q_0 = \frac{\partial J^{\theta}}{\partial\epsilon_0}\,, \qquad Q_{\pm} = \frac{\partial J^{\theta}}{\partial \epsilon_{\mp}}\,,
\eeq
which we package into a matrix as
\beq
	Q = \begin{pmatrix} Q_0 & Q_+ \\ Q_- & -Q_0\end{pmatrix}\,.
\eeq
Under constant $PSL(2;\mathbb{R})$ transformations $h$ this matrix of charges transforms by conjugation,
\beq
	Q \to h Q h^{-1}\,.
\eeq
The quadratic Casimir is
\beq
	\mathcal{H} = \frac{6\pi}{C}\text{tr}'(Q^2) = \frac{12\pi}{C}\left( Q_0^2 + Q_+Q_-\right)\,.
\eeq
However, under time-dependent transformations $Q$ does not transform covariantly. Under an infinitesimal transformation $h=I_2 + \varepsilon$ we have
\beq
	\delta_\varepsilon Q = [\varepsilon,Q]+\frac{C}{12\pi}\,\dot{\varepsilon}\,.
\eeq
Consequently there are no ``gauge-invariant'' observables constructed from these charges.

This matrix of charges is nothing more than the current conjugate to the variation $\delta_{\varepsilon} \bar{g} = \varepsilon \bar{g}$ of the group-valued field $\bar{g}$, where we now allow $\varepsilon$ to be a function of $t$ and $\theta$. Under the variation the chiral WZW action transforms as
\beq
	\delta_{\varepsilon} S_+ = \frac{C}{6\pi }\int d^2x\, \text{tr}'\left( \varepsilon ((\partial_+ \bar{g} )\bar{g}^{-1})'\right) =\int d^2x \,\text{tr}'\left( \varepsilon Q'\right)\,,
\eeq
which gives
\beq
	Q = \frac{C}{6\pi}(\partial_+\bar{g}) \bar{g}^{-1}\,.
\eeq
Using either the expression~\eqref{E:noether} or the variation of $S_+$ above, we find
\begin{align}
\begin{split}
	Q_- & = -\frac{C}{12\pi}\left( \frac{\partial_+F''}{F'^2}-\frac{F''\partial_+F'}{F'^3}\right)\,,
	\\
	Q_0 & =\frac{C}{12\pi}\left( \frac{\partial_+F'}{F'}-F\left(\frac{\partial_+F''}{F'^2}- \frac{F''\partial_+F'}{F'^3}\right)\right) \,,
	\\
	Q_+ & =\frac{C}{12\pi}\left( 2\partial_+F-\frac{2F\partial_+F'}{F'}+F^2\left( \frac{\partial_+F''}{F'^2}-\frac{F''\partial_+F'}{F'^3}\right)  \right) \,,
	\\
	\mathcal{H} & = \frac{C}{12\pi}\left( \frac{(\partial_+F')^2}{F'^2}-2\partial_+F\left( \frac{\partial_+F''}{F'^2}-\frac{F''\partial_+F'}{F'^3}\right)\right)\,.
\end{split}
\end{align}
For a Euclidean AdS$_3$ geometry with torus boundary of complex structure $\tau$ as discussed in the previous Subsection, choosing the $\theta$ circle to be contractible, these charges evaluate to
\beq
	Q_+ = -Q_- = \frac{C}{24\pi}\frac{i \tau}{\text{Im}(\tau)}\,, \qquad Q_0=0\,.
\eeq

\subsection{Conical defects and the Poincar\'e patch}
\label{S:deficits}

Consider a conical defect, described by the Lorentzian metric
\beq
	 -(r^2+\alpha^2) dt^2 + r^2 d\theta^2 + \frac{dr^2}{r^2+\alpha^2}\,, 
\eeq
with $\alpha \in (0,1)$. This geometry is supported by a massive probe particle at $r=0$, and the total energy is $-\frac{\alpha^2}{8G} = - \frac{C \alpha^2}{12}$. Taking the dreibein to be
\beq
	e^0 = \sqrt{r^2+\alpha^2}\,dt\,, \qquad e^1 = r d\theta\,, \qquad e^r = \frac{dr}{\sqrt{r^2+\alpha^2}}\,,
\eeq
the connection $A$ may be written as $A=g^{-1}dg$ with
\beq
	g = \begin{pmatrix} \rho \cos\left(\frac{\alpha x^+}{2}\right) & - \rho^{-1} \sin\left(\frac{\alpha x^+}{2}\right) \\ \rho \sin\left(\frac{\alpha x^+}{2}\right) & \rho^{-1} \cos\left(\frac{\alpha x^+}{2}\right)\end{pmatrix}\,, \qquad \rho = \sqrt{\sqrt{\frac{r^2}{\alpha^2}+1}+\frac{r}{\alpha}}\,.
\eeq
There are similar expressions for $\bar{A}$. The holonomy around the $\theta$-circle is now non-trivial. As above, we proceed by gauge-fixing $A_0$ and integrating over locally flat spatial connections $\tilde{A}$ with this holonomy. In terms of the Gauss parameterization, one finds that $\lambda$ and $\Psi$ are constrained as in~\eqref{E:constraint} with $F$ finite near the boundary. However, on account of the holonomy, $F$ behaves on the boundary as  $F|_{\partial} = \tan \left( \frac{\alpha \phi}{2}\right)$ where $\phi$ is at fixed time a Diff($\mathbb{S}^1)$ field obeying $\phi(\theta+2\pi,t)=\phi(\theta,t)+2\pi$. Further, only a $U(1)$ subgroup of the $PSL(2;\mathbb{R})$ redundancy is consistent with this boundary condition, and so $\phi$ is in fact a Diff$(\mathbb{S}^1)/U(1)$ field at fixed time. The end result is that the boundary action is
\beq
	S = S_-[\phi] + S_+[\bar{\phi}]\,, \qquad S_{\pm}[\phi] =  -\frac{C}{24\pi}\int d^2x\left( \frac{\phi'' \partial_{\pm}\phi'}{\phi'^2}-\alpha^2 \phi' \partial_{\pm}\phi\right)\,.
\eeq
We recognize this as the quantization of the normal orbits of the Virasoro group, Eq.~\eqref{E:ourTheory}, with the identification $ \alpha^2=-\frac{b_0}{48\pi C}$.

Near the conical deficit the metric is approximately
\beq
	-\alpha^2 dt^2 + \frac{1}{\alpha^2}(dr^2+r^2 d\tilde{\theta}^2)\,, \qquad \tilde{\theta} \sim \tilde{\theta} + 2\pi\alpha\,.
\eeq
The periodicity of $\tilde{\theta}$ is also the periodicity of $\alpha\phi$, which is $2\pi \alpha$. Effectively, the bulk deficit leads to a deficit in the radius of the Diff$(\mathbb{S}^1)$ field on the boundary.

For $\alpha^2<0$ we have the exterior of a BTZ black hole rather than a conical defect. This geometry is geodesically incomplete and one should instead consider an extension. In Section~\ref{S:BTZ} we adapt our analysis to the two-sided eternal BTZ black hole following a morally similar derivation for nearly AdS$_2$ gravity described in~\cite{Maldacena:2016upp} and elaborated on in~\cite{Harlow:2018tqv}.

The Poincar\'e patch of AdS$_3$ is described by the metric
\beq
	r^2(- dt^2 + dx^2) + \frac{dr^2}{r^2}\,,
\eeq
where $x\in \mathbb{R}$ and $r>0$. The conformal boundary is at $r\to\infty $ and it has the topology of the plane, while constant-time slices are given by half-space. Running through the same sort of argument as above one finds that the fields $\lambda$ and $\Psi$ are constrained as before but now $F$ behaves on the boundary as $F|_{\partial} = \phi$ where at fixed time $\phi \in \text{Diff}(\mathbb{R})/PSL(2;\mathbb{R})$. To be more explicit, $\phi$ is monotone and obeys the boundary condition
\beq
	\lim_{x\to\pm \infty} \frac{\phi(x,t)}{x} = 1\,,
\eeq
and moreover is subject to the quasi-local quotient
\beq
	\phi(x,t) \sim \frac{a(t) \phi(x,t)+b(t)}{c(t)\phi(x,t)+d(t)}\,.
\eeq
The chiral actions are now
\beq
\label{E:poincareS}
	S_{\pm}[\phi] =  -\frac{C}{24\pi} \int d^2x \,\frac{\phi'' \partial_{\pm}\phi'}{\phi'^2}\,.
\eeq

Unsurprisingly, this action is merely the decompactification limit of the theory~\eqref{E:finalS} on the circle. To see this, we restore the radius of the circle, $\theta \sim \theta + 2\pi L$. The boundary theory~\eqref{E:finalS} describing fluctuations around global AdS$_3$ is then
\beq
	S_{\pm}[\phi] =  -\frac{C}{24\pi} \int dt \int_0^{2\pi L}d\theta \left( \frac{\phi'' \partial_{\pm}\phi'}{\phi'^2}-L^{-2}\phi' \partial_{\pm}\phi\right)\,,
\eeq
where $\phi(\theta+2\pi L,t) = \phi(\theta,t)+2\pi$ and we identify
\beq
	L \tan\left( \frac{\phi}{2L}\right) \sim \frac{a(t)L \tan\left( \frac{\phi}{2L}\right)+b(t)}{c(t)L \tan\left(\frac{\phi(t)}{2L}\right)+d(t)}\,.
\eeq
Clearly the $L\to\infty$ limit gives~\eqref{E:poincareS}.

\subsection{Coupling to an external geometry}
\label{S:curved}

In the remainder of this Section we put the boundary model~\eqref{E:finalS} on a non-trivial curved space. This is tricky because the theory is chiral, and so not manifestly Lorentz-invariant. However, using the gravity dual and the AdS/CFT dictionary we can repeat the same steps we used to arrive at the action~\eqref{E:finalS}, starting from bulk coordinates where the conformal boundary is naturally equipped with a non-trivial zweibein. 

We are not aware of a reference which develops the AdS/CFT dictionary for the Chern-Simons description of AdS$_3$ gravity with a non-trivial boundary geometry. In what follows we simply follow our noses.

Our starting point is a bulk dreibein,
\beq
	e^a = r E^a + O(r^{-1})\,, \qquad e^r = \frac{dr}{r}\,,
\eeq
where $E^a$ is the zweibein on the boundary as $r\to\infty$. The spin connection is
\beq
	\omega_{ABM} = (e^{-1})^N_{[A} \eta_{B]C}\left( \partial_N e^C_M-\Gamma^P{}_{MN}e^C_P\right)\,, \qquad T_{[AB]} = \frac{1}{2}(T_{AB}-T_{BA})\,,
\eeq
from which we find that the asymptotic behavior of $A$ and $\bar{A}$ are
\beq
\label{E:newAsy}
	A = \begin{pmatrix} \frac{1}{2}\left(\Omega + \frac{dr}{r}\right)  & O(r^{-1}) \\ rE^+ & -\frac{1}{2}\left( \Omega + \frac{dr}{r}\right) \end{pmatrix}\,, \qquad \bar{A} = \begin{pmatrix} \frac{1}{2}(\Omega - \frac{dr}{r}) & -r E^- \\ O(r^{-1}) & -\frac{1}{2}(\Omega-\frac{dr}{r})\end{pmatrix}\,,
\eeq
where $\Omega = \frac{1}{2}\eta^{ab}\Omega_{ab\mu}dx^{\mu}$ is the spin-connection one form corresponding to the boundary zweibein $E^a$ and we have defined $E^{\pm} = E^1 \pm E^0$\,.

Separating time and space as before, $A = A_0 dt + \tilde{A}$, the action is given by~\eqref{E:CSaction} where the boundary term is now modified to be
\beq
	S_{\rm bdy} = - \frac{k}{4\pi}\int_{\partial\mathcal{M}} d^2x \left( \frac{E^+_t}{E^+_{\theta}}\text{tr}'\left( A_{\theta}^2\right) - \frac{E^-_t}{E^-_{\theta}}\text{tr}'\left(\bar{A}_{\theta}^2\right)\right)\,,
\eeq
again to ensure a good variational principle with the asymptotic behavior~\eqref{E:newAsy}. The bulk functional integral reduces to one over flat spatial connections. Using a Gauss parameterization~\eqref{E:Gauss} the boundary conditions~\eqref{E:newAsy} fix the asymptotic form of $\lambda$ and $\Psi$ to be
\beq
\label{E:newConstraints}
	\lambda = \sqrt{\frac{r}{\mathcal{F}}}\,, \qquad \Psi =- \frac{1}{2r}\left( (\ln \mathcal{F})' - \Omega_{\theta}\right)\,, \qquad \mathcal{F} = \frac{F'}{E^{+}_{\theta}}\,,
\eeq
with $F$ finite on the boundary. The boundary action is now
\beq
\label{E:curvedS}
	S = S_-[\phi]+S_+[\bar{\phi}] \,, \qquad S_{\pm}[\phi] = -\frac{C}{6\pi}\int_{\partial\mathcal{M}} d^2x \left( \frac{\lambda'D_{\pm}\lambda}{\lambda^2}+ \lambda^2 F' D_{\pm}\Psi\right)\,,
\eeq
where 
\beq
	D_{\pm} \equiv \pm \frac{1}{2}\left(\partial_t + \frac{E^{\pm}_t}{E^{\pm}_{\phi}}\partial_{\theta}\right)\,.
\eeq
Plugging in the constrained value of the fields~\eqref{E:newConstraints} we obtain the curved space generalization of the right-moving half $S_-$ in~\eqref{E:finalS}. (To obtain the left-moving half we plug in the same constraints~\eqref{E:newConstraints}, but with $E^+_{\theta}\to E^-_{\theta}$ and $\Omega_{\theta}\to-\Omega_{\theta}$.) As before, the decomposition $A = g^{-1}\tilde{d}g$ is redundant and we identify
\beq
	F(\theta,t) \sim \frac{a(t) F(\theta,t)+b(t)}{c(t)F(\theta,t)+d(t)}\,.
\eeq

Observe that, whereas in flat space the term involving $\Psi$ in the action was a total derivative and so dropped out, this term is present in curved space. From the point of view of the flat space theory it improves the stress tensor as we now discuss.

Let us compute the stress tensor of the model, which we define by
\beq
	T^{\mu}_a = \frac{1}{\text{det}(e)} \frac{\delta S}{\delta E^a_{\mu}}\,.
\eeq
The usual stress tensor with two curved indices is $T^{\mu\nu} = T^{\mu}_A \eta^{AB} (E^{-1})^{\nu}_B$. A straightforward but tedious computation shows that the flat space stress tensor is traceless with
\beq
	T_{--} = -\frac{C}{24\pi}\left\{ \tan\left(\frac{\bar{\phi}}{2}\right),\theta\right\} \,, \qquad T_{++} = - \frac{C}{24\pi}\left\{ \tan\left(\frac{\phi}{2}\right),\theta\right\}\,,
\eeq
where we have substituted $F = \tan\left(\frac{\phi}{2}\right)$ and $\bar{F} = \tan\left(\frac{\bar{\phi}}{2}\right)$. So the stress tensor is indeed the Schwarzian derivative as advertised in the previous Section.

The curved space action~\eqref{E:curvedS} is not invariant under infinitesimal Weyl rescalings or local Lorentz rotations, which act on $E^a_{\mu}$ as
\beq
	\delta_{\sigma} E^a_{\mu} =\sigma E^a_{\mu}\,, \qquad \delta_v E^a_{\mu}= -v\epsilon^a{}_b E^b_{\mu}\,.
\eeq
This is ultimately due to the anomalies of the chiral theory~\eqref{E:curvedS}, whose central charge at large $C$ is approximately $C$. Recall that in the AdS/CFT dictionary, Chern-Simons terms in the bulk action correspond to anomalies for the global symmetries of the CFT dual, essentially by the inflow mechanism. See e.g.~\cite{Kraus:2006wn} for a discussion in AdS$_3$ gravity. (The Weyl anomaly usually arises from a different mechanism~\cite{zbMATH03971640,Henningson:1998gx}.) In the present setting, the Chern-Simons form of the bulk action, combined with the AdS/CFT dictionary, guarantee that the classical limit of the boundary theory $S_+$ has the anomalies of a left-moving CFT with central charge $C = 6k=\frac{3}{2G}$. 

That is, the action $S_{\pm}$ may be viewed as a Wess-Zumino term for the anomalies of a chiral CFT with central charge $C$. 

However, we have been unable to ascertain the scheme in which the anomalies are matched. In quantizing the bulk Chern-Simons theory we have separated time from space, and in doing so we have broken manifest covariance under coordinate transformations. As a result the curved space action~\eqref{E:curvedS} accounts for the anomalies in a non-standard way, differing by a presently unknown, non-covariant, local counterterm from the usual textbook presentation of the anomaly~\cite{bertlmann2000anomalies}.

\subsection{Sphere partition function}

As an illustrative example we put our theory on a sphere of radius $L$. The partition function of a non-chiral CFT with central charge $c$ on a sphere of radius $L$ is fixed by the Weyl anomaly to be
\beq
\label{E:ZS2}
	\mathcal{Z}_{\mathbb{S}^2} = (\mu L)^{\frac{c}{3}}Z_0\,,
\eeq
where $\mu$ is an energy scale which must be introduced. The constant $Z_0$ is unphysical, since it may be rescaled by a redefinition of $\mu$. The coefficient of the logarithm, i.e. $c$, is physical. A practical way to compute it is to evaluate
\beq
\label{E:RGZ}
	\mu \frac{\partial F_{\mathbb{S}^2}}{\partial\mu} = \frac{c}{3}\,, \qquad F_{\mathbb{S}^2} = \ln \mathcal{Z}_{\mathbb{S}^2}\,.
\eeq

The gravitational dual at large $c$ is well-known. The bulk metric where the boundary is a sphere of radius $L$ is
\beq
	g=L^2\left(r^2  g_{\mathbb{S}^2} + \frac{dr^2}{r^2L^2+1}\right)\,,
\eeq
where $g_{\mathbb{S}^2}=d\psi^2+\sin^2\psi \,d\theta^2$ is the metric of a unit sphere. The classical bulk action
\beq
	S_{\rm EH} = \frac{1}{16\pi G}\int d^3x \sqrt{g}(R+2)
\eeq
evaluated on this solution gives the sphere partition function at large $C$ via
\beq
	\ln \mathcal{Z}_{\mathbb{S}^2} \approx  S_{\rm EH}\,.
\eeq
This classical action is logarithmically divergent: the integrand is
\beq
	\sqrt{g}(R+2) = -\frac{4L^3 r^2\sin\psi}{\sqrt{r^2L^2+1}} = -4L^2r\sin\psi +2 \frac{\sin\psi}{r} + O(r^{-3})\,,
\eeq
and so its integral divergence near the boundary. Regulating the divergence by integrating up to $r=\Lambda$, one finds a logarithmic term in the action
\beq
	S_{\rm EH} =  \frac{1}{2G}\ln \Lambda + \cdots\,.
\eeq
Using $C = \frac{3}{2G}$ we find
\beq
	\Lambda\frac{\partial S_{\rm EH} }{\partial \Lambda}\approx \Lambda \frac{\partial F_{\mathbb{S}^2}}{\partial\Lambda}= \frac{C}{3} \,,
\eeq
which by~\eqref{E:RGZ} gives $c\approx C$.

In the last Subsection we showed how to put the boundary graviton theory in curved space. We will now show that the sphere partition function of that theory recovers this logarithmic divergence, and we will use it to compute the one-loop correction to $c$.

Our starting point is the curved space action we derived in~\eqref{E:curvedS}. Wick-rotating to Euclidean signature and taking the zweibein to be
\beq
	E^{\psi} = d\psi\,, \qquad E^{\theta} = \sin\psi \,d\theta\,,
\eeq
the action~\eqref{E:curvedS} evaluates to
\beq
	S_{\pm}[\phi] = \frac{C}{24\pi}\int d^2x \left( \frac{\mathcal{F}'D_{\pm}\mathcal{F}}{\mathcal{F}^2}-2\sin\psi D_{\pm} \left( \csc\psi \frac{F''}{F'} \pm i \cot \psi\right)\right)\,,
\eeq
with
\beq
	\mathcal{F} = \csc\psi \,F'\,, \qquad F = \tan\left(\frac{\phi}{2}\right)\,, \qquad D_{\pm} =\frac{1}{2}\left( \pm i \frac{\partial}{\partial\psi}+\csc\psi\,\frac{\partial}{\partial\theta}\right)\,.
\eeq
The classical trajectory is
\beq
	\phi_0 = \theta\,,
\eeq
and the classical action is logarithmically divergent:
\beq
	S_{\pm}[\phi_0] = -\frac{C}{12}\int_0^{\pi} d\psi \,\csc\psi\,.
\eeq
Introducing a position-space cutoff near the poles $\psi = 0,\pi$, i.e. integrating over $\psi \in [\Lambda^{-1},\pi-\Lambda^{-1}]$, we find that the classical action $S_0 = S_+[\phi_0]+S_-[\phi_0]$ diverges logarithmically as
\beq
	\Lambda \frac{\partial S_0}{\partial\Lambda} = - \frac{C}{3}\,.
\eeq
The sphere partition function is at large $C$
\beq
	\ln \mathcal{Z}_{\mathbb{S}^2} \approx - S_0\,,
\eeq
so that
\beq
	\Lambda \frac{\partial F_{\mathbb{S}^2}}{\partial\Lambda}\approx  \frac{C}{3}\,,
\eeq
as it should be.

We now compute the one-loop correction.  The harmonic analysis for fluctuations around the classical trajectory is a bit tricky for the choice of coordinates we used above. Instead it is convenient to parameterize the sphere as 
\beq
	E^y = \text{sech} \,y\, dy\,, \qquad E^{\theta} = \text{sech}\, y \,d\theta\,,
\eeq
where $y = \text{arcsech}(\sin\psi)$. Here $y\in \mathbb{R}$, with $y\to \pm \infty$ corresponding to $\psi\to 0,\pi$. Note that we are effectively parameterizing the sphere as a cylinder.

We proceed by evaluating the action in this background and expanding around the classical trajectory $\phi = \theta + \epsilon(\theta,y)$. The quadratic part of $S_+$ is, after an integration by parts,
\beq
	S_2 = \frac{C}{24\pi} \int d^2x \left( \epsilon'' \bar{\partial}\epsilon' - \epsilon' \bar{\partial}\epsilon\right)\,, \qquad \bar{\partial} = \frac{1}{2}(\partial_{\theta}+i \partial_y)\,.
\eeq
This is the same quadratic action we find on a torus in Section~\ref{S:torus}. Fourier transforming,
\beq
	\epsilon(\theta,y) = \int_{-\infty}^{\infty} \frac{d\omega}{(2\pi)^2} \sum_{n} \tilde{\epsilon}(n,\omega) e^{i n \theta + i \omega y}\,, \qquad \tilde{e}(n,\omega)^*= \tilde{\epsilon}(-n,-\omega)\,,
\eeq
we use the quasi-local $PSL(2;\mathbb{R})$ redundancy to set an infinite number of modes to vanish,
\beq
	\tilde{\epsilon}(n,\omega) = 0\,, \qquad \forall n = -1,0,+1\,.
\eeq
The quadratic part of $S_+$ then reads
\beq
	S_2 = \frac{iC}{24\pi}\int_{-\infty}^{\infty} \frac{d\omega}{(2\pi)^2} \sum_{n \neq -1,0,1} n (n^2-1)(\omega -i n)|\tilde{\epsilon}(n,\omega)|^2\,.
\eeq

Recalling the classical part of the computation, we introduce a UV cutoff by integrating up to a positive-space cutoff near the poles of the sphere. Translated to the $y$-coordinates, we take the $y$-direction to be compact with $y \in [-\ln \Lambda,\ln \Lambda]$, which turns the integral over $\omega$ into a sum. We also sum over modes which obey Dirichlet boundary conditions at $y=\pm \ln \Lambda$, as well as modes which obey Neumann boundary conditions there. Equivalently, we sum over modes which are periodic in $y$ with a periodicity $4 \ln \Lambda$. Thus, the one-loop determinant appearing here is that on a long torus of complex structure $\tau = \frac{2i}{\pi} \ln \Lambda$. In Section~\ref{S:torus} we compute that determinant and find that the chiral half is
\beq
	q^{-\frac{13}{24}} \prod_{n=2}^{\infty}\frac{1}{1-q^n}\,, \qquad q=e^{2\pi i \tau}\,.
\eeq
Multiplying by the contribution from the right-movers, inserting $\tau = \frac{2i}{\pi} \ln \Lambda$, taking $\Lambda\to\infty$, and including the classical contribution from $S_0$\,, we find
\beq
	\ln \mathcal{Z}_{\mathbb{S}^2,\rm 1-loop} = -S_0+\frac{13}{3}\ln \Lambda  \,,
\eeq
so that
\beq
	\Lambda \frac{\partial F_{\mathbb{S}^2}}{\partial \Lambda} = \frac{C+13}{3}\,.
\eeq
In other words, the one-loop approximation to the central charge is $c=C+13$.

\section{Doubling and eternal BTZ black holes}
\label{S:BTZ}

\begin{figure}[t]
\begin{center}
\includegraphics[width=2.5in]{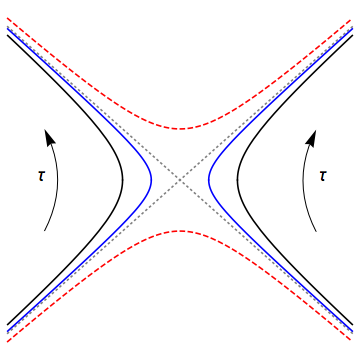}
\end{center}
\caption{
	\label{F:BTZ}
	A graphical representation of the Kruskal extended BTZ black hole~\eqref{E:kruskal}. Here $u$ and $v$ are the standard light-cone coordinates in the plane and $\theta$ is suppressed. The solid black lines are the asymptotic AdS boundaries, the dashed grey lines are the horizons, and the dotted red lines are the past and future singularities. The blue lines indicate curves of constant $z$.
}
\end{figure}

Pure AdS$_3$ gravity famously has black hole solutions, whose exterior region is described by the metric
\beq
	g = -(r^2-r_h^2)dt^2 + r^2 d\theta^2 + \frac{dr^2}{r^2-r_h^2}\,,
\eeq
with a horizon located at $r=r_h$ and conformal boundary at $r\to\infty$. This patch of spacetime is geodesically incomplete, and we instead consider its maximal Kruskal extension, the two-sided eternal black hole described by the metric (see e.g.~\cite{Fidkowski:2003nf})
\beq
\label{E:kruskal}
	g = \frac{-4 du dv + r_h^2(1-uv)^2 d\theta^2}{(1+uv)^2}\,.
\eeq
See Fig.~\ref{F:BTZ}. This geometry has two horizons located at $u=0$ and $v=0$, with the exteriors located in the region $uv<0$. There are two asymptotically AdS$_3$ regions, with conformal boundaries at $uv\to -1$. The past and future singularities are at $uv \to 1$. More useful coordinates are Rindler-like coordinates in the two exterior regions,
\beq
	- uv = z^2\,, \qquad \text{arctanh}\left(\frac{u+v}{v-u}\right) = r_h\tau\,\text{sgn}(z)\,,
\eeq
in terms of which the metric is
\beq
\label{E:gBTZ}
	g = \frac{4r_h^2}{(1-z^2)^{2}}\left(- z^2 d\tau^2 +\frac{(1+z^2)^2}{4}d\theta^2\right) + \frac{4dz^2}{(1-z^2)^2}\,.
\eeq
In this choice of coordinates the boundaries are at $z\to \pm 1$, and the horizons at $z=0$. The right exterior region corresponds to $\tau,\theta$ real with $z\in (0,1)$, and the left exterior region has $z\in (-1,0)$. In both exteriors the time $\tau$ increases as one moves up in Fig.~\ref{F:BTZ}. The coordinates $(\tau,\theta)$ parameterize time and space on the two boundaries. The standard interpretation of the two-sided black hole in the AdS/CFT correspondence is that it is dual to the thermofield double state of the boundary CFT~\cite{Maldacena:2001kr} at an inverse temperature $\beta = \frac{2\pi}{r_h}$.

In this Section we repeat the procedure of the previous Section for these two-sided black holes. A different approach to quantizing AdS$_3$ gravity on single-sided BTZ backgrounds was taken in~\cite{Kim:2015qoa,Maloney:2015ina}. Our first step is to obtain the $SO(2,1)\times SO(2,1)$ gauge fields which parameterize the background, and to write these in term in terms of gauge transformation parameters $g$ and $\bar{g}$. Taking the bulk dreibein in the two exterior regions to be
\beq
	e^0 = \pm \frac{2z}{(1-z^2)}r_hd\tau\,, \qquad e^1 = \frac{1+z^2}{1-z^2}r_hd\theta\,,\qquad e^z = \frac{4dz}{1-z^2}\,,
\eeq
we find 
\beq
	A = \begin{pmatrix} \frac{dz}{1-z^2} & \frac{1-z}{(1+z)} r_h dx^{\pm}\\ \frac{1+z}{2(1-z)}r_h dx^{\pm}& - \frac{dz}{1-z^2}\end{pmatrix}\,, \qquad \bar{A} = \begin{pmatrix}-\frac{dz}{1-z^2}  & -\frac{1+z}{2(1-z)}r_h dx^{\mp} \\ -\frac{1-z}{2(1+z)}r_h dx^{\mp} & \frac{dz}{1-z^2}\end{pmatrix}\,,
\eeq
where we have defined the boundary light-cone coordinates $x^{\pm} = \theta \pm \tau$. Being flat connections they may be written locally as
\beq
	A = g^{-1}dg\,,\qquad \bar{A} = \bar{g}^{-1}d\bar{g}\,.
\eeq
We find the representatives
\beq
	g = \begin{pmatrix} \rho \cosh\left( \frac{r_h x^{\pm}}{2}\right) & \rho^{-1} \sinh\left(\frac{r_h x^{\pm}}{2}\right) \\ \rho \sinh\left( \frac{r_h x^{\pm}}{2}\right) & \rho^{-1}\cosh\left( \frac{r_hx^{\pm}}{2}\right) \end{pmatrix}\,, 
	\,\,\,
	\bar{g} = \begin{pmatrix} \rho^{-1} \cosh\left(\frac{r_hx^{\mp}}{2}\right) & -\rho \sinh\left( \frac{r_hx^{\mp}}{2}\right) \\ -\rho^{-1} \sinh\left(\frac{r_hx^{\mp}}{2}\right) & \rho \cosh\left(\frac{r_hx^{\mp}}{2}\right)\end{pmatrix}\,,
\eeq
with
\beq
	\rho = \sqrt{\frac{1+z}{1-z}}\,.
\eeq

The asymptotic behavior of $A$ and $\bar{A}$ near the AdS boundaries $z^2 = 1-2\epsilon$ is, as $z\to 1$,
\beq
\label{E:BTZbc}
	A = \begin{pmatrix}- \frac{d\epsilon}{2\epsilon} & O(\epsilon) \\ \frac{r_hdx^+}{\epsilon} & \frac{d\epsilon}{2\epsilon}\end{pmatrix}\,, \qquad \bar{A} = \begin{pmatrix} \frac{d\epsilon}{2\epsilon} & -\frac{r_hdx^-}{\epsilon} \\ O(\epsilon) & -\frac{d\epsilon}{2\epsilon} \end{pmatrix}\,,
\eeq
and as $z\to -1$,
\beq
	A = \begin{pmatrix} \frac{d\epsilon}{2\epsilon} & \frac{r_h dx^-}{\epsilon} \\ O(\epsilon) & -\frac{d\epsilon}{2\epsilon} \end{pmatrix}\,, \qquad \bar{A} = \begin{pmatrix} -\frac{d\epsilon}{2\epsilon} & O(\epsilon) \\ -\frac{r_h dx^+}{\epsilon} & \frac{d\epsilon}{2\epsilon}\end{pmatrix}\,,
\eeq
which we then enforce as a boundary condition. We now quantize the bulk Chern-Simons theory on this spacetime. We  do so by again separating one direction from the others. Rather than separating time from space, we elect to separate
\beq
	A = A_{\theta}d\theta + \tilde{A}\,, \qquad \bar{A} = \bar{A}_{\theta} d\theta + \tilde{\bar{A}}\,.
\eeq
This is physically reasonable. The exterior of a BTZ geometry is the Wick-rotation of thermal global AdS, in which we analytically continue the boundary spatial circle to real time. In our quantization on thermal global AdS, we gauge-fixed the component of $A$ and $\bar{A}$ along the non-contractible circle, which after the Wick-rotation to BTZ becomes the $\theta$-direction. Then the bulk gravitational action is
\beq
	S_{\rm grav} = S[A] - S[\bar{A}] + S_{\rm bdy,+}+S_{\rm bdy,-}\,,
\eeq
where
\begin{align}
\begin{split}
	S[A]&=-\frac{k}{2\pi}\int_{\mathcal{M}} d\theta \wedge \text{tr}'\left(- \frac{1}{2}\tilde{A}\wedge \partial_{\theta}\tilde{A} + A_{\theta} \tilde{F}\right)\,,
	\\
	S_{\rm bdy,\pm} &= \frac{k}{4\pi}\int_{\partial\mathcal{M}_\pm} d^2x \left( \text{tr}'\left( A_{\tau}^2\right) + \text{tr}'\left(\bar{A}_{\tau}^2\right)\right)\,.
\end{split}
\end{align}
Here $\partial\mathcal{M}_{\pm}$ refers to the two boundaries at $z\to \pm 1$. We gauge-fix $A_{\theta}$ and $\bar{A}_{\theta}$ to their values for the BTZ solution. The residual integral is over flat connections $\tilde{A}$ and $\tilde{\bar{A}}$, which we parameterize as
\beq
	\tilde{A} = g^{-1}\tilde{d}g\,,\qquad \tilde{\bar{A}} = \bar{g}^{-1}\tilde{d}\bar{g}\,.
\eeq
We use separate Gauss decompositions~\eqref{E:Gauss} for $A$ near the two asymptotic boundaries. Near the right exterior we find that the boundary conditions~\eqref{E:BTZbc} fix the fields $\lambda$ and $\Psi$ appearing there to be constrained near the boundary $z\to1$ as
\beq
	\lambda^2 = \frac{1}{1-z}\frac{r_h}{\dot{F_{+}}}\,, \qquad \Psi = -\frac{1-z}{2}\frac{\ddot{F}_+}{r_h \dot{F_{+}}}\,.
\eeq
In the other region we parameterize
\beq
	g = \begin{pmatrix} 1 & -F \\ 0 & 1\end{pmatrix}\begin{pmatrix}\lambda^{-1} & 0 \\ 0 & \lambda \end{pmatrix} \begin{pmatrix} 1 & 0 \\ \Psi & 1 \end{pmatrix}\,,
\eeq
and near the other boundary $z\to -1$ the AdS boundary conditions constrain
\beq
	\lambda^2 = \frac{1}{1+z}\frac{r_h}{\dot{F}_-}\,, \qquad \Psi = \frac{1+z}{2}\frac{\ddot{F}_-}{r_h\dot{F}_-}\,,
\eeq
where $F_{\pm} = \lim_{z\to\pm 1} F$ are the boundary values of the field $F$. We see that the boundary dynamics are now doubled: the Chern-Simons theory may be rewritten as a boundary model in which the $F_+=\tanh \left(\frac{r_h\phi_+}{2}\right)$ and $F_- = -\coth\left( \frac{r_h \phi_-}{2}\right)$ are the dynamical fields.

For the two-sided black hole metric~\eqref{E:gBTZ} we have
\beq
	\phi_+ = \phi_-= \tau\,,
\eeq
so that $F_- = -1/F_+$. More generally we expect that the $\phi_{\pm}$ (and their barred cousins) obey the boundary condition
\beq
	\lim_{\tau\to \infty} \frac{\phi_{+}}{\tau} = \lim_{\tau\to\infty} \frac{\phi_-}{\tau}\,, \qquad \lim_{\tau\to- \infty} \frac{\phi_{+}}{\tau} = \lim_{\tau\to-\infty} \frac{\phi_-}{\tau}\,,
\eeq
and are monotone in $\tau$, $\dot{\phi}_{\pm}>0$. Note that these boundary conditions do not fix the value of the horizon radius $r_h$. Up to a quotient, the $\phi_{\pm}$ are Diff$(\mathbb{R})$ fields at fixed $\theta$. As in our analysis of global AdS$_3$, we have introduced a quasi-local $PSL(2;\mathbb{R})$ redundancy in writing out $A = g^{-1}\tilde{d}g$.  Namely, both $g(\tau,\theta,z)$ and $h(\theta)g(\tau,\theta,z)$ lead to the same $A$ for any $h(\theta)\in PSL(2;\mathbb{R})$. Since the boundary fields $\phi_{\pm}$ both come from the same $g$, the quotient by  $PSL(2;\mathbb{R})$ acts on both $\phi_{\pm}$ simultaneously. For
\beq
	h = \begin{pmatrix} d & c \\ b & a \end{pmatrix}\,,
\eeq
we have
\beq
	F_+ \to \frac{a F_++b}{cF_++d}\,, \qquad F_- \to \frac{dF_--c}{-bF_-+a}\,,
\eeq
and we identify those configurations in the remaining integral over $\phi_{\pm}$. The second transformation is equivalent to
\beq
	\mathcal{F}_- \to \frac{a\mathcal{F}_-+b}{c\mathcal{F}_-+d}\,, \qquad \mathcal{F}_- = -\frac{1}{F_-}\,.
\eeq
Similar statements apply for the barred fields.

The final result for the boundary action is the quadrupled theory
\beq
\label{E:quadruple}
	S = -S_-[\phi_+]+S_+[\phi_-] + S_+[\bar{\phi}_+]-S_-[\bar{\phi}_-]\,,
\eeq
where
\beq
	S_{\pm}[\phi] =- \frac{C}{24\pi}\int d\tau d\theta \left( \frac{\ddot{\phi} \partial_{\pm} \dot{\phi}}{\dot{\phi}^2} + \left(\frac{2\pi}{\beta}\right)^2 \dot{\phi} \partial_{\pm}\phi\right)\,.
\eeq
The fields $\phi_{\pm}$ are, at fixed $\theta$, elements of
\beq
	\faktor{ \text{Diff}(\mathbb{R})\times \text{Diff}(\mathbb{R})}{PSL(2;\mathbb{R})}\,,
\eeq
as are the $\bar{\phi}_{\pm}$. The final result for the boundary theory is reminiscent of the doubled Schwarzian theory described in~\cite{Maldacena:2016upp} and~\cite{Harlow:2018tqv}. Note that the fields $\phi_-$ and $\bar{\phi}_-$ on the left boundary have opposite chirality to the fields $\phi_+$ and $\bar{\phi}_+$ on the right boundary. 

However, this theory is not a doubled version of the boundary graviton action we obtained for global AdS$_3$ in Section~\ref{S:AdS}. The theory given above differs by a double Wick-rotation (along with an analytic continuation in field space, whose combined effect sends Diff$(\mathbb{S}^1)\to \text{Diff}(\mathbb{R})$), but more importantly the constraints tie the fields on the left and right boundaries together. Correspondingly, the Hilbert space of this model does not factorize into a tensor product of left and right states, and the two-sided black hole does not correspond to the thermofield double state of the ``single-sided'' model.

We expect that the path integral has an additional non-local constraint between the fields on the two boundaries, distinct from the simultaneous $PSL(2;\mathbb{R})$ quotient. Recall that a single copy of the quantization of the coadjoint orbit has a $PSL(2;\mathbb{R})$ Noether current. See Subsection~\ref{S:noether} for details. The quadrupled theory~\eqref{E:quadruple} has four sets of $PSL(2;\mathbb{R})$ currents, which can all be defined so as to have zero angular component. Let us consider the first two terms, $-S_-[\phi_+]$ and $S_+[\phi_-]$, coming from the first $SO(2,1)$ factor of the Chern-Simons theory. There is a $PSL(2;\mathbb{R})$ charge associated with $-S_-[\phi_+]$ living on the right boundary, which we call $Q_R$\,, and a $PSL(2;\mathbb{R})$ charge associated with $S_+[\phi_-]$ on the left boundary, which we call $Q_L$\,. In terms of the asymptotic values $g_{\pm}= \lim_{z\to \pm 1} g$ of the field $g$ as $z$ approaches the two boundaries, the charges $Q_R$ and $Q_L$ are given by
\beq
	Q_R = -\frac{C}{6\pi}(\partial_- g_+)g_+^{-1}\,, \qquad Q_L = \frac{C}{6\pi}(\partial_+g_-)g_-^{-1}\,,
\eeq
each satisfying
\beq
	\frac{\partial Q_R}{\partial t}  = 0\,, \qquad \frac{\partial Q_L}{\partial t }=0\,,
\eeq
on-shell. The two-sided black hole~\eqref{E:gBTZ} has $F_+ = \tanh\left( \frac{\pi \tau}{\beta}\right) = - 1/F_- = F_+(-\tau + \frac{i\beta}{2})$, so that the charges evaluate to
\beq
	Q_R = -Q_L^T = - \frac{C}{6\pi\beta}\begin{pmatrix} 0 & 1 \\ 1 & 0 \end{pmatrix} \,.
\eeq

One might wonder whether this charge matching condition should hold more generally, and if it should, how it is imposed in the path integral. Here we simply make two observations, leaving the complete resolution of this point to future study. First, the condition 
\beq
\label{E:Qmatch}
	Q_R = - Q_L^T\,,
\eeq
would be consistent with the simultaneous $PSL(2;\mathbb{R})$ quotient. Both sides transform in the same way under $\theta$-dependent $PSL(2;\mathbb{R})$ transformations. Indeed, the matching condition~\eqref{E:Qmatch} is nothing more than the statement that the simultaneous $PSL(2;\mathbb{R})$ charge vanishes. Second, the $PSL(2;\mathbb{R})$ quotient does not require that this charge vanishes. The situation here is somewhat different than in ordinary gauge theory, where in addition to the quotient the time-component of the gauge field $A_0$ acts as a Lagrange multiplier setting the total gauge charge to vanish, for the simple reason that here there is no gauge field. It is presently unclear whether, in the path integral, the classical condition~\eqref{E:Qmatch} arises from (i.) boundary conditions in the far past and future, perhaps by (ii.) an additional Lagrange multiplier term in the action $\sim \text{tr}'[A_0(Q_R + Q_L^T)]$, or (iii.) a different mechanism entirely.

\section{Torus partition function and $\langle TT\rangle$}
\label{S:torus}

In this Section we compute the exact chiral central charge and operator content of the path integral quantization of Diff$(\mathbb{S}^1)/PSL(2;\mathbb{R})$ and of Diff$(\mathbb{S}^1)/U(1)$. We do so by calculating the torus partition function and the two-point function of the stress tensor on the cylinder.

The partition function of a chiral CFT on a torus of complex structure $\tau$ may be written as a sum over states,
\beq
\label{E:Ztau}
	\mathcal{Z}(\tau) = \text{tr}\left( q^{L_0 - \frac{c}{24}}\right)\,, \qquad q=e^{2\pi i \tau}\,,
\eeq
where $c$ is the exact central charge.\footnote{Here we work in the usual convention for $L_0$, namely $[L_n,L_m] = (n-m)L_{n+m} + \frac{c}{12}(n^3 - n) \delta_{n+m}$, rather than that in~\eqref{E:VirasoroAlgebra}.} The partition function may be decomposed into a sum of Virasoro characters,
\beq
	\mathcal{Z}(\tau) = q^{-\frac{c}{24}}\prod_{n=2}^{\infty} \frac{1}{1-q^n} + \sum_h q^{h-\frac{c}{24}}\prod_{n=1}^{\infty} \frac{1}{1-q^n}\,,
\eeq
where the first term is the character of the vacuum module and the other terms represent a sum over Virasoro primaries of dimension $h$. The quantizations under consideration are weakly coupled as $C\to \infty$ with $1/C$ playing the role of the weak coupling, and so we expect to find $c = C + O(C^0)$.

For the quantization of Diff$(\mathbb{S}^1)/PSL(2;\mathbb{R})$ we find that its partition function is the vacuum character with an exact central charge
\beq
	c = C + 13\,,
\eeq
which we establish by a localization argument.

The quantization of the normal orbits Diff$(\mathbb{S}^1)/U(1)$ with $b_0 > -\frac{C}{48\pi}$ is secretly quadratic. Its Hilbert space has a single Verma module with highest-weight state $|h\rangle$ and
\beq
	h-\frac{c-1}{24} = 2\pi b_0\,.
\eeq
To solve for $h$ and $c$, we compute the connected two-point function of the stress tensor on the cylinder. In the state $|h\rangle$ it takes the form
\beq
	\langle \hspace{-.1cm} \langle T(w) T(0) \rangle \hspace{-.1cm} \rangle = \frac{c}{32\sin^4\left( \frac{w}{2}\right)} - \frac{h}{2\sin^2\left( \frac{w}{2}\right)}\,, \qquad w = \theta + i y\,,
\eeq
from which we find
\beq
	c= C+1\,, \qquad h = 2\pi b_0+\frac{C}{24}>0\,.
\eeq
We begin by computing the one-loop approximation to $\mathcal{Z}(\tau)$, and then use a localization argument to show that the partition function is one-loop exact. We also demonstrate that the quantization of the normal orbits is a free theory in the appropriate variables, and compute $\langle \hspace{-.1cm} \langle T(w)T(0)\rangle \hspace{-.1cm} \rangle $.

These results match nicely with the literature. In the context of path integral methods, there is an expectation that the torus partition functions of the geometric models are one-loop exact, although we have not found a reference which explicitly demonstrates this. Witten has also performed a K\"ahler quantization of the orbits Diff$(\mathbb{S}^1)/PSL(2;\mathbb{R})$ and Diff($\mathbb{S}^1)/U(1)$ in~\cite{Witten:1987ty}. He showed from a variety of viewpoints that the Hilbert space of the quantization of Diff$(\mathbb{S}^1)/PSL(2;\mathbb{R})$ is the vacuum module, and that the Hilbert space of the quantization of Diff($\mathbb{S}^1)/U(1)$ is a single Verma module. This first result was crucial to Maloney and Witten in their argument for the partition function~\eqref{E:ZMW1}. Below we give a direct path integral derivation of this result, which as far as we know is the first to compute the exact relations for $c$ and $h$ mentioned above.

\subsection{One-loop determinants}

We begin with the action~\eqref{E:ourTheory} appropriate for the path integral quantization of Diff$(\mathbb{S}^1)/PSL(2;\mathbb{R})$, which we reprise here:
\beq
\label{E:Sv1}
	S =- \frac{C}{24\pi}\int dt \int_0^{2\pi} d\theta \left( \frac{(\partial_+\phi')\phi''}{\phi'^2}  - (\partial_+\phi)\phi'\right)\,.
\eeq
Wick-rotating to imaginary time, $t = -i y$, and putting the theory on a torus of complex structure $\tau$, the Euclidean action is
\beq
\label{E:Seuclidean}
	S_E = \frac{C}{24\pi}\int d^2x \left( \frac{(\bar{\partial}\phi')\phi''}{\phi'^2} - (\bar{\partial}\phi)\phi'\right)\,,
\eeq
where we have defined the complex coordinate $z = \theta + i y$ subject to the identifications $z \sim z + 2\pi$ and $z \sim z + 2\pi \tau$. The field $\phi$ obeys the boundary conditions
\begin{align}
\begin{split}
	\phi(\theta+2\pi,y) & = \phi(\theta,y) + 2\pi\,,
	\\
	\phi(\theta+2\pi \text{Re}(\tau),y+2\pi \text{Im}(\tau)) & = \phi(\theta,y)\,,
\end{split}
\end{align}
and is subject to the local $PSL(2;\mathbb{R})$ redundancy
\beq
\label{E:redundancy}
	\tan \left(\frac{\phi(\theta,y)}{2}\right) \sim \frac{a(y) \tan\left(\frac{\phi(\theta,y)}{2}\right)+b(y)}{c(y) \tan\left(\frac{\phi(\theta,y)}{2}\right)+d(y)}\,, \qquad \begin{pmatrix} a(y) & b(y) \\ c(y) & d(y) \end{pmatrix} \in PSL(2;\mathbb{R})\,,
\eeq
which includes $\phi \sim \phi+ a$ as a subgroup.

As we mentioned in Subsection~\ref{S:phasespace}, the real part of the Euclidean action is $\text{Re}(S_E) = \int dy\, H(y)$ with
\beq
H = -\frac{C}{24\pi}\int_0^{2\pi} d\theta\left( \{\phi,\theta\}+\frac{\phi'^2}{2}\right)\,,
\eeq
which we showed is bounded below as $H \geq - \frac{C}{24\pi}$. It follows that 
\beq
\label{E:bound}
	\text{Re}(S_E) \geq - \frac{\pi C}{12}\text{Im}(\tau)\,,
\eeq
and so we expect that the Euclidean functional integral is well-defined.

We now take $C$ to be very large and compute the partition function to one-loop order in the small coupling $1/C$. We begin by classifying the saddle points of the Euclidean action. The field equation of the model is
\beq
	\frac{\delta S_E}{\delta \phi} = \frac{C}{48\pi \phi'} \bar{\partial}\left( \{\phi,\theta\} + \frac{\phi'^2}{2}\right)=0\,,
\eeq
which modulo the redundancy~\eqref{E:redundancy} possesses a unique saddle consistent with the boundary conditions, namely
\beq
\label{E:torusSaddle}
	\phi_0 = \theta - \frac{\text{Re}(\tau)}{\text{Im}(\tau)}y\,.
\eeq
The saddle point action is
\beq
	S_0 = \frac{\pi C}{12}i \tau \,,
\eeq
whose real part saturates the bound~\eqref{E:bound}. Expanding $\phi$ in fluctuations around the saddle,
\beq
	\phi = \phi_0 + \sum_{m,n} \frac{\epsilon_{m,n}}{(2\pi)^2} e^{\frac{i m y}{\text{Im}(\tau)} + i n \left( \theta - \frac{\text{Re}(\tau)}{\text{Im}(\tau)}y\right) }\,, \qquad \epsilon_{m,n}^* = \epsilon_{-m,-n}
\eeq
we may use the local redundancy~\eqref{E:redundancy} to set an infinite number of the modes to vanish:
\beq
	 \epsilon_{m,n=-1,0,+1} = 0\,.
\eeq
The quadratic action is
\beq
S_E = S_0 +\frac{i C}{96\pi^3} \sum_{m=-\infty}^{\infty}\sum_{n\neq -1,0,1}n (n^2 -1)(m-n \tau) |\epsilon_{m,n}|^2 +O(\epsilon_{m,n}^3)\,,
\eeq
so that the one-loop approximation to the torus partition function is
\beq
	\mathcal{Z}_{\rm 1-loop} = N q^{-\frac{C}{24}} \prod_{m,n} (m-n\tau)^{-\frac{1}{2}}\,, \qquad q=e^{2\pi i \tau}\,.
\eeq
The infinite product is, excepting the modes $n=-1,0,+1$, the determinant for a chiral boson, $\text{det}^{-\frac{1}{2}}(\bar{\partial}) $. Differentiating $\ln \mathcal{Z}_{\rm 1-loop}$ with respect to $\tau$ and performing the sum over $m$, we find
\beq
	\frac{\partial \ln \mathcal{Z}_{\rm 1-loop}}{\partial\tau} = -\frac{\pi i C}{12} - \pi \sum_{n=2}^{\infty} n \cot(\pi n \tau)\,,
\eeq
This sum diverges, and we regularize it by writing
\beq
	\sum_{n=2}^{\infty} n \cot(\pi n \tau) \to \sum_{n=2}^{\infty} n\left( \cot(\pi n \tau) + i \right) -i \sum_{n=2}^{\infty}n\,.
\eeq
The first sum converges for Im$(\tau)>0$, and we evaluate the second sum via zeta-function regularization. Integrating with respect to $\tau$ and exponentiating, we recover $\mathcal{Z}_{\rm 1-loop}$ up to an overall normalization constant. It is
\beq
\label{E:identityZ}
	b_0 = - \frac{C}{48\pi}: \qquad \mathcal{Z}_{\rm 1-loop} = q^{-\frac{c}{24}} \prod_{n=2}^{\infty} \frac{1}{1-q^n}\,, \qquad c = C+13\,, 
\eeq
which is the Virasoro character of the identity representation with central charge $c$. The factor of $13$ arises from the zeta-regularization,
\beq
\sum_{n=2}^{\infty} n \to \zeta(-1)-1 = -\frac{13}{12}\,.
\eeq
It is a one-loop renormalization of the central charge, $c = C +13 + O(1/C)$.

Writing $\mathcal{Z}$ as a sum over states as in~\eqref{E:Ztau}, we see that the Hilbert space of the model is merely the vacuum and its Virasoro descendants, as advertised. Perturbation theory at higher orders in $1/C$ cannot introduce new states, and so may only further renormalize the central charge $c$. However, as we will soon see, the one-loop result $c=C+13$ is exact.

The same arguments can be adapted to compute the partition function of the quantization of Diff$(\mathbb{S}^1)/U(1)$ for $b_0 > - \frac{C}{48\pi}$. We find
\beq
	\mathcal{Z}_{\rm 1-loop} = q^{h-\frac{c}{24}} \prod_{n=1}^{\infty} \frac{1}{1-q^n}\,, \qquad h-\frac{c-1}{24} = 2\pi b_0\,.
\eeq
So the Hilbert space is a single Verma module, and it remains to solve for $h$ and $c$.

\subsection{Revisiting the gravitational one-loop partition function}
\label{S:oneLoopGravity}

Before going on to discuss the one-loop exactness of the partition function, we pause to revisit the direct gravitational computation of~\cite{Giombi:2008vd}. The authors of that paper used the heat kernel to compute the one-loop determinant of pure gravity on both Euclidean AdS$_3$ and on a Euclidean BTZ black hole, which may be understood as an orbifold $\mathbb{H}^3/\mathbb{Z}$. They thereby obtained the one-loop approximation to the torus partition function of pure AdS$_3$ gravity to be the character of the vacuum module,
\beq
	\mathcal{Z}_{\rm 1-loop}^{\rm grav}(\tau,\bar{\tau}) = \left| q^{-\frac{c}{24}} \prod_{n=2}^{\infty} \frac{1}{1-q^n}\right|^2\,, 
\eeq
where $c$ is the one-loop renormalized central charge,
\beq
	c = \frac{3L}{2G}\left( 1 + O\left( \frac{G}{L}\right)\right)\,.
\eeq

In their paper, the authors of~\cite{Giombi:2008vd} focused on the infinite product in $\mathcal{Z}_{\rm 1-loop}$\,, and did not focus on the renormalization of the central charge. However, it turns out that this one-loop renormalization is implicitly contained in their calculation, and moreover they find the same one-loop shift of $13$ that we found above.

To see this we begin by parameterizing the partition function as
\beq
	-\ln \mathcal{Z}_{\rm 1-loop}^{\rm gravity} =  S^{(0)} - S^{(1)}\,,
\eeq
where $S^{(0)}$ is the on-shell Euclidean action of pure AdS$_3$ gravity, proportional to $\frac{L}{G}\gg 1$, and $S^{(1)}$ is the $O(1)$ contribution which arises from fluctuation determinants. They (and we) ignore further corrections to the right-hand-side with inverse powers of $\frac{L}{G}$\,. On a Euclidean spacetime $X=\mathbb{H}^3$ or $\mathbb{H}^3/\mathbb{Z}$, a standard computation gives
\beq
	S^{(0)} = \frac{L}{4\pi G}\text{vol}(X)\,.
\eeq
This expression is divergent, owing to the infinite volume near the conformal boundary of hyperbolic space. We will come back to this shortly.

Evaluating $S^{(1)}$ for pure gravity on $\mathbb{H}^3$ via the heat kernel technique gives their Eq. (3.8),
\beq
	S^{(1)} = \text{vol}(\mathbb{H}^3)\int_0^{\infty} \frac{dt}{t} \frac{1}{(4\pi t)^{3/2}} (e^{-t} (1+8t)-e^{-4t}(1+2t))\,.
\eeq
There are two divergences here. The first is the same infinite volume appearing above, and the second is a divergence in the integral at small $t$. The latter is ubiquitous in the heat kernel literature and can be treated by any of the standard techniques. One technique, used in~\cite{Giombi:2008vd}, is to analytically continue the integral in $t$ (equivalently, to rewrite the integral using the integral representation for the Gamma function). Another approach, familiar from the computation of anomalies, is to integrate in $t$ to a small cutoff $\epsilon$ and to take the $O(\epsilon^0)$ coefficient of the expansion. Either way one finds
\beq
	S^{(1)} = -\frac{13}{6\pi}\text{vol}(\mathbb{H}^3)\,,
\eeq
which still diverges, albeit in the same way as the saddle point action $S^{(0)}$.

The authors of~\cite{Giombi:2008vd} evaluated $S^{(1)}$ on the orbifold $\mathbb{H}^3/\mathbb{Z}$ using the heat kernel on $\mathbb{H}^3$ and the method of images. The result is
\beq
	S^{(1)} = \text{vol}(\mathbb{H}^3/\mathbb{Z})\int_0^{\infty} \frac{dt}{t} \frac{1}{(4\pi t)^{3/2}} (e^{-t} (1+8t)-e^{-4t}(1+2t)) - \sum_{n=2}^{\infty}\ln|1-q^n|^2\,.
\eeq
Treating the first term (which was dropped in~\cite{Giombi:2008vd}) in the same way as on $\mathbb{H}^3$, the one-loop approximation to the partition function on $\mathbb{H}^3/\mathbb{Z}$ then reads
\beq
\label{E:1loopGravity}
	\mathcal{Z}_{\rm 1-loop}^{\rm grav} = e^{-\left( \frac{3L}{2G}+13\right)\frac{1}{6\pi} \text{vol}(\mathbb{H}^3/\mathbb{Z})}\left| \prod_{n=2}^{\infty} \frac{1}{1-q^n}\right|^2\,.
\eeq
It remains to regularize the volume of $\mathbb{H}^3/\mathbb{Z}$. This is a familiar problem in the AdS/CFT correspondence, an example of what is often called holographic renormalization. One first regularizes the infinite volume by integrating the volume form on the bulk spacetime up to a ``cutoff slice'' near the conformal boundary, adds various covariant boundary terms to the ``cutoff slice,'' and then takes the cutoff to the conformal boundary. In any case, the holographically renormalized volume is
\beq
	\text{vol}(\mathbb{H}^3/\mathbb{Z}) = - \text{Im}(\tau) \pi^2\,.
\eeq
Inserting into~\eqref{E:1loopGravity} gives
\beq
	\mathcal{Z}_{\rm 1-loop}^{\rm grav} = \left| q^{-\frac{c}{24}} \prod_{n=2}^{\infty} \frac{1}{1-q^n}\right|^2\,, \qquad c = \frac{3L}{2G}+13\,,
\eeq
whose chiral half is the partition function of the geometric model that we found in the previous Subsection. The gravitational result displays the same one-loop shift of $13$. 

\subsection{Localization}
\label{S:localization}

It is perhaps an underappreciated result that (up to boundary conditions) any phase space path integral is invariant under a ``hidden supersymmetry,'' and that under modest conditions this ``supersymmetry'' can be used to localize the path integral. See e.g.\! Section 4 of~\cite{Szabo:1996md}. When this is the case, the one-loop exactness of the path integral follows from the localization formula of equivariant cohomology, and is in a sense the quantization of the Duistermaat-Heckman theorem. Here we give a low-brow derivation of the result. Our approach parallels that of Stanford and Witten~\cite{Stanford:2017thb} in their ``physicist's proof'' of the Duistermaat-Heckman theorem. It also shares several features with~\cite{Szabo:1996md}.

Consider a classical Hamiltonian system on a phase space $\mathcal{M}$ with Hamiltonian $H$ and symplectic form $\omega$ (keeping in mind the coadjoint orbits of a Lie group). Let $x^i$ denote coordinates on the phase space, and $\alpha = \alpha_i dx^i$ the presymplectic potential locally satisfying $\omega = d\alpha$. Promoting the variables $x^i$ to functions of time $x^i(t)$, an action which leads to the Hamiltonian $H$ and symplectic form $\omega$ is~\eqref{E:action},
\begin{equation*}
	S = -\int dt \big( \dot{x}^i \alpha_i + H\big)\,.
\end{equation*}

Consider the thermal partition function of the quantized system,
\beq
\mathcal{Z} = \text{tr}\left( e^{-\beta H}\right)\,,
\eeq
which can be written as a path integral after Wick-rotating time as $t = -iy$,
\beq
\label{E:thermalZ}
	\mathcal{Z} = \int [dx^i] \text{Pf}(\omega) e^{-S_E}\,, \qquad S_E = \int_0^{\beta} dy \left( i\frac{\partial x^i}{\partial y} \alpha_i +H\right)\,,
\eeq
and the $\int [dx^i]$ integral is taken over closed loops $x^i(y) = x^i(y+\beta)$. The factor of $\text{Pf}(\omega)$ plays the role of the measure on the phase space at each time. It can be exponentiated by introducing Grassmann-odd fields $\psi^i$ which obey the same boundary conditions as the $x^i$, namely $\psi^i(y) = \psi^i(y+\beta)$. The $\psi^i$ are better thought of as ghosts rather than fermions. In any case, we write
\beq
	\mathcal{Z} = \int [dx^i][d\psi^j] e^{-S_E'}\,, \qquad S_E' = \int_0^{\beta} dy \left( i \frac{\partial x^i}{\partial y}\alpha_i + H - \frac{1}{2}\omega_{ij} \psi^i \psi^j\right)\,.
\eeq
This action is invariant under a ``hidden supersymmetry'' generated by a Grassmann-odd supercharge $Q$ (which does not carry a Lorentz index and so resembles a BRST supercharge). Its action on $(x^i,\psi^j)$ is
\begin{align}
\begin{split}
	Qx^i &= \psi^i\,,
	\\
	Q\psi^i & = v^i - i \frac{\partial x^i}{\partial y}\equiv V^i\,,
\end{split}
\end{align}
where we have defined $v^i$ to be the flow generated by the Hamiltonian,
\beq
	v^i = \omega^{ij} \partial_j H\,,
\eeq
and $\omega^{ij}$ is the inverse of the symplectic form satisfying $\omega^{ik}\omega_{jk} = \delta^i_j$. The variation of $\psi^i$ is nothing but the classical equation of motion for $x^i$ in imaginary time, 
\beq
	V^i = - \omega^{ij} \frac{\delta S_E}{\delta x^i}\,.
\eeq
It is straightforward to demonstrate that the action $S_E'$ is invariant under $Q$, upon using that $\omega_{ij} = \partial_i \alpha_j - \partial_j \alpha_i$ and that $\partial_{[i}\omega_{jk]}=0$, i.e. $d\omega = 0$. Acting on $x^i$,  $Q$ squares to the equation of motion,
\beq
	Q^2 x^i = v^i - i \frac{\partial x^i}{\partial y}\,,
\eeq
and more generally 
\beq
	Q^2 = \delta_v - i \frac{\partial}{\partial y}\,,
\eeq
where $\delta_v$ is the flow generated by $H$. So $Q^2=0$  is simply the classical equation of motion of the model.  It follows that the Ward identities of the ``hidden supersymmetry'' are nothing more than the Schwinger-Dyson equations, and, in a sense, the invariance under $Q$ is content-free. It is then no surprise that $Q$ can be used to localize the path integral only under special conditions.

One such condition is if there is a metric $g_{ij}$ on the phase space which is invariant under the flow $\delta_v$, that is if $g_{ij}$ satisfies
\beq	
	\delta_v g_{ij} = v^k\partial_k g_{ij} + g_{ik}\partial_j v^k + g_{kj}\partial_i v^k=0\,.
\eeq
When that is the case, not only is
\begin{align}
\begin{split}
	V &= \int_0^{\beta} dy \,Q\left( g_{ij} (Q\psi^i)\psi^j\right) 
	\\
	&= \int_0^{\beta} dy \Big( g_{ij}V^i V^j  + \psi^k (\partial_k g_{ij}) V^i \psi^j + g_{ij} (\psi^k\partial_k v^i - i\partial_y \psi^i)\psi^j\Big)
\end{split}
\end{align}
$Q$-exact, but it is also $Q$-closed. After some computation we find
\beq
	QV = \int_0^{\beta}dy\left\{ (\delta_v g_{ij})V^i\psi^j + i \frac{\partial}{\partial y}(g_{ij}V^i\psi^j)\right\} =0\,.
\eeq
Combined with the fact that its bosonic part is also positive-definite, $V$ may then be used as a localizing term. Adding it to the action with an arbitrary coefficient,
\beq
	S_E' \to S_E' + t V\,,
\eeq
and barring an anomaly in $Q$, the partition function is independent of $t$. Sending $t\to\infty$, the path integral localizes onto trajectories near the classical solutions $V^i = 0$ and $\psi^j=0$ and is rendered one-loop exact. A standard calculation demonstrates that there is a relative cancellation between the ghost and bosonic determinants, so that the full partition function equals the one-loop approximation of the original theory~\eqref{E:thermalZ} without the ghosts, i.e.
\beq
\mathcal{Z}=\sum e^{-S_E[x_c]}\,\text{det}^{-1/2}\left(\frac{\delta^2 S_E}{\delta x^2}\right)\,,
\eeq
with $x_c$ the classical trajectories satisfying $V^i[x_c]=0$.

Now recall that the first exceptional orbit (and the normal orbits with $b_0>-\frac{C}{48\pi}$) of the Virasoro group are K\"ahler and additionally possess an invariant K\"ahler form. See~\eqref{E:kahler} and the surrounding discussion. This K\"ahler form gives a metric invariant under Hamiltonian flow on the orbit, which for our theory is the group action generated by $L_0$. The existence of an invariant metric on the phase space is precisely the criterion we need to localize the partition function. We conclude that the thermal partition functions of the quantizations of Diff$(\mathbb{S}^1)/PSL(2;\mathbb{R})$ and Diff$(\mathbb{S}^1)/U(1)$ (with $b_0 > -\frac{C}{48\pi}$) are one-loop exact. Running through another version of the localization argument above with ``twisted'' boundary conditions shows that the torus partition functions of these theories at general $\tau$ are one-loop exact.

Before going on, we note that there is a natural guess for the measure of our path integrals that appears to differ from $[d\phi]\text{Pf}(\omega)$, namely~\cite{Alekseev:1990mp}
\beq
\label{E:naiveMeasure}
	\prod_{\theta,y} \frac{dF}{F'}\,, \qquad F = \tan\frac{\phi}{2}\,.
\eeq
This guess is natural insofar as it is invariant under local $PSL(2;\mathbb{R})$ transformations. By $[d\phi] \text{Pf}(\omega)$, we mean that the measure of the phase space path integral is
\beq
	\prod_t \left(\text{Pf}(\omega(\phi)) \prod_{\theta} d\phi \right)\,.
\eeq
Stanford and Witten~\cite{Stanford:2017thb} showed that the expression in parantheses is equivalent to $\prod_{\theta} \frac{dF}{F'}$, so that $[d\phi]\text{Pf}(\omega)$ is in fact equivalent to the naive measure~\eqref{E:naiveMeasure}.

\subsection{$\langle TT \rangle$}

The partition function for the normal orbits informs us that the corresponding Hilbert space is a single Verma module with highest-weight state $|h\rangle$ satisfying
\begin{equation*}
	h-\frac{c-1}{24} = 2\pi b_0\,.
\end{equation*}
Here we compute the exact stress tensor two-point function on the cylinder and solve for $h$ and $c$.

The key step is that the action for the normal orbits is secretly quadratic. Using the field redefinition
\beq
	f = \ln \left( e^{\sqrt{B}\phi}\right)'\,, \qquad B = \frac{48\pi b_0}{C}\,,
\eeq
the Euclidean action and stress tensor become
\beq
\label{E:redefinedS}
	S_E = \frac{C}{24\pi}\int d^2x (\bar{\partial} f)f'\,, \qquad T = -\frac{C}{12}\left( f''-\frac{f'^2}{2}\right)\,.
\eeq
The redundancy $\phi(x,y)\sim \phi(x,y)+a(y)$ persists for the $f$-field,
\beq
\label{E:fredundancy}
	f(\theta,y) \sim f(\theta,y) +\sqrt{B} a(y)\,,
\eeq
and the boundary conditions for $\phi$ on the Euclidean cylinder become
\beq
	f(\theta+2\pi,y) = f(\theta,y)+2\pi \sqrt{B}\,.
\eeq
Moreover, in these variables the symplectic form becomes the canonical one corresponding to the action~\eqref{E:redefinedS},
\beq
	\omega = \frac{C}{48\pi}\int_0^{2\pi}\, df \wedge df'\,.
\eeq
Since both the action and measure are quadratic in the $f$-variable, this appears to be a free theory. The only potential problem is the quotient. But for the normal orbits the redundancy is linear in the field $f$, acting as~\eqref{E:fredundancy}, and so this theory is indeed free.

While a similar redefinition exists for the first exceptional orbit, rendering the action quadratic, that theory is not free. The culprit is the $PSL(2;\mathbb{R})$ quotient. Whereas the $U(1)$ redundancy acts linearly on $f$, the $PSL(2;\mathbb{R})$ quotient acts in a non-trivial, non-linear way. However it is easy to account for the quotient at one-loop level. The quotient simply removes the $n=-1,0,+1$ spatial Fourier modes, and the ensuing determinant immediately yields the character of the vacuum module.

Moving on, the unique saddle (modulo the $U(1)$ redundancy) is $f=\sqrt{B}\theta$. Expanding around it,
\beq
	f=\sqrt{B}\theta + \epsilon\,,
\eeq
one finds that the $\epsilon$-propagator is
\beq
	\langle \epsilon(w) \epsilon(0)\rangle = -\frac{12}{C}\ln(1-u)\,, \qquad u = e^{i \text{sgn}(y)w}\,, \qquad w = \theta + i y\,.
\eeq
The stress tensor is $T = 2\pi b_0-\frac{C}{12}\left( \epsilon'' -\sqrt{B} \epsilon' -\frac{\epsilon'^2}{2}\right)$, and its connected two-point function is given by the sum of an exchange diagram and a one-loop diagram. It reads
\beq
	\langle \hspace{-.1cm} \langle T(w) T(0)\rangle \hspace{-.1cm}\rangle = \frac{C+1}{32\sin\left( \frac{w}{2}\right)^4} - \frac{2\pi \left( b_0+\frac{C}{48\pi}\right)}{2\sin^2\left(\frac{w}{2}\right)}\,,
\eeq
from which we infer
\beq
\label{E:cAndh}
	c = C+1\,, \qquad h = 2\pi b_0 + \frac{C}{24}\,.
\eeq

\section{Virasoro blocks}
\label{S:blocks}

The geometric actions considered in this paper also compute the atomic ingredients of correlation functions of a 2d CFT, namely Virasoro blocks. Consider the four-point function of local operators $\mathcal{O}_i$\,. Suppressing the anti-holomorphic dependence, the four-point function is a function of a single conformally invariant cross-ratio $u$ and can be written as
\beq
	\langle \mathcal{O}_1(z_1)\mathcal{O}_2(z_2) \mathcal{O}_3(z_3)\mathcal{O}_4(z_4)\rangle = \frac{1}{(z_{12})^{h_1+h_2}(z_{34})^{h_3+h_4}}\,\mathcal{C}(u)\,, \qquad u = \frac{z_{12}z_{34}}{z_{13}z_{24}}\,,
\eeq
with $z_{ij} = z_i - z_j$. The function $\mathcal{C}(u)$ may be decomposed into a sum of conformal blocks. These blocks are intimately related to the operator production expansion (OPE): a single block corresponds to the exchange of a single primary operator $\mathcal{O}_h$ of dimension $h$ along with its descendants between the fusions $\mathcal{O}_1\mathcal{O}_2$ and $\mathcal{O}_3\mathcal{O}_4$. Equivalently, using the operator/state correspondence, consider the completeness relation for the CFT Hilbert space. Associating $\mathcal{O}_h$ to the primary state $|h\rangle=\mathcal{O}_h(0)|0\rangle$, one may write the identity operator as a sum over irreducible representations of the conformal group as labeled by the primary states,
\beq
	\mathbb{1} =\sum_h\Big(  |h\rangle\langle h| + (\text{descendants})\Big)\,,
\eeq
where the ``descendants'' refers to the properly normalized descendant states of $|h\rangle$. The global descendants of $|h\rangle$ are
\beq
	|h;n\rangle = \frac{L_{-1}^n |h\rangle}{\sqrt{\langle h|L_1^nL_{-1}^n|h\rangle}}\,,
\eeq
while the Virasoro descendants include (these states are not all orthogonal and so give an overcomplete basis for the space of descendant states)
\beq
	|h;\{n_i\}\rangle = \frac{\left(\prod_{i=1}^{\infty} L_{-i}^{n_i}\right)|h\rangle}{\sqrt{\langle h|\left(\prod_{i=1}^{\infty} L_{-i}^{n_i} \right)^{\dagger}\left(\prod_{j=1}^{\infty} L_{-j}^{n_j}\right)|h\rangle}}\,.
\eeq
We then write 
\begin{align}
\begin{split}
\label{E:block}
	\mathcal{C}(u) &= \sum_hc_{12h}c_{34h} \, \mathcal{F}(h_i;h;u)\,,
	\\
	 \mathcal{F}(h_i;h;u)&= \lim_{z\to\infty}\frac{z^{2h_1}u^{h_3+h_4}}{c_{12h}c_{34h}} \, \langle 0|\mathcal{O}_1(z)\mathcal{O}_2(1)\Big( |h\rangle \langle h |+(\text{descendants})\Big)\mathcal{O}_3(u)\mathcal{O}_4(0)|0\rangle \,,
\end{split}
\end{align}
where $c_{ijk}$ is an OPE coefficient and $\mathcal{F}$ is the block. If the second line of~\eqref{E:block} contains the global descendants of $|h\rangle$, then $\mathcal{F}$ is a global conformal block. If the second line contains Virasoro descendants, then it is a Virasoro block and we notate it as $\mathcal{V}$. In either case, the block is an eigenfunction of the corresponding conformal Casimir.

The global blocks are known in closed form and are given by~\cite{Dolan:2000ut,Dolan:2003hv}
\beq
	\mathcal{F}(h_i;h;u) =u^{h}\,_2F_1(h-h_1+h_2,h+h_3-h_4;2h;u)\,.
\eeq
The Virasoro blocks are not known in closed form. They are approximately known in various limits, mostly at large central charge $c\gg 1$. See e.g.~\cite{Zamolodchikov:1985ie,osti_6447081,Harlow:2011ny,Hartman:2013mia,Fitzpatrick:2014vua,Fitzpatrick:2015zha,Perlmutter:2015iya}. 

In this Section we use the quantization of $\text{Diff}(\mathbb{S}^1)/PSL(2;\mathbb{R})$ to compute the identity Virasoro blocks at large central charge in two limits. The first is when the external dimensions are ``light'' compared to $c$, with $h_i \lesssim \sqrt{c}$. The second is in the ``heavy-light'' limit of~\cite{Fitzpatrick:2015zha}, in which one operator is ``heavy'' with $h_H=O(c)$ and the other is ``light'' with $h_L=O(1)$. In both cases our computations have a simple diagrammatic interpretation. Our results reproduce those of~\cite{Fitzpatrick:2014vua,Fitzpatrick:2015zha}, including the ``quantum'' corrections of~\cite{Beccaria:2015shq,Fitzpatrick:2015dlt}.

The identity block $\mathcal{V}_0$ appears in the four-point function of two identical operators $\mathcal{O}_1=\mathcal{O}_2=V$, $\mathcal{O}_3=\mathcal{O}_4=W$. The contribution of $\mathcal{V}_0$ to the four-point function is
\beq
	\langle V(z_1)V(z_2)W(z_3)W(z_4)\rangle = \frac{1}{(z_{12})^{2h_V}(z_{34})^{2h_W}}\big( \mathcal{V}_0 + (\text{other blocks})\big)\,,
\eeq
and from~\eqref{E:block} it is given by
\beq
	\mathcal{V}_0(h_V,h_W;u) = 1 + O\left(\frac{1}{c}\right)\,,
\eeq
in the $c\to\infty$ limit (with $h_i$ fixed). The ``1'' comes from the vacuum state $|0\rangle$ and the corrections arise from the Virasoro descendant states of the vacuum $\propto \cdots L_{-3}^{n_3}L_{-2}^{n_2}|0\rangle$.

We compute the identity block in the following way. In addition to the stress tensor $T\propto \{ \phi,\theta\}$, the path integral quantization of Diff$(\mathbb{S}^1)/PSL(2;\mathbb{R})$ contains nonlocal operators invariant under the local $PSL(2;\mathbb{R})$ symmetry. These include bilocal operators, which can be thought of as reparameterized two-point functions,\footnote{These bilocal operators are familiar from the Schwarzian theory~\cite{Maldacena:2016hyu,Jensen:2016pah,Maldacena:2016upp,Engelsoy:2016xyb,Mertens:2017mtv,Mertens:2018fds}.}
\beq
\label{E:bilocalPlane}
	\mathcal{B}(h;z_1,z_2) = \left(\frac{\phi'(z_1)\phi'(z_2)}{(\phi(z_1)-\phi(z_2))^2}\right)^h\,,
\eeq
where $z_1$ and $z_2$ are at the same time and \,$'$ denotes a spatial derivative. (We require that $z_1$ and $z_2$ are at the same time so that the bilocal is invariant under time-dependent $PSL(2;\mathbb{R})$ transformations.) Our proposal is that the identity block is proportional to the two-point function of bilocals,
\beq
\label{E:twoPtBi}
	\mathcal{V}_0(h_V,h_W;u) = \frac{\langle \mathcal{B}(h_V;z_1,z_2)\mathcal{B}(h_W;z_3,z_4)\rangle}{\langle \mathcal{B}(h_V;z_1,z_2)\rangle \langle \mathcal{B}(h_W,z_3,z_4)\rangle}\,.
\eeq
Recalling that in our geometric theory the total central charge is $c=C+13$ and that the theory is weakly coupled as $C\to\infty$, we immediately recover that the identity block is $1+O(1/c)$ in the large $C$ limit. The various $1/c$ corrections are then obtained by evaluating~\eqref{E:twoPtBi} in perturbation theory in $1/C$. Roughly speaking, the Fourier modes of the reparameterization field $\phi$ correspond to the Virasoro generators $L_{n}$\,, and we obtain the full identity block by ``dressing'' the identity's contribution $\frac{1}{(z_{12})^{2h_V}(z_{34})^{2h_W}}$ to the four-point function by the reparameterization field. Throughout we find it convenient to first evaluate the block on the cylinder, and then to take the decompactification limit.

The quantization of Diff$(\mathbb{S}^1)/U(1)$ may be used to compute the Virasoro blocks corresponding to the exchange of other operators $\mathcal{O}_h$\,. However we do not pursue this further here.

\subsection{The light-light limit}

We first work in the limit where we take $C\to\infty$ and the operator dimensions are relatively light, scaling no larger than $h\lesssim \sqrt{c}$. In this limit the identity block exponentiates and is given by~\cite{Fitzpatrick:2014vua}
\beq
\label{E:exponentiation}
	\mathcal{V}_0(h_V,h_W;u) = \exp\left( \frac{2 h_V h_W}{c}\,u^2\,_2F_1(2,2;4;u)\right) \left( 1 + O\left(\frac{1}{\sqrt{c}}\right)\right)\,.
\eeq
The argument of the exponential is just the global $h=2$ block corresponding to the exchange of the stress tensor.

We recover this result in steps. First, let us treat $\frac{h_Vh_W}{c}$ as a small expansion parameter and work to first order in it. This is equivalent to evaluating the leading $1/c$ correction to $\mathcal{V}_0$ in the limit that $h_V$ and $h_W$ are held fixed as $c\to\infty$. 

Our starting point is the two-point function of a dimension--$h$ operator on the cylinder with the insertions at the same time. It is
\beq
	\langle \mathcal{O}(\theta_1)\mathcal{O}(\theta_2)\rangle = \frac{1}{\left(2\sin\left(\frac{\theta_{12}}{2}\right)\right)^{2h}}\,,
\eeq
which may be obtained from the two-point function in the plane at fixed time, $1/x_{12}^{2h}$\,, after conformally transforming from the plane to the cylinder. In our geometric theory we consider the bilocal operator, which is a reparameterized two-point function. On the cylinder it is given by
\beq
\label{E:bilocalCyl}
	\mathcal{B}(h;\theta_1,\theta_2) = \left( \frac{\phi'(\theta_1)\phi'(\theta_2)}{4\sin^2\left(\frac{\phi(\theta_1)-\phi(\theta_2)}{2}\right)^2}\right)^h\,,
\eeq
which is also related to the bilocal operator on the plane,~\eqref{E:bilocalPlane}, upon substituting $\phi \to \tan\left(\frac{\phi}{2}\right)$ into~\eqref{E:bilocalPlane}. Crucially, the two insertions in~\eqref{E:bilocalCyl} must be at the same Euclidean time in order for this operator to be $PSL(2;\mathbb{R})$ invariant. 

\begin{figure}[t]
	\begin{center}
	\includegraphics[width=2.5in]{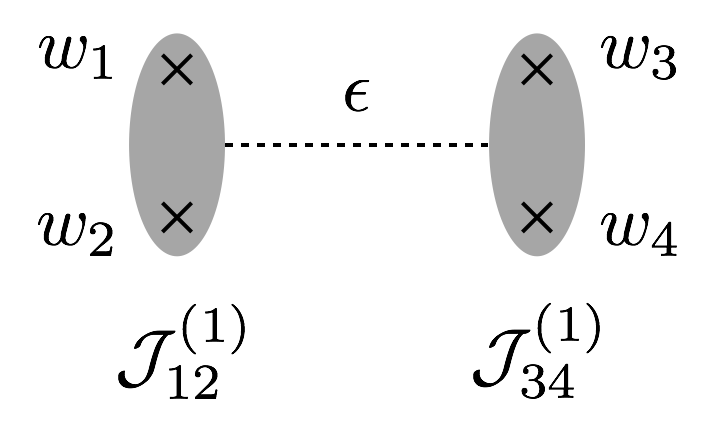}
	\caption{
		\label{F:exchange} The exchange diagram which computes the $O(1/C)$ correction to the identity block. The blobs refer to the bilocal operators, and the vertex factors $\mathcal{J}_{12}^{(1)}$ and $\mathcal{J}_{34}^{(1)}$ are the linear couplings of the reparameterized two-point function to the $\epsilon$ field and are given in~\eqref{E:J1}. The $\epsilon$ propagator is given in~\eqref{E:eps2}.
	}
	\end{center}
\end{figure}

Expanding around the saddle, $\phi = \theta + \epsilon(\theta,y)$, we find
\beq
\label{E:bilocalLO}
	\mathcal{B}(h;\theta_1,\theta_2)  = \frac{1}{\left(2\sin\left(\frac{\theta_{12}}{2}\right)\right)^{2h}}\Big( 1 + h\,\mathcal{J}_{12}^{(1)}\cdot \epsilon + O(\epsilon^2)\Big)\,,
\eeq
where
\beq
\label{E:J1}
	\mathcal{J}_{12}^{(1)}\cdot\epsilon = \epsilon'_1+\epsilon_2' -\cot\left(\frac{\theta_{12}}{2}\right) \epsilon_{12}\,,
\eeq
is the linear coupling of the bilocal to the reparamaterization field. By~\eqref{E:twoPtBi} our proposal is that the identity block on the cylinder is given by
\beq
	\mathcal{V}_0 = \frac{\langle \mathcal{B}(h_V;\theta_1,\theta_2) \mathcal{B}(h_W;\theta_3,\theta_4)\rangle}{\langle \mathcal{B}(h_V;\theta_1,\theta_2)\rangle\langle \mathcal{B}(h_W;\theta_3,\theta_4)\rangle}\,,
\eeq
where the insertions $(\theta_1,\theta_2)$ are at the same Euclidean time, as are $(\theta_3,\theta_4)$. Without loss of generality we take $(\theta_1,\theta_2)$ to be inserted at Euclidean time $y$ and $(\theta_3,\theta_4)$ at Euclidean time $0$, with $y>0$. That is, the four insertions are at
\beq
	w_1=\theta_1 + i y\,, \qquad w_ 2= \theta_2 + i y\,, \qquad w_3 = \theta_3\,, \qquad w_4 =\theta_4\,.
\eeq
To evaluate the block at arbitrary times we simply compute it for these values of $w_i$\,, and then analytically continue the final result exploiting that the block is holomorphic in its arguments. In any case, we have
\beq
\label{E:LOblock}
	\mathcal{V}_0 = 1 + h_V h_W\langle \hspace{-.1cm} \langle  (\mathcal{J}_{12}^{(1)}\cdot \epsilon)(\mathcal{J}_{34}^{(1)}\cdot \epsilon)\rangle \hspace{-.1cm}\rangle + O\left( \frac{1}{C^2}\right)\,,
\eeq
whose $O(1/C)$ part is a tree-level exchange diagram of the $\epsilon$ field between the two bilocals, represented in Figure~\ref{F:exchange}.
(Since in our proposal we divide by the expectation value of the bilocals, the identity block is given by the sum of connected diagrams between the two bilocals, and so at $O(1/C)$ we do not include the one-loop renormalization of the bilocals.)

To proceed we require the $\epsilon$ propagator. The quadratic action on the cylinder is
\beq
	S_2 = \frac{C}{24\pi}\int d^2x \Big( (\bar{\partial}\epsilon')\epsilon'' -(\bar{\partial}\epsilon)\epsilon'\Big)\,.
\eeq
Fourier transforming as
\beq
	\epsilon(w) = \int_{-\infty}^{\infty} \frac{d\omega}{(2\pi)^2}\sum_{n=-\infty}^{\infty}  e^{in \theta + i \omega y}\tilde{\epsilon}(p)\,,
\eeq
where $p=(n,\omega)$, the Fourier-space propagator is
\beq
	\langle \tilde{\epsilon}(p_1)\tilde{\epsilon}(p_2) \rangle= \frac{24\pi}{C}\frac{1}{i n_1(n_1^2-1)(\omega_1-i n_1)} (2\pi)^2 \delta^{(2)}(p_1+p_2)\,.
\eeq
The zero modes $n=-1,0,+1$ for all $\omega$ are pure ``gauge,'' and may be set to zero by a suitable local $PSL(2;\mathbb{R})$ transformation. Removing those modes and Fourier-transforming back to position space, we find the position-space propagator
\begin{align}
\begin{split}
\label{E:eps2}
	\langle \epsilon(w)\epsilon(0)\rangle &= \int_{-\infty}^{\infty} \frac{d\omega_1 d\omega_2}{(2\pi)^4} \sum_{n_1,n_2\neq -1,0,+1} e^{i n_1 \theta+i \omega_1 y} \langle \tilde{\epsilon}(p_1)\tilde{\epsilon}(p_2)\rangle
	\\
	& = \frac{6}{C}\left( -1+\frac{3\zeta}{2}-\frac{(1-\zeta)^2}{\zeta}\ln(1-\zeta)\right)\,, \qquad \zeta = e^{i \text{sgn}(y)w}\,,
\end{split}
\end{align}
where in going from the first line to the second we have used Cauchy's theorem. Evaluating the exchange diagram in~\eqref{E:LOblock}, taking the decompactification limit by rescaling $w_i=\alpha z_i$ and sending $\alpha\to 0$, and then putting the four insertions at $z_1 = \infty$, $z_2 = 1$, $z_3=u$, and $z_4 = 0$, we find
\begin{align}
\begin{split}
\label{E:LOblock2}
	\mathcal{V}_0 &= 1 + \frac{12 h_V h_W}{C}\left( -2+\left( 1-\frac{2}{u}\right)\ln(1-u)\right) + O\left( \frac{1}{C^2}\right) 
	\\
	& = 1+\frac{2h_Vh_W}{C}u^2 \,_2F_1(2,2;4,u) + O\left(\frac{1}{C^2}\right)\,,
\end{split}
\end{align}
which matches~\eqref{E:exponentiation} to $O(1/C)$.

\begin{figure}[t]
\begin{center}
	\includegraphics[width=6in]{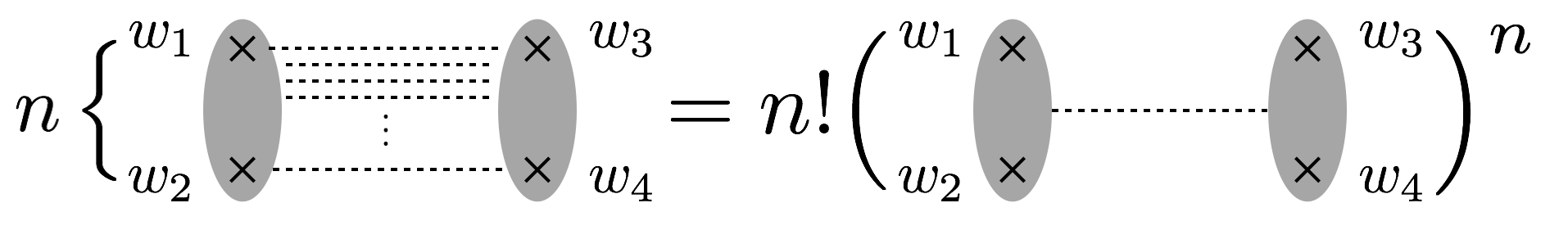}
	\caption{
		\label{F:exponentiation} The diagram with $n$ exchanges of the $\epsilon$ field between two bilocals.
	}
\end{center}
\end{figure}

Now we wish to recover the exponentiation~\eqref{E:exponentiation} in the limit where $h/\sqrt{c}$ is held fixed. To do so we rescale the $\epsilon$ field as $\epsilon \to \frac{\epsilon}{\sqrt{C}}$ so that the $\epsilon$-propagator is $O(1)$ and the interaction vertices with $n$ reparameterization fields scale as $O(C^{-n/2})$. We also write $h_i = \sqrt{C} \,\mathfrak{h}_i$\,. In the large $C$ limit with $\mathfrak{h}$ fixed, the corrections to the bilocal in~\eqref{E:bilocalLO} exponentiate as
\beq
	\left(2\sin\left( \frac{\theta_{12}}{2}\right)\right)^{2h}\mathcal{B}(h;w_1,w_2) = \exp\left( \mathfrak{h} \left( \mathcal{J}_{12}^{(1)}\cdot \epsilon+\frac{1}{\sqrt{C}} \mathcal{J}_{12}^{(2)} \cdot \epsilon + O\left(\frac{1}{C}\right)\right)\right)\,,
\eeq
where
\beq
\label{E:J2}
	\mathcal{J}_{12}^{(2)}\cdot \epsilon = -\frac{1}{2}\left( \epsilon_1'^2+\epsilon_2'^2-\frac{1}{2\sin^2\left( \frac{w_{12}}{2}\right)}\epsilon_{12}^2\right)\,,
\eeq
is the quadratic coupling to the $\epsilon$-field. The block is then approximately given by
\beq
\label{E:VpreExp}
	\mathcal{V}_0 = \langle \hspace{-.1cm}\langle \exp\left( \mathfrak{h}_V \mathcal{J}_{12}^{(1)}\cdot \epsilon\right)\exp\left( \mathfrak{h}_W \mathcal{J}_{12}^{(1)}\cdot \epsilon\right)\rangle \hspace{-.1cm}\rangle \left(1 + O\left( \frac{1}{\sqrt{C}}\right)\right)\,.
\eeq
As $C\to\infty$ we neglect self-interactions of the $\epsilon$-field as well as the higher order couplings $\mathcal{J}_{12}^{(n>1)}$ of the bilocal to the reparameterization field. The remaining terms in~\eqref{E:VpreExp} that contribute have $n$ powers of $\mathcal{J}_{12}^{(1)}\cdot \epsilon$ which are contracted with $n$ powers of $\mathcal{J}_{34}^{(1)}\cdot \epsilon$, and are represented in Figure~\ref{F:exponentiation}. There are $n!$ such contractions and each is weighted by a factor of $\frac{\mathfrak{h}_V^n\mathfrak{h}_W^n}{n!^2}$ arising from the exponentials. Thus the contribution with $n$ exchanges is $\frac{1}{n!}$ times the single-exchange to the $n^{\rm th}$ power. So the single exchange~\eqref{E:LOblock} exponentiates as
\begin{align}
\begin{split}
\label{E:expBlock}
	\mathcal{V}_0 &= \left( \sum_{n=0}^{\infty}\frac{1}{n!}\left(  \mathfrak{h}_V\mathfrak{h}_W\langle \hspace{-.1cm}\langle (\mathcal{J}_{12}^{(1)}\cdot \epsilon)(\mathcal{J}_{34}^{(1)}\cdot \epsilon)\rangle \hspace{-.1cm}\rangle\right)^n\right)\left( 1 + O\left( \frac{1}{\sqrt{C}}\right)\right)
	\\
	&= \exp\left( \frac{2h_Vh_W}{C}u^2\,_2F_1(2,2;4,u)\right)\left( 1 + O\left(\frac{1}{\sqrt{C}}\right)\right)\,,
\end{split}
\end{align}
recovering~\eqref{E:exponentiation} as we sought.

The subleading corrections in $1/C$ to the block are given by the sum over connected diagrams. In the limit where $h_V$ and $h_W$ are held fixed, there are several diagrams which contribute to the block at $O(1/C^2)$. These include the one-loop renormalization of the $\epsilon$-propagator, as well as tree-level diagrams. The latter also determine the $O(1/\sqrt{C})$ correction to~\eqref{E:expBlock}. 

\subsection{The heavy-light limit}

The identity block is also known in the ``heavy-light'' limit, in which one of the operators is ``heavy'' with dimension $h_H=O(c)$, while the other is ``light with dimension $h_L = O(1)$. This block is most naturally written in a different kinematic limit than the one we considered above, namely
\beq
	\langle \mathcal{O}_H(\infty) \mathcal{O}_L(1) \mathcal{O}_L(u) \mathcal{O}_H(0)\rangle\,.
\eeq
Conformally transforming to the cylinder, so that the heavy operators put us into the state $|h_H\rangle$, this limit is equivalent to the two-point function of the light operator in the heavy state, which is known~\cite{Fitzpatrick:2015zha} to be
\beq
\label{E:heavylightLO}
	\langle h_H| \mathcal{O}_L(w)\mathcal{O}_L(0)|h_H\rangle = \left( \frac{\alpha}{2 \sin\left( \frac{\alpha w}{2}\right)}\right)^{2h}\left( 1 + O\left(\frac{1}{c}\right)\right)\,, \qquad \alpha = \sqrt{1-\frac{24h_H}{c}}\,.
\eeq
The $O(1/c)$ corrections were computed in~\cite{Beccaria:2015shq,Fitzpatrick:2015dlt}, and higher order corrections were computed in~\cite{Chen:2016cms}.

There is quite a bit of physics already in the leading order result. As the dimension of the heavy operator is increased from $0$, there is a transition at $h_H = \frac{c}{24}$, which corresponds to the mass of the lightest BTZ black hole. Below this threshold, $\alpha \in (0,1)$,  the leading order result is periodic around the circle with a new periodicity $\theta \sim \theta + \frac{2\pi}{\alpha}$. Above this threshold it is periodic in imaginary time with $y \sim y + \frac{2\pi}{|\alpha|}$, i.e. it obeys a KMS relation with an effective inverse temperature $\beta = \frac{|\alpha|}{2\pi}$. Of course the full correlator does not obey the KMS condition, and indeed the $1/c$ corrections are not periodic in imaginary time. We refer the reader to~\cite{Fitzpatrick:2015zha} for further discussions of the physics of this quasi-periodicity and its relation to the Eigenstate Thermalization Hypothesis.

In the notation of the previous Subsection with $\mathcal{O}_1=\mathcal{O}_2=\mathcal{O}_H$, the leading order result~\eqref{E:heavylightLO} corresponds to
\beq
	\mathcal{V}_0(u) = \left( \frac{\alpha u (1-u)^{\frac{\alpha-1}{2}}}{1-(1-u)^{\alpha}}\right)^{2h_L}\left( 1 + O\left( \frac{1}{c}\right)\right)\,.
\eeq
(Recall that we normalize the identity block so that $\mathcal{V}_0\to 1$ as $u\to 0$.)
Conformally transforming to the thermal cylinder at inverse temperature $\beta$ and performing an analytic continuation to the second sheet, this gives the identity's contribution to the out-of-time-ordered four-point function which has been used as a definition for quantum Lyapunov growth~\cite{Shenker:2013pqa,kitaev}.  This contribution well-approximates the full answer for the four-point function in the limit of large central charge and a large higher-spin gap, and when both of those conditions hold true the four-point function grows with a maximal~\cite{Maldacena:2015waa} Lyapunov exponent $2\pi\beta$~\cite{Roberts:2014ifa}
\beq
	 \frac{\langle \mathcal{O}_H(t) \mathcal{O}_L(0) \mathcal{O}_H(t) \mathcal{O}_L(0)\rangle_{\beta}}{\langle \mathcal{O}_H\mathcal{O}_H\rangle_{\beta} \langle \mathcal{O}_L \mathcal{O}_L\rangle_{\beta}}   = \left(\frac{1}{1+ 6 \pi h_H \, e^{\frac{2\pi}{\beta}(t - t_*)}}\right)^{2 h_L}\left( 1 + O\left( \frac{1}{c}\right)\right)\,,
\eeq
where $t_* = \frac{\beta}{2\pi} \,\log(c)$ and we have written the denominator to first order in $\frac{h_H}{c}$. (This result is presented in a convention in which the heavy insertions are at imaginary times $0,\frac{\beta}{2}$ and the light insertions are at $\frac{\beta}{4}, \frac{3\beta}{4}$.)

In the remainder of this Section we obtain the leading order result for the heavy-light block~\eqref{E:heavylightLO} as well as its $O(1/c)$ corrections. In principle we could compute it via the two-point function of bilocal operators on the cylinder, with one bilocal corresponding to the insertions of the heavy operator and the other to the light insertions. The heavy operator strongly ``backreacts'' on the reparameterization field, and to proceed one must solve for the latter in the presence of the heavy bilocal. To efficiently proceed we place the heavy insertions in the infinite past and future. The classical trajectory is then modified to
\beq
\label{E:HLsaddle}
	\phi_c = \alpha_0 \theta\,, \qquad \alpha_0 = \sqrt{1-\frac{24 h_H}{C}}\,.
\eeq
One way to obtain this is to use that the classical stress tensor on the cylinder is now modified to be
\beq
	T_c = -\frac{C}{12}\left\{\tan\left(\frac{\phi_c}{2}\right),\theta\right\} = h_H - \frac{C}{24} \,,
\eeq
which may be integrated to give~\eqref{E:HLsaddle}. The field equation of the model is just the conservation of the stress tensor, and so this configuration is indeed a saddle.

The modified saddle~\eqref{E:HLsaddle} no longer obeys $\phi(\theta+2\pi) = \phi(\theta) + 2\pi$. The insertion of the heavy operator has twisted the boundary conditions for the reparameterization field, and in fact~\eqref{E:HLsaddle} is the unique saddle consistent with the new boundary conditions. This twisting also breaks the local $PSL(2;\mathbb{R})$ redundancy down to a local $U(1)$ redundancy, so that we identify $\phi(\theta,y) \sim \phi(\theta,y) + a(y)$. Both this, and the fact that we are now working in a state $|h_H\rangle$ rather than the vacuum $|0\rangle$, evoke the quantization of the normal orbits of the Virasoro group Diff$(\mathbb{S}^1)/U(1)$. Indeed, if we redefine $\phi_{\rm old} = \alpha_0\phi_{\rm new}$ so that the redefined $\phi$ is a standard Diff$(\mathbb{S}^1)$ field obeying $\phi(\theta+2\pi) = \phi(\theta)+2\pi$, the action is redefined as
\beq
	S_E = \frac{C}{24\pi}\int d^2x \left( \frac{(\bar{\partial}\phi')\phi''}{\phi'^2} - (\bar{\partial}\phi)\phi'\right)\to \frac{C}{24\pi}\int d^2x \left( \frac{(\bar{\partial}\phi')\phi''}{\phi'^2} - \alpha_0^2 (\bar{\partial}\phi)\phi'\right)\,.
\eeq
(Recall that precisely this mechanism -- the radius of the Diff$(\mathbb{S}^1)$ being modified by a change in the energy -- was at work in the description of conical defects in AdS$_3$ presented in Subsection~\ref{S:deficits}.)
This is the action for the quantization of a normal orbit with $b_0 = -\frac{C\alpha_0^2}{48\pi} $~\eqref{E:Seuc}. Solving for $h_H$ gives
\beq
	h_H = 2\pi b_0 + \frac{C}{24}\,,
\eeq
which matches our exact relation between the weight $b_0$ of the primary state, and the $C$ that we found in Eq.~\eqref{E:cAndh}. It will be important later that in this setting the exact central charge is
\beq
	c = C+1\,,
\eeq
rather than $c=C+13$ as it is for the quantization of Diff$(\mathbb{S}^1)/PSL(2;\mathbb{R})$. 

In this setting our proposal is that the four-point function~\eqref{E:heavylightLO} is given by an appropriately renormalized expectation value of a bilocal operator corresponding to the insertion of the light operators in the state $|h_H\rangle$,
\beq
\label{E:heavylightProposal}
	\langle h_H| \mathcal{O}_L(w)\mathcal{O}_L(0)|h_H\rangle = \langle \mathcal{B}(h_L;w,0)\rangle_{\alpha}\,.
\eeq
Evaluating the bilocal~\eqref{E:bilocalCyl} in the reparameterized background~\eqref{E:HLsaddle} sourced by the heavy operator gives the classical approximation as $C\to\infty$,
\beq
\label{E:heavylightTree}
	\langle \mathcal{B}(h_L;\theta,0)\rangle_{\alpha} = \left( \frac{\alpha_0}{2\sin\left( \frac{\alpha_0 \theta}{2}\right)}\right)^{2h_L}\left( 1 + O\left( \frac{1}{C}\right)\right)\,.
\eeq
Analytically continuing $\theta \to w$, this matches the leading order result~\eqref{E:heavylightLO} using that $\alpha_0$ becomes $\alpha$ as $C\to\infty$.

\begin{figure}[t]
\begin{center}
\includegraphics[width=2in]{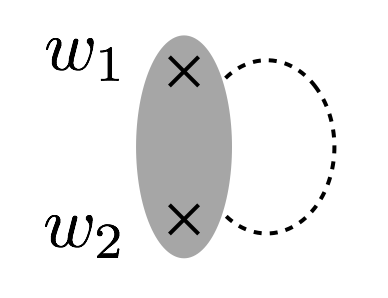}
\caption{\label{F:1loop} The one-loop diagram which gives the $O(1/C)$ correction to the identity block in the heavy-light limit. The coupling to the reparameterization field is given by $\frac{h_L^2}{2C}(\mathcal{J}_{12}'^{1}\cdot\epsilon)^2 + h_L \mathcal{J}_{12}'^{(2)}\cdot\epsilon$.}
\end{center}
\end{figure}

The $O(1/C)$ corrections to the block are computed by the one-loop approximation to~\eqref{E:heavylightProposal}. In the background~\eqref{E:HLsaddle}, the coupling of the bilocal operator~\eqref{E:bilocalCyl} to the reparameterization field $\phi = \alpha_0  \theta + \frac{\epsilon}{\sqrt{C}}$ is given by
\beq
	\mathcal{B}(h;\theta_1,\theta_2) = \left( \frac{\alpha_0}{2\sin\left( \frac{\alpha_0 \theta_{12}}{2}\right)}\right)^{2h}\left( 1 + \frac{h}{\sqrt{C}}\mathcal{J}_{12}'^{(1)}\cdot \epsilon + \frac{1}{C}\left(\frac{h^2}{2}(\mathcal{J}_{12}'^{(1)}\cdot \epsilon)^2 + h \mathcal{J}_{12}'^{(2)}\cdot \epsilon\right) + O\left( \frac{1}{C^{3/2}}\right)\right)\,,
\eeq
where
\begin{align}
\begin{split}
	\mathcal{J}_{12}'^{(1)} \cdot \epsilon & = \frac{\epsilon_1'+\epsilon_2'}{\alpha_0}-\cot\left( \frac{\alpha_0 \theta_{12}}{2}\right)\epsilon_{12}\,,
	\\
	\mathcal{J}_{12}'^{(2)} \cdot \epsilon & = -\frac{1}{2}\left( \frac{\epsilon_1'^2 + \epsilon_2'^2}{\alpha_0^2} - \frac{1}{2}\csc^2\left( \frac{\alpha_0 \theta_{12}}{2}\right)\epsilon_{12}^2\right)\,.
\end{split}
\end{align}
It then follows that
\beq
\label{E:HL1loop}
	\langle \mathcal{B}(h_L;w,0)\rangle_{\alpha} = \left( \frac{\alpha}{2\sin\left(\frac{\alpha w}{2}\right)}\right)^{2h_L} \left( 1 + \frac{h_L}{c}\mathcal{V}_{h/c}(w) + \frac{h_L^2}{c}\mathcal{V}_{h^2/c}(w) + O\left( \frac{1}{c^2}\right)\right)\,,
\eeq
with
\begin{align}
\begin{split}
	\mathcal{V}_{h^2/c}(w) &=\frac{1}{2} \langle (\mathcal{J}_{12}'^{(1)}\cdot \epsilon)^2 \rangle \,,
		\\
	\mathcal{V}_{h/c}(w) & = \langle \mathcal{J}_{12}'^{(2)}\cdot \epsilon \rangle -\frac{1-\alpha^2}{\alpha^2}\left( 1 -\frac{\alpha w}{2} \cot\left( \frac{\alpha w}{2}\right)\right)\,.
\end{split}
\end{align}
The last term in $\mathcal{V}_{h/c}$ arises from expanding the tree-level result~\eqref{E:heavylightTree} to $O(1/c)$,
\beq
	\left( \frac{\alpha_0}{2\sin\left( \frac{\alpha_0 w}{2}\right)}\right)^{2h_L} = \left( \frac{\alpha}{2\sin\left( \frac{\alpha w}{2}\right)}\right)^{2h_L} \left( 1 - \frac{h_L(1-\alpha^2)}{c\alpha^2}\left( 1 -\frac{ \alpha w}{2} \cot\left( \frac{\alpha w}{2}\right)\right)+ O\left( \frac{1}{c^2}\right)\right) \,,
\eeq
using that $\alpha_0 = \sqrt{1-\frac{24 h_H}{C}}$ and $C=c-1$. The self-interactions of the $\epsilon$ field are suppressed as $O(1/\sqrt{C})$, and so the $O(1/C)$ approximation to the block is given by the one-loop bubble diagram represented in Figure~\ref{F:1loop}.

We require the $\epsilon$ propagator. The Fourier space propagator is now
\beq
	\langle \tilde{\epsilon}(p_1)\tilde{\epsilon}(p_2)\rangle= \frac{24\pi\alpha^2}{i n_1(n_1^2-\alpha^2) (\omega - i n_1)}(2\pi)^2 \delta^{(2)}(p_1+p_2)\,,
\eeq
which implies
\begin{align}
\begin{split}
	\langle \epsilon(w)\epsilon(0)\rangle & = \int \frac{d\omega_1 d\omega_2}{(2\pi)^4}\sum_{n_1,n_2\neq 0} e^{in_1 \theta+i \omega_1 y} \langle \tilde{\epsilon}(p_1)\tilde{\epsilon}(p_2)\rangle
	\\
	& = \frac{6}{C}\left( 2\ln (1-\zeta) + \Phi(\zeta,1,\alpha)+\Phi(\zeta,1,-\alpha)\right) \,, \qquad \zeta = e^{i \text{sgn}(y)w}\,,
\end{split}
\end{align}
where $\Phi(w,s,a)$ is the Lerch transcendant
\beq
	\Phi(w,s,a) = \sum_{n=0}^{\infty} \frac{w^n}{(n+a)^s}\,.
\eeq
For $s=1$ it is related to a certain incomplete Beta function as
\beq
\label{E:fromLerchToBeta}
	B(w,a,0) = w^a \Phi(w,1,a)\,.
\eeq

We are now in a position to evaluate the $O(h_L^2/c)$ and $O(h_L/c)$ corrections in~\eqref{E:HL1loop}. Both are logarithmically divergent, and so must be appropriately renormalized. We do so via background subtraction, defining the renormalized expectation value of the bilocal by
\beq
	\langle \mathcal{B}(h_L;w_1,w_2)\rangle_R = \langle \mathcal{B}(h_L;w_1,w_2)\rangle_{\alpha} - \left(\langle \mathcal{B}(h_L;w_1,w_2)\rangle_{\alpha\to 1}\right)_{w_i \to \alpha w_i}\,.
\eeq
With this scheme, the expectation value of the bilocal is exactly the two-point function of a dimension $h_L$ operator in the limit $\alpha \to 1$. In evaluating the coincident limits of the propagator in~\eqref{E:HL1loop} we also use a point-splitting procedure in which we take one of the insertions to be slightly after the other in Euclidean time. This gives
\begin{align}
\begin{split}
    	\langle \epsilon_1^2 \rangle_{\alpha} &\to \langle \epsilon( \delta)\epsilon(0)\rangle_{\alpha} = - \frac{6}{C}(H_{\alpha}+H_{-\alpha})+O(\delta)\,,
	\\
	\langle \epsilon_1'^2\rangle_{\alpha} & \to \langle \epsilon'(\delta)\epsilon'(0)\rangle_{\alpha} = -\frac{6\alpha^2}{C}\left( H_{\alpha}+H_{-\alpha} + 2 \ln (-i \delta)\right) + O(\delta)\,,
\end{split}
\end{align}
where Im$(\delta)\neq 0$ and $H_{\alpha}$ is the harmonic number. We also perform a point-splitting procedure even for the contributions at nonzero separation, taking the two insertions of the bilocal to be at infinitesimally different Euclidean times. Putting the pieces together, we find
\begin{align}
\nonumber
	\mathcal{V}_{h^2/c}  =& 6 \Big (-\csc\left( \frac{\alpha w}{2}\right)^2 \Big( \frac{B(e^{i w},\alpha,0) + B(e^{-iw},\alpha,0)+B(e^{iw},-\alpha,0)+B(e^{-iw},-\alpha,0)}{2} 
	\\
	\nonumber
	& \qquad + H_{\alpha}+H_{-\alpha} + 2 \ln \left( 2 \sin\left(\frac{w}{2}\right)\right)\Big) + 2 \ln \left( \alpha \sin\left(\frac{w}{2}\right)\csc\left( \frac{\alpha w}{2}\right)\right)+1\Big)\,,
	\\
	\nonumber
	\mathcal{V}_{h/c} = &-\frac{1}{2}\csc^2\left(\frac{\alpha w}{2}\right) \Big( 3 (\Phi(e^{i w},1,\alpha)+\Phi(e^{-i w},1,\alpha)+\Phi(e^{iw},1,-\alpha)+\Phi(e^{-iw},1,-\alpha))
	\\
	\nonumber
	& \qquad +\cos(\alpha w)\left(6(H_{\alpha}+H_{-\alpha}+i \pi)-5\right) +12 \ln\left( -2 i \sin\left(\frac{w}{2}\right)\right)+5\Big)
	\\
	& \qquad \qquad -\frac{1}{\alpha^2}-\frac{13\alpha^2-1}{2\alpha}w \cot\left(\frac{\alpha w}{2}\right)+12 \ln\left(-\frac{2i}{\alpha}\sin\left(\frac{\alpha w}{2}\right)\right)\,,
\end{align}
where we have used~\eqref{E:fromLerchToBeta}. This result precisely matches the $O(1/c)$ correction computed in~\cite{Fitzpatrick:2015dlt}, upon replacing $t$ in Eq. (1.4) of their manuscript with $-i w$.

We expect that further subleading corrections to the block in $1/c$ are computed by higher loop corrections. In particular, the $O(1/C^n)$ corrections should be computed by the $n$-loop diagrams. It would be interesting to test if this is indeed the case using the known $O(1/c^2)$ and $O(1/c^3)$ corrections computed in~\cite{Chen:2016cms}.

\section{Supersymmetry}
\label{S:generalizations}

There are a number of obvious generalizations to the geometric models considered in this work. One may consider the quantization of the coadjoint orbits of a Virasoro-Kac-Moody group, or of super-Virasoro groups with varying amounts of supersymmetry. In gravitational terms these quantizations respectively correspond to (the chiral half of) pure gravity and Chern-Simons theory on AdS$_3$, or to (the chiral half of) pure supergravity on AdS$_3$. Upon reduction, one expects these theories to respectively yield the version of the Schwarzian theory enhanced with global symmetries obtained in~\cite{Davison:2016ngz,Choudhury:2017tax}, or the super-Schwarzian theory of~\cite{Fu:2016vas}.

In this Section we consider the coadjoint orbits of the $\mathcal{N}=1$ super-Virasoro group and set up their path integral quantization, focusing on the analogue of the exceptional orbit Diff$(\mathbb{S}^1)/PSL(2;\mathbb{R})$ with AdS$_3$ supergravity in mind. Most of this Section is not new, but we have found the literature to be a bit scattered and misleading in places. Thus our goal here is to summarize the main results, and to pave the way for future computations like those considered in this work, e.g. of super-Virasoro blocks in the heavy-light limit.

We now turn our attention to the $\mathcal{N}=1$ super-Virasoro group. See e.g.~\cite{Friedan:1986rx,Bakas:1988mq,Aratyn:1989qq,Delius:1990pt}. We have found~\cite{Aratyn:1989qq} especially helpful. The quantum mechanical models we so obtain will be 2d chiral theories with $\mathcal{N}=1$ superconformal symmetry. The $\mathcal{N}=1$ generalization of Diff($\mathbb{S}^1)$ is the diffeomorphism group of the supercircle Diff$_{\pm}(\mathbb{S}^{1|1})$. The supercircle $\mathbb{S}^{1|1}$ is parameterized by a bosonic periodic variable $\varphi \sim \varphi+2\pi$ and a Grassmann-odd coordinate $\theta$,\footnote{Beware that we are switching notation in this Subsection relative to the rest of the manuscript: $\theta$ is now the fermionic coordinate, rather than the coordinate along the $\mathbb{S}^1$.} which we sometimes group together into a supercoordinate $X=(\varphi,\theta)$. The $\pm$ subscript refers to the spin structure. The supercircle is equipped with a superderivative
\beq
	D = \frac{\partial}{\partial\theta}+\theta \frac{\partial}{\partial\varphi}\,, \qquad D^2 = \frac{\partial}{\partial\varphi}\,.
\eeq
A superreparametrization of the supercircle is a change of super-coordinates $(\tilde{\varphi},\tilde{\theta})$ constrained so that the superderivative rescales homogeneously,
\beq
	\tilde{D} = \frac{\partial}{\partial\tilde{\theta}}+\tilde{\theta}\frac{\partial}{\partial\tilde{\varphi}}\propto D\,.
\eeq
Solving this constraint leads to the condition
\beq
\label{E:superreparam}
	D\tilde{\varphi} - \tilde{\theta} D\tilde{\theta} = 0\,.
\eeq
Purely bosonic solutions are given by $\tilde{\varphi} = f(\varphi)$, $\tilde{\theta} = \sqrt{f'(\varphi)}\theta$. The most general solution is~\cite{Fu:2016vas} given in terms of a Grassmann-even function $f$ and Grassmann-odd one $\eta$ as
\beq
	\tilde{\varphi} = f(\varphi + \theta \eta(\varphi))\,, \qquad \tilde{\theta} = \sqrt{f'(\varphi)} \left( \theta + \eta(\varphi) + \frac{1}{2}\theta \eta(\varphi)\eta'(\varphi)\right)\,.
\eeq
The bosonic function $f$ is a standard Diff$(\mathbb{S}^1)$ field obeying $f(\varphi+2\pi)=f(\varphi)+2\pi$, while $\eta(\varphi+2\pi) = \pm \eta(\varphi)$ depending on the spin structure. In any case, under superreparameterizations, the superderivative, super-one-form $d\varphi+\theta d\theta$, and super-measure $dX=d\varphi d\theta$ rescale homogeneously as
\beq
	\tilde{D} = (D\tilde{\theta})^{-1} D\,, \qquad d\tilde{\varphi} + \tilde{\theta}d\tilde{\theta} = (D\tilde{\theta})^2(d\varphi+\theta d\theta) \,, \qquad d\tilde{X} = (D\tilde{\theta}) dX\,,
\eeq
so that $D\tilde{\theta}$ plays the role of the Jacobian of the transformation. More generally, one defines $h$-superdifferentials, which are superfields $A$ that transform as
\beq
	\tilde{A}(\tilde{X}) = (D\tilde{\theta})^{-2h}A(X)\,.
\eeq

An infinitesimal superreparameterization
\beq
	\tilde{\phi} = \phi + \delta \phi\,, \qquad \tilde{\theta} = \theta + \delta \theta\,,
\eeq
may be packaged into a superfield as
\beq
	V = \delta \phi + \theta \delta \theta\,.
\eeq
Here $V$ is a supervector field, an adjoint vector of the algebra of Diff$_{\pm}(\mathbb{S}^{1|1})$. In terms of $V$ we have $\delta\theta = \frac{1}{2}DV$ and $\delta D\tilde{\theta} = \frac{1}{2}\partial V$. A straightforward computation demonstrates that $V$ transforms as a $-1$ differential.  The infinitesimal form of this transformation gives the supercommutator of supervector fields,
\beq
	\delta_{V_1}V_2 = [V_1,V_2] = V_1 \partial V_2 - V_2\partial V_1 + \frac{1}{2}DV_1 DV_2\,,
\eeq
which defines the commutator of the algebra of Diff$_{\pm}(\mathbb{S}^{1|1})$. In more detail, an infinitesimal superreparameterization,
\beq
	\delta \phi = \xi_2(\varphi)+\theta \eta_2(\varphi)\,, \qquad \delta\theta = \eta_2(\varphi) + \frac{\theta}{2}\xi_2'(\varphi)\,,
\eeq
corresponds to a supervector
\beq
	V_2 = \xi_2(\varphi)  + 2\theta \eta_2(\varphi)\,,
\eeq
whose components transform under another infinitesimal superreparamterization $V_1$ as
\beq
	\delta_{1} \xi_2 = \xi_1 \xi_2' - \xi_2 \xi_1'+2\eta_1\eta_2\,, \qquad \delta_{1} \eta_2 = \xi_1 \eta_2' - \xi_2 \eta_1' -\frac{1}{2}\left( \eta_2 \xi_1' - \eta_1\xi_2'\right)\,.
\eeq

The $\mathcal{N}=1$ super-Virasoro group $\widehat{\text{Diff}}_{\pm}(\mathbb{S}^{1|1})$ is the central extension of the diffeomorphism group of the supercircle. Introducing a central element $c$, vectors are now the combination of a supervector field $V$ and a number $a$ and we group them as $(V,a)$, or equivalently $V-i a c$. The commutator is now deformed by
\beq
\label{E:superVirasoro}
	\left[ (V_1,a_1),(V_2,a_2)\right] = \left(V_1 \partial V_2 - V_2\partial V_1+\frac{1}{2}DV_1 DV_2 ,  -\frac{1}{48\pi}\int dX (V_1 \partial^2 DV_2 - V_2 \partial^2DV_1)\right)\,.
\eeq
From this we recover the standard presentation of the $\mathcal{N}=1$ super-Virasoro algebra. We define the generators $L_n$ to correspond to the bosonic transformation $\xi = i e^{i n \varphi}$ and the Grassmann-odd generators $G_{\mu}$ correspond to the Grassmann-odd ones $\eta = \sqrt{i} \gamma e^{i \mu \varphi}$, with $\gamma^2 = 1$ and $\mu \in \mathbb{Z} $ or $\mathbb{Z}+\frac{1}{2}$ depending on the spin structure. With this identification Eq.~\eqref{E:superVirasoro} then implies
\begin{align}
\begin{split}
	[L_n,L_m] & = (n-m)L_{n+m} +\frac{c}{12}n^3 \delta_{n+m}\,,
	\\
	[L_n,G_{\mu}] & = \left(\frac{n}{2}-\mu\right)G_{\mu+n}\,,
	\\
	\{G_{\mu},G_{\nu}\} & = 2L_{\mu+\nu}+\frac{c}{3}\mu^2 \delta_{\mu+\nu}\,,
\end{split}
\end{align}
which differs from the usual algebra by $L_{0,\rm usual} = L_{0,\rm here}+\frac{c}{24}$.

Coadjoint vectors of Diff$_{\pm}(\mathbb{S}^{1|1})$ are Grassmann-odd, $\frac{3}{2}$-differentials $B$. Coadjoint vectors of its central extension are the combination of a Grassmann-odd superfield $B$ and an ordinary number $t$, which we denote as $(B,t)$. The pairing between $(B,t)$ and an adjoint vector $(V,a)$ is
\beq
	\langle (B,t),(V,a)\rangle = \int dX \, B V +  ta\,.
\eeq
This pairing must be super-Virasoro-invariant, which fixes the variation of the coadjoint vector under a transformation generated by $(V,a)$ to be
\begin{align}
\begin{split}
\label{E:deltaB}
	\delta B &= V \partial B + \frac{3}{2}B\partial V + \frac{1}{2}DV DB-\frac{t\partial^2 DV}{24\pi}\,,
	\\
	\delta a & = 0\,.
\end{split}
\end{align}

Consider the orbit of a constant coadjoint vector $(B_0 = \theta b_0,C)$. The finite form of the transformation~\eqref{E:deltaB} under a superreparameterization is
\beq
\label{E:superCoadjointTransformation}
	(B_0,C) \to (B(X),C) = \left( \tilde{\theta} b_0 (D\tilde{\theta})^{3} - \frac{C}{24\pi}S(\tilde{X},X),C\right)\,,
\eeq
where
\beq
	S(\tilde{X},X) = 2\left(\frac{D^4\tilde{\theta}}{D\tilde{\theta}}-2\frac{D^3\tilde{\theta} D^2\tilde{\theta}}{D\tilde{\theta}^2}\right)
\eeq
is the $\mathcal{N}=1$ super-Schwarzian derivative. (We use a slightly different normalization for the super-Schwarzian than is common in the literature. Ours reduces to the ordinary Schwarzian $\{f(\varphi),\varphi\}$ under a bosonic transformation $\tilde{\varphi} = f(\varphi), \tilde{\theta} = \sqrt{f'(\varphi)}\theta$.) These orbits are isomorphic to $\widehat{\text{Diff}}_{\pm}(\mathbb{S}^{1|1})/S$, where $S$ is the stabilizer of the orbit. To gain some intuition about the orbits, note that at a point $(B(X),C)$ on the orbit, the generator $L_0$ corresponds to the function
\beq
\label{E:superL0}
	L_0 = - \int dX \left( \frac{C}{24\pi}S(\tilde{X},X) - \tilde{\theta} b_0 (D\tilde{\theta})^3\right)
\eeq
so that at the origin $(B_0,C)$ one has $L_0 = 2\pi b_0$. The quantization of the orbit leads to a quantum mechanical model whose Hilbert space is composed of a single super-Verma module with a highest weight state $|h\rangle$ satisfying $h=2\pi b_0+\frac{C}{24}$, where $C$ plays the role of the central charge up to an $O(C^0)$ one-loop exact correction.

Calling the periodic spin structure ``Ramond'' and the anti-periodic structure ``Neveu-Schwarz,'' the analogues of the  first exceptional orbit of the Virasoro group are
\begin{enumerate}
\item Ramond: $b_0 =0$. For this value the quantization leads to $h=\frac{C}{24}$, which is the dimension of the Ramond vacuum at large $C$. This orbit is exceptional in that the superreparameterizations $(f(\varphi),\eta(\varphi))$ and $(f(\varphi)+\alpha,\eta(\varphi)+\beta)$ lead to the same coadjoint vector for any $(\alpha,\beta)$. The stabilizer is generated by $L_0$ and $G_0$, i.e. by a constant supervector $V$.
\item Neveu-Schwarz: $b_0 = - \frac{C}{48\pi}$. At this value one has $h=0$, which is the correct dimension of the Neveu-Schwarz vacuum. For this value the stabilizer is $OSp(1|2)$, which is generated by the supervector $V = \alpha_0 +\alpha_1e^{i\varphi}+\alpha_{-1} e^{-i\varphi} + 2\theta( \beta_{1/2} e^{i\varphi/2}+\beta_{-1/2}e^{-i\varphi/2})$, i.e. by $(L_{-1},L_0,L_1,G_{-1/2},G_{1/2})$. 
\end{enumerate}
In each case the quantization leads to the vacuum module, as we will demonstrate shortly. Orbits with larger $b_0$ are ``normal,'' and are stabilized by $L_0$ alone.

The Kirillov-Kostant supersymplectic form at a point $(B,C)$, evaluated on two supervectors $(V_1,a_1)$ and $(V_2,a_2)$ is given by
\begin{align}
\label{E:superomega}
	\omega((V_1,a_1),(V_2,a_2)) &=- \langle (B,C),[(V_1,a_1),(V_2,a_2)]\rangle 
	\\
	\nonumber
	&= -\int dX \left\{ B \left( V_1 \partial V_2 - V_2\partial V_1 + \frac{1}{2}DV_1 DV_2\right) - \frac{C}{48\pi}(V_1\partial^2 DV_2 - V_2\partial^2DV_1)\right\}\,.
\end{align}
We would like to write this as a two-form. To do so we need to write the infinitesimal variation of a superreparameterization $\tilde{X}$ generated by a vector $V$ as a formal one-form. That variation is nicely packaged into the super one-form $\tilde{U}=d\tilde{\phi}+\tilde{\theta} d\tilde{\theta}$ with
\beq
\tilde{U}(V) = \delta_V \tilde{\phi} + \tilde{\theta} \delta_V \tilde{\theta} = (D\tilde{\theta})^2 V\,,
\eeq
where $\delta_V \tilde{\phi}$ and $\delta_V\tilde{\theta}$ denote the variations of $\tilde{\phi}$ and $\tilde{\theta}$ generated by $V$. (In our review of the coadjoint orbits of the Virasoro group, the analogue of $\tilde{U}$ was $d\phi$, which satisfied $d\phi(F) = \delta_F \phi = f \phi'$. However note that unlike $d\phi$, $\tilde{U}$ is not closed: $d\tilde{U} = d\tilde{\theta} \wedge d\tilde{\theta}$.) Using this relation to eliminate $V_1$ and $V_2$ in~\eqref{E:superomega}, 
and 
substituting~\eqref{E:superCoadjointTransformation}, we 
then perform the superspace integral using
\begin{align}
\begin{split}
	\tilde{U}| & = df + f' \eta d\eta\,,
	\\
	D\tilde{U}| &= \eta \left( df' + f' \eta' d\eta\right) + 2f' d\eta\,,
\end{split}
\end{align}
allowing us to write $\omega$ in terms of the fields $f$ and $\eta$. We find after a suitable integration by parts
\begin{align}
\begin{split}
	\omega = -\int_0^{2\pi} d\varphi \Big\{& \frac{C}{48\pi}\left(\frac{df'\wedge df''}{f'^2} +4 d\eta'\wedge d\eta'-d(\eta \eta' \eta'' d\eta+2\{f,\varphi\} \eta d\eta)\right) 
	\\
	&  + b_0 (df \wedge df' + 2 f' \eta df' \wedge d\eta + f'^2 d\eta \wedge d\eta)\Big\}\,,
\end{split}
\end{align}
which is clearly closed. The bosonic part is just the symplectic form on the Virasoro orbits in~\eqref{E:symplecticForm}. In this presentation we also read off the presymplectic potential $\omega = d\alpha$ as
\begin{align}
\begin{split}
	\hspace{-.3in}\alpha =- \int_0^{2\pi} d\varphi &\left\{\frac{C}{48\pi}\left( -\frac{f''df'}{f'^2} + 4 \eta'd\eta' -\eta \eta'\eta''d\eta -2\{f,\varphi\}\eta d\eta\right)- b_0 (f' df - f'^2 \eta d\eta)\right\}\,.
\end{split}
\end{align}
This last expression can be recast as a superspace integral
\beq
	\alpha = \int dX \left\{ \frac{C}{24\pi}\frac{D^3\tilde{\theta}}{(D\tilde{\theta})^3}D\tilde{U}+ b_0 \tilde{\theta} (D\tilde{\theta}) \tilde{U}\right\}\,.
\eeq

To perform a path integral quantization we must do two things. First, we promote the superreparameterization to be a function of time. Second, we are interested in the Hamiltonian corresponding to $L_0$~\eqref{E:superL0}, which may be written as
\begin{align}
\begin{split}
	L_0 &=-\int_0^{2\pi}d\varphi \left\{ \frac{C}{24\pi}\left( \{f,\varphi\} (1-\eta\eta')+ \eta \eta'''+3\eta'\eta''\right) - b_0 f'^2(1-\eta \eta')\right\}
	\\
	&= \int_0^{2\pi}d\varphi \left\{ \frac{C}{48\pi}\left( \frac{f''^2}{f'^2}+2\{f,\varphi\} \eta\eta' -4\eta'\eta''\right) -b_0f'^2(1-\eta \eta')\right\}
\end{split}
\end{align}
following superspace integration. The action $-\int dt( \alpha_i dx^i +H)$ is then
\beq
	\hspace{-.1in}S^{\mathcal{N}=1}   = -\int d\varphi dt \left\{ \frac{C}{24\pi}\left( \frac{f''\partial_+f'}{f'^2}+2\{f,\varphi\} \eta \partial_+\eta -4 \eta'\partial_+\eta'+\eta \eta'\eta'' \partial_+\eta\right) + 2b_0(f'\partial_+f - f'^2 \eta \partial_+\eta)\right\}\,.
\eeq
Its bosonic part coincides with~\eqref{E:ourTheory}. Its Euclidean continuation $t = - i y$ is
\beq
	\hspace{-.1in} S^{\mathcal{N}=1}_E = \int d^2x \left\{ \frac{C}{24\pi}\left( \frac{f''\bar{\partial}f'}{f'^2}+2\{f,\varphi\}\eta \bar{\partial}\eta - 4 \eta' \bar{\partial}\eta' + \eta \eta' \eta'' \bar{\partial}\eta\right) + 2 b_0 (f' \bar{\partial}f - f'^2 \eta \bar{\partial} \eta )\right\}\,.
\eeq

Let us briefly consider these models on a torus of complex structure $\tau$, as in our analysis in Section~\ref{S:torus}. We have four choices of spin structure, depending on whether the fermion $\eta$ is (anti-)periodic around each of the two cycles. In all cases the unique saddle point consistent with the boundary conditions is
\beq
\label{E:SUSYsaddle}
	f_0 = \varphi - \frac{\text{Re}(\tau)}{\text{Im}(\tau)}y\,, \qquad \eta_0 = 0\,.
\eeq
Expanding around this saddle,
\beq
	f = f_0 + \epsilon\,, \qquad \eta = \chi\,,
\eeq
the quadratic action is
\beq
	S_2 = \int d^2x \left\{ \frac{C}{24\pi}\left( \epsilon'' \bar{\partial}\epsilon' -4 \chi' \bar{\partial}\chi'\right) + 2b_0(\epsilon' \bar{\partial}\epsilon - \chi \bar{\partial}\chi)\right\}\,.
\eeq

Following our discussion above Eq.~\eqref{E:superomega}, consider the special case of the Ramond model with $b_0 = 0$. In this setting the ``gauge'' redundancies identify
\beq
	\epsilon(\varphi,y) \sim \epsilon(\varphi,y) + a(y)\,, \qquad \chi(\varphi,y) \sim \chi(\varphi,y) + \alpha(y)\,,
\eeq
so that the $n=0$ spatial Fourier modes of $\epsilon$ and $\eta$ are ``pure gauge'' and may be consistently set to zero. The saddle point action vanishes, and the Fourier space quadratic action is then
\beq
	S_{\rm R} = \frac{C\pi}{6}\left( i\sum_{m=-\infty}^{\infty}\sum_{n\neq 0} n^3(m-n\tau) |\epsilon_{m,n}|^2 +4\sum_{\nu} \sum_{n\neq 0}n^2(\nu-n\tau)\chi_{n,m}^*\chi_{n,m}\right)\,,
\eeq
where $\nu \in \mathbb{Z}$ if $\chi$ is periodic around the ``time'' circle and $\nu \in \mathbb{Z}+\frac{1}{2}$ if $\chi$ is anti-periodic. Let us call the former case R-R and the latter NS-R. The one-loop approximation to the partition function is
\beq
\mathcal{Z}_{\rm 1-loop}(\tau) = \sqrt{\frac{\prod_{\nu,n}(\nu-n\tau)}{\prod_{m,p}(m-p\tau)}}\,.
\eeq
Evaluating the determinant gives
\beq
	\mathcal{Z}_{\rm 1-loop}(\tau) = 
		\prod_{n=1}^{\infty}\frac{1\mp q^n }{1-q^n}\,,
\eeq
where $\pm$ refers to the fermion boundary condition around the ``time'' circle. These are of course the vacuum characters
\beq
	\mathcal{Z}_{\rm R-R}(\tau) = \text{tr}_{\rm R}\left((-1)^Fq^{L_0-\frac{c}{24}}\right)\,, \qquad \mathcal{Z}_{\rm NS-R}(\tau) = \text{tr}_{\rm R} \left(q^{L_0-\frac{c}{24}}\right)\,,
\eeq
with the usual convention for $L_0$, using that $L_0 = \frac{c}{24}$ in the (non-degenerate) Ramond vacuum. This computation does not give the central charge directly, but a computation of the two-point function of the stress tensor on the cylinder reveals that the chiral central charge is not renormalized,
\beq
	c_{\rm R} = C\,.
\eeq
Here, the ``R'' in $c_{\rm R}$ is short for ``Ramond.'' 

It might be surprising that the R-R partition function is nonzero as the fermion $\chi$ has a zero mode. However, for this orbit the zero mode is pure ``gauge.'' For more general Ramond orbits, this zero mode persists but is no longer pure gauge, leading to the usual result that the non-vacuum R-R character vanishes.

Now consider the Neveu-Schwarz model at $b_0 = - \frac{C}{48\pi}$, which we expect to produce the Neveu-Schwarz vacuum module. In this case the ``gauge redundancy'' identifies
\begin{align}
\begin{split}
	\epsilon(\varphi,y) &\sim \epsilon(\varphi,y) + a_0(y) + a_{-1}(y) e^{-i \varphi}+a_1(y)e^{i\varphi}\,,
	\\
	\chi(\varphi,y)&\sim \chi(\varphi,y) + b_{-\frac{1}{2}}(y)e^{-i \varphi/2} + b_{\frac{1}{2}}(y)e^{i\varphi/2}\,,
\end{split}
\end{align}
and so removes the $n=-1,0,+1$ spatial Fourier modes of $\epsilon$ and the $\nu = -\frac{1}{2},+\frac{1}{2}$ modes of $\chi$. The quadratic action is modified to
\begin{align}
\begin{split}
	S_{\rm NS} =\frac{\pi C}{12}i\tau+ \frac{C\pi}{6}&\left( i\sum_{m=-\infty}^{\infty}\sum_{n\neq -1,0,+1} in(n^2-1)(m-n\tau) |\epsilon_{m,n}|^2\right.
	\\
	 & \qquad \qquad +\left.4\sum_{\nu} \sum_{n\neq \pm \frac{1}{2}}\left(n^2-\frac{1}{4}\right)(\nu-n\tau)\chi_{n,m}^*\chi_{n,m}\right)\,,
\end{split}
\end{align}
where $\nu \in \mathbb{Z}$ if we impose periodic boundary conditions around the ``time'' circle and $\mathbb{Z}+\frac{1}{2}$ with antiperiodic boundary conditions. The one-loop approximation to the torus partition function gives
\beq
	\mathcal{Z}_{\rm 1-loop} = q^{-\frac{C}{24}} \sqrt{\frac{\prod_{\nu,n}(\nu-n \tau)}{\prod_{m,p}(m-p\tau)}}\,,
\eeq
which leads to
\beq
	\mathcal{Z}_{\rm 1-loop} = q^{-\frac{c_{\rm NS}}{24}} \prod_{n=2}^{\infty} \frac{1\mp q^{n-\frac{1}{2}}}{1-q^n}\,, \qquad c_{\rm NS} = C+\frac{15}{2}\,,
\eeq
where $\pm$ refers to the fermion boundary condition around the ``time'' circle. These are the characters of the Neveu-Schwarz vacuum with a renormalized central charge $c_{\rm NS}$,
\beq
	\mathcal{Z}_{\rm R-NS} = \text{tr}_{\rm NS}\left((-1)^F q^{L_0-\frac{c}{24}}\right)\,, \qquad \mathcal{Z}_{\rm NS-NS}=\text{tr}_{\rm NS}\left( q^{L_0-\frac{c}{24}}\right)\,.
\eeq

Recall that an important observable in the quantization of the orbits of the Virasoro group was the bilocal operator, which we viewed as a reparameterized two-point function. In Section~\ref{S:blocks} we gave significant evidence that correlation functions of these bilocals are Virasoro blocks.  The natural guess for bilocal probes of the supersymmetric theory studied here is, on the infinite plane,
\beq
	\mathcal{B}(h;z_i,\theta_i) = \left( \frac{\mathcal{V}_1\mathcal{V}_2}{(\tilde{\phi}_1 - \tilde{\phi}_2 -\tilde{\theta}_1\tilde{\theta}_2)^2}\right)^h\,, \qquad \mathcal{V} = \tilde{\phi}' + \tilde{\theta} \tilde{\theta}'\,,
\eeq
although we do not know for sure if this guess is correct.

There is some literature on the quantization of coadjoint orbits of super-Virasoro groups with extended SUSY, e.g.~\cite{0253-6102-16-3-295,Aoyama:2018lfc,Aoyama:2018voj}. The quantization of the orbits of the $\mathcal{N}=2$ super-Virasoro group in particular is related to the $\mathcal{N}=2$ supersymmetric version of the SYK model presented in~\cite{Fu:2016vas}. Although it is rather interesting we do not pursue it further here.

\section{Discussion}
\label{S:discussion}

There is common lore that the dual to pure AdS$_3$ gravity is Liouville theory. This lore can be traced back to~\cite{Coussaert:1995zp}. The claim cannot be correct as stated, as we discussed in the Introduction. We have revisited the matter, and after careful treatment have shown that pure gravity on global AdS$_3$ can be rewritten as a cousin of Liouville theory, as anticipated by various authors~\cite{Barnich:2017jgw,Mertens:2018fds}. This cousin is two copies of the path integral quantization of a certain coadjoint orbit of the Virasoro group. This path integral quantization has appeared in the literature before, going back nearly 30 years to a paper of Alekseev and Shatashvili~\cite{Alekseev:1988ce}. However, relatively little has been computed with it using standard path integral techniques. 

In this manuscript we have carefully obtained this theory from gravity, paying close attention to the quotient and boundary conditions. Our analysis complements that of Maloney and Witten, who arrived at a theory of boundary gravitons which is a geometric quantization of the same coadjoint orbit.  We derived related models for gravity on other spacetimes, including conical deficits and two-sided BTZ geometries. These models are all weakly coupled at large central charge, and one virtue of our reformulation is that it may be used to compute loop-level physics in the dual quantum gravity. We have developed in detail the technology to compute observables in perturbation theory in $1/c$, namely the sphere and torus partition functions and Virasoro blocks. Along the way we computed several effects that are one-loop in the bulk, including the ``heavy-light'' block to $O(1/c)$. Our calculations appear to be substantially simpler than those of previous approaches, like the heat kernel computation of the one-loop torus partition function in~\cite{Giombi:2008vd} or the algebraic approach to the heavy-light block in~\cite{Fitzpatrick:2015dlt}.

In the remainder of this discussion we comment on some of the curious aspects of our results, and outline some speculations and directions for future research.

It has long been known that the quantization of the coadjoint orbits of a group provides a computational engine for calculating group-theoretic quantities.  Since the defining data of two-dimensional CFT, like blocks or characters, is determined by representation theory, it is not surprising that the technology of quantized coadjoint orbits computes them.  What is surprising to us is that the geometric actions in this paper allow for much easier calculations of CFT data, such as Virasoro blocks at large central charge, than those in the existing literature. This diagrammatic approach to computing CFT data begs further exploration. Concretely, in Section~\ref{S:blocks} we computed the ``heavy-light'' block to $O(1/c)$ from a one-loop computation. We expect that higher loops compute higher-order terms in the $1/c$ expansion. Besides verifying this explicitly, by comparing with known results for the blocks~\cite{Chen:2016cms}, there must be ``non-perturbative'' contributions to the blocks that are suppressed as $\exp(-c)$. Are there instanton configurations in our theory which account for these?

Relatedly, our theory should prove useful for computing the large $c$ limits of Virasoro blocks for higher-point functions, about which relatively little is known. For example, the three-point function of bilocal operators should compute the Virasoro six-point identity block.

The rewriting of AdS$_3$ gravity as a boundary field theory is analogous to the rewriting of Jackiw-Teitelboim theory on nearly-AdS$_2$ spacetimes as the Schwarzian theory.  In both cases, the natural way to parameterize the boundary excitations is as reparameterization fields, and there are bilocal probes which, at least in nearly-AdS$_2$, correspond to matter fields in the gravity dual.  Furthermore, certain partition functions of the geometric actions considered here as well as that of the Schwarzian theory are one-loop exact~\cite{Stanford:2017thb}. As emphasized in~\cite{Harlow:2018tqv}, Jackiw-Teitelboim theory on nearly-AdS$_2$ spacetime is a self-consistent theory of quantum gravity which is capable of supporting wormholes but is devoid of black hole microstates. Our geometric actions behave analogously for AdS$_3$ gravity, and indeed our analysis seems to verify the conjectures of~\cite{Harlow:2018tqv} regarding non-factorization in pure AdS$_3$ gravity.

Another shared feature of nearly-AdS$_2$ and AdS$_3$ gravity is the following. The Lorentzian version of the Schwarzian quantum mechanics has exponentially growing modes which are pure ``gauge'': they are removed by the $PSL(2;\mathbb{R})$ quotient of the model.  These modes do not appear in time-ordered correlation functions, but they do appear in out-of-time-ordered correlation functions and are responsible for the maximal Lyapunov exponent $2\pi/\beta$. Similar exponentially growing modes are also present in our geometric actions -- these modes are likewise pure ``gauge,'' but are unleashed by out-of-time-ordered correlators, leading to the maximal Lyapunov exponent one finds from the identity block~\cite{Roberts:2014ifa}.

The $PSL(2;\mathbb{R})$-invariant bilocal operators were crucial in this manuscript. For example, we presented significant evidence that the connected two-point function of bilocals is the Virasoro identity block. What is the gravity dual of a bilocal probe? The bilocal operators in the Schwarzian theory are well-understood in nearly-AdS$_2$ gravity: they correspond to boundary insertions of sources for operators dual to bulk matter fields. Our intuition is that, in our geometric model, the probe is dual to a bulk Wilson line which ends on the AdS boundary at the two points of the bilocal. However we do not yet have a firm argument that this is the case. If this proposal is correct, then perhaps the machinery we developed in Section~\ref{S:blocks} can be used to clarify previous speculations (e.g.~\cite{Fitzpatrick:2016mtp}) regarding bulk Wilson lines and blocks. We expect the analysis of the recent work~\cite{Blommaert:2018oro} to be especially useful.

Using the AdS/CFT dictionary we obtained the curved space version of the quantization of Diff($\mathbb{S}^1)/PSL(2;\mathbb{R})$, which we then used to compute the one-loop approximation to the sphere partition function of AdS$_3$ gravity. It would be interesting to use this curved space version to study higher-genus Euclidean observables and quantum corrections thereof. One such observable is the position-space R\'enyi entropy, interpreted as the partition function on a branched cover of a Riemann surface.

In a sense, the quantization of Diff$(\mathbb{S}^1)/PSL(2;\mathbb{R})$ considered in this text is an exotic sort of hydrodynamics, at least according to a particle-physics-oriented definition of hydrodynamics as a theory of conserved quantities. The only local operators of the model are built from its stress tensor, and the field equation is merely its conservation. So the action of the model is, in that precise sense, an action for (non-dissipative) hydrodynamics. However this action has little obvious relation to other effective actions for fluids recently developed in e.g.~\cite{Crossley:2015evo,Haehl:2015uoc,Jensen:2017kzi}. Those systems are also theories of reparameterizations, however the symmetry principles underlying them are rather different. In those theories the analogue of the $\phi$-space has a fixed thermal vector. There is no such geometric interpretation for our model.

However, it may still be possible to arrive at the actions in this manuscript by some other effective field theory-like arguments. After all, our theory is minimal: it is just a Wess-Zumino term for the anomalies of a chiral CFT, and the quantum theory has the minimum possible spectrum consistent with conformal (but not modular) invariance. Indeed, the authors of~\cite{Blake:2017ris} have recently proposed effective actions for energy fluctuations beyond the gradient expansion which seem capable of hosting the quasi-local $PSL(2;\mathbb{R})$ symmetry so crucial in this manuscript. Can our theory be understood as an example of the ``hydrodynamics'' considered there? More generally, since our theory describes the dynamics of pure AdS$_3$ gravity, one expects on physical grounds that our theory serves as an effective description of two-dimensional CFTs satisfying all of the requisite properties required for a semiclassical gravity dual, namely large central charge, a sparse spectrum, and a large higher-spin gap. This is the outlook taken by the authors of~\cite{Haehl:2018izb}. From that point of view it would be useful to quantify the applicability of our geometric model as an effective description.

The BMS$_3$ asymptotic symmetry of flat-space gravity has been successfully obtained from a suitable flat-space limit of AdS$_3$ gravity~\cite{Bagchi:2013toa,Barnich:2012aw,Andrade:2015fna}. This raises the question: what is the flat space limit of our construction, and what does it have to do with soft limits in three dimensional Minkowski space? Or for that matter with coadjoint orbits of BMS$_3$? We expect that the quantization of Diff$(\mathbb{S}^1)/PSL(2;\mathbb{R})$ suitably encodes the dressing of matter $S$-matrix elements by soft quanta, reminiscent of the dressing of bilocal probes that gave us Virasoro blocks in Section~\ref{S:blocks}. We note that there has already been significant effort devoted toward this end, including~\cite{Barnich:2012rz,Oblak:2016eij,Barnich:2017jgw}.

There has been a recent explosion of activity on the so-called $T\bar{T}$ deformation of Zamolodchikov and Smirnov~\cite{Zamolodchikov:2004ce,Smirnov:2016lqw}, including the question of its holographic dual (see e.g.~\cite{McGough:2016lol}). Here we simply note that, as far as pure Einstein gravity is concerned, one can study the $T\bar{T}$ deformation directly by a suitable deformation of the theory of boundary gravitons obtained in this manuscript, just as one can study the Lorentz-breaking $J\bar{T}$ deformation~\cite{Guica:2017lia} of Einstein gravity with a $U(1)$ Chern-Simons theory.

One loose end left by our work is a puzzle regarding the quantization of parameters. In the path integral quantization of coadjoint orbits, well-definedness of the path integral requires quantized periods of the symplectic flux. For the quantization of affine Lie groups, which leads to chiral WZW models, this condition is the usual one that the level must be integer. We did not find such a quantization condition when treating the orbits of the Virasoro group. We are not sure if this should be surprising. After all, we find the quantization of Diff$(\mathbb{S}^1)/PSL(2;\mathbb{R})$ directly from the Chern-Simons formulation of AdS$_3$ gravity. The latter has gauge group $SO(2,1)\times SO(2,1)$ (and not a cover thereof, as we discussed in Section~\ref{S:AdS}). The bare central charge $C$ characterizing the coadjoint orbit is related to the $SO(2,1)$ Chern-Simons level as $ C=24k$, and one expects the latter to be integral~\cite{Witten:2007kt}. Here we note that if $C$ is indeed a multiple of 24, then the exact chiral central charge $c = C+13$ is not.

Another loose end concerns an analogy with nearly AdS$_2$ gravity. Kitaev~\cite{Kitaev:2017awl} (see also~\cite{Maldacena:2017axo}) has shown that one way to arrive at the Schwarzian theory is to consider the problem of a charged particle on the hyperbolic disc in the presence of a magnetic field, in a limit were the mass is tuned so that the particle remains near the boundary of the disc. The na\"{i}ve computation of the particle's action gives a Liouville quantum mechanics, rather than the Schwarzian theory. However, after appropriately quotienting out by the $PSL(2;\mathbb{R})$ isometry of the disc, one may use a field redefinition to rewrite the quantum mechanics as the Schwarzian theory. Is there a similar way to arrive at our geometric model from a string propagating on AdS$_3$ in the presence of a uniform NS flux? A na\"{i}ve computation leads to Liouville theory, rather than our action. Perhaps, after passing to Hamiltonian form and quotienting out by the $PSL(2;\mathbb{R})\times PSL(2;\mathbb{R})$ isometry of AdS$_3$, one lands on the geometric model instead.

\acknowledgments

We would like to thank N.~Benjamin, A.~Fitzpatrick, D.~Grumiller, T.~Hartman, S.~Kachru, J.~Kaplan, P.~Kraus, J.~Maldacena, A.~Maloney, G.~Mandal, S.~Rezchikov, M.~Rozali, N.~Seiberg, H.~Verlinde, and E.~Witten for useful and enlightening discussions. J.~C. is supported by the Fannie and John Hertz Foundation and the Stanford Graduate Fellowship program.  K.~J. is supported in part by the Department of Energy under grant number DE-SC0013682.

\bibliographystyle{JHEP}
\bibliography{refs}

\end{document}